\documentclass{aa}
\usepackage[utf8]{inputenc} 
\usepackage{graphicx} 
\usepackage[varg]{txfonts}
\usepackage{xcolor} 
\usepackage{natbib}
\usepackage{rotating}
\usepackage{hyperref}
\usepackage{textcomp}
\usepackage[export]{adjustbox}

\pdfpageattr{/Group << /S /Transparency /I true /CS /DeviceRGB>>}
\bibpunct{(}{)}{;}{a}{}{,} 

\newcommand{\emm}[1]{\ensuremath{#1}}
\newcommand{\emr}[1]{\emm{\mathrm{#1}}}

\newcommand{\nH}{\emr{n_H}} 
\newcommand{\Av}{\emr{A_\mathrm{V}}}

\newcommand{\jp}[1]{{\color{magenta} #1}}

\definecolor{ochre}{rgb}{0.8, 0.47, 0.13}

\newcommand{\FigRankingSingleRatiosTranslucent}{
  \begin{figure*}
    \includegraphics[width=0.5\linewidth,valign=t]{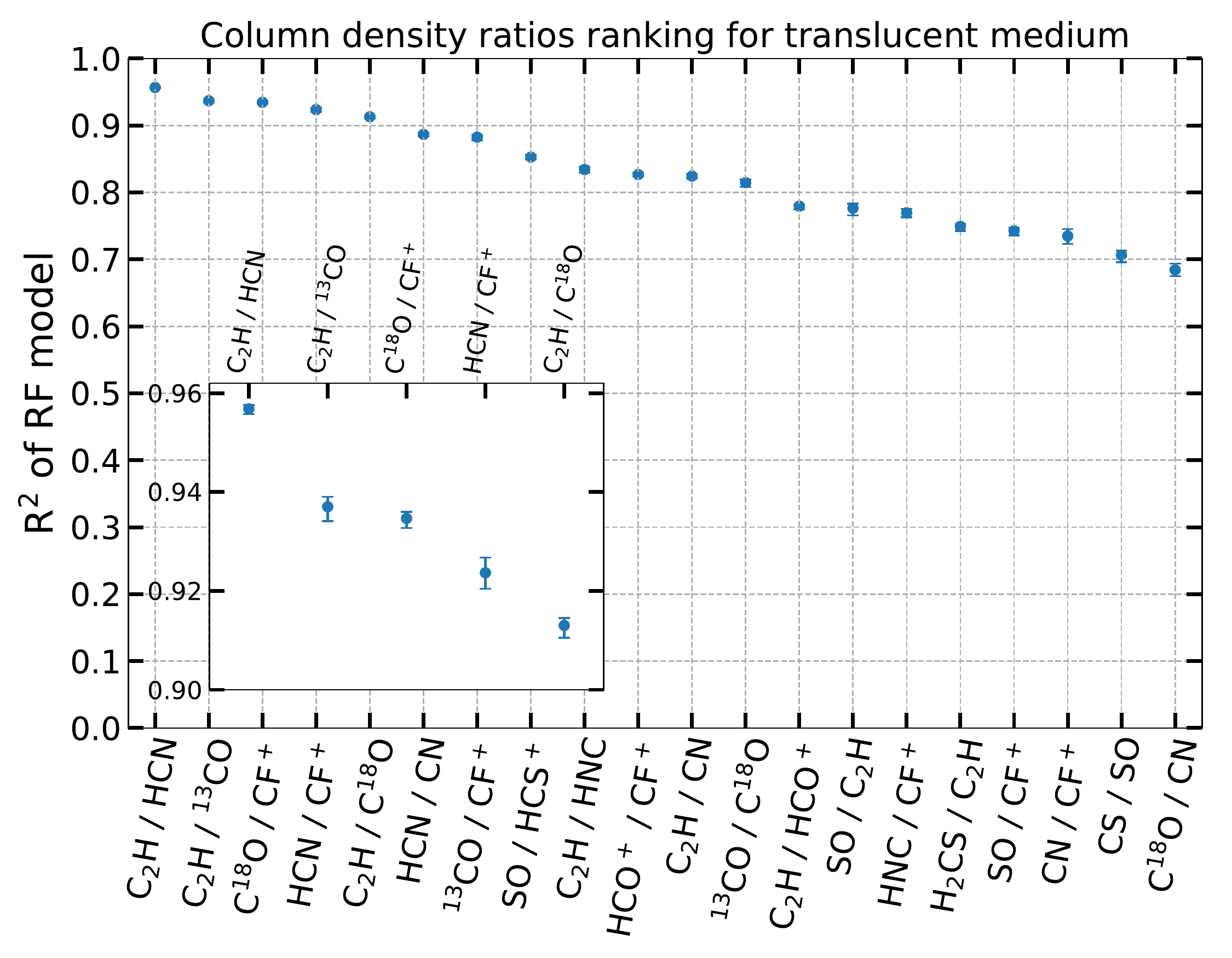}
    \includegraphics[width=0.5\linewidth,valign=t]{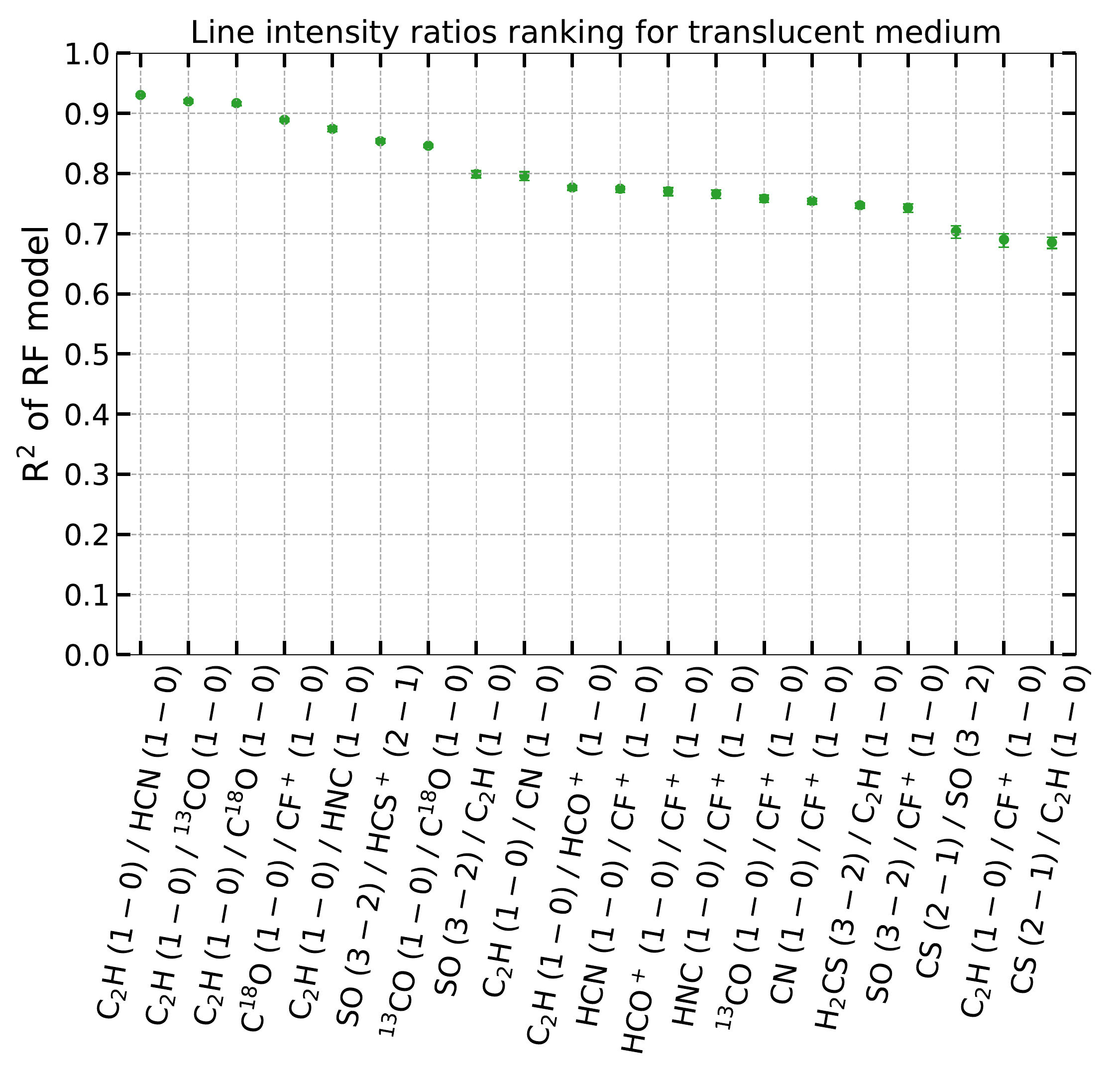}
    \caption{Ranking of column density ratios (left) or line intensity ratios (right) of observable tracers by order of the predictive power for predicting the ionization fraction (measured by the $R^2$ coefficient), in the case of translucent medium conditions (showing only the first 20). 
Errorbars of the $R^2$ estimates are computed by cross-validation (see text for explanations). The inset in the left panel shows a zoom on the first five ratios in order to make the magnitude of the errorbars visible.}
    \label{fig:RankingSingleRatiosTranslucent}
  \end{figure*}
}

\newcommand{\FigRFmodelBestRatiosTranslucent}{
  \begin{figure*}
    \includegraphics[width=0.5\linewidth]{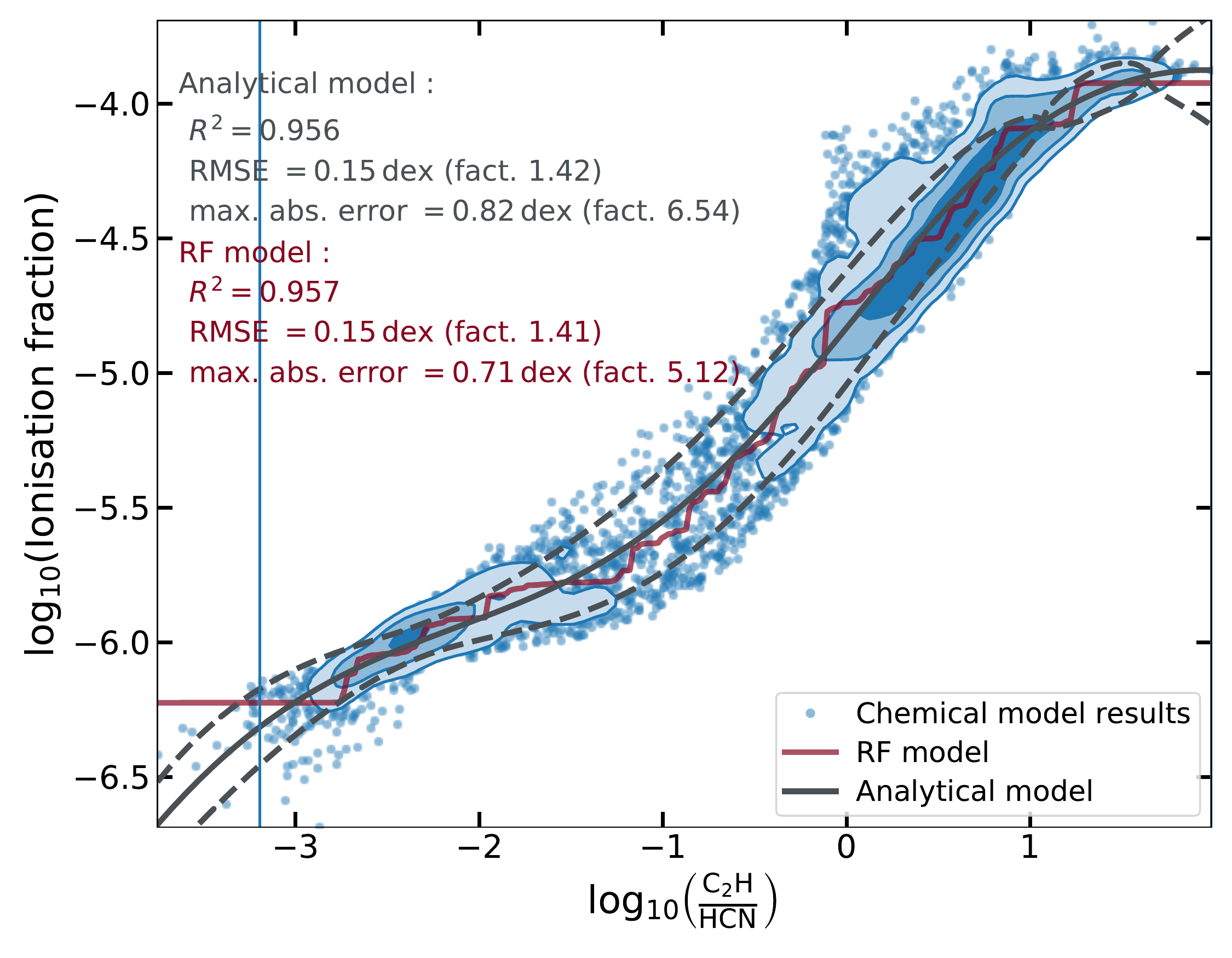}
    \includegraphics[width=0.5\linewidth]{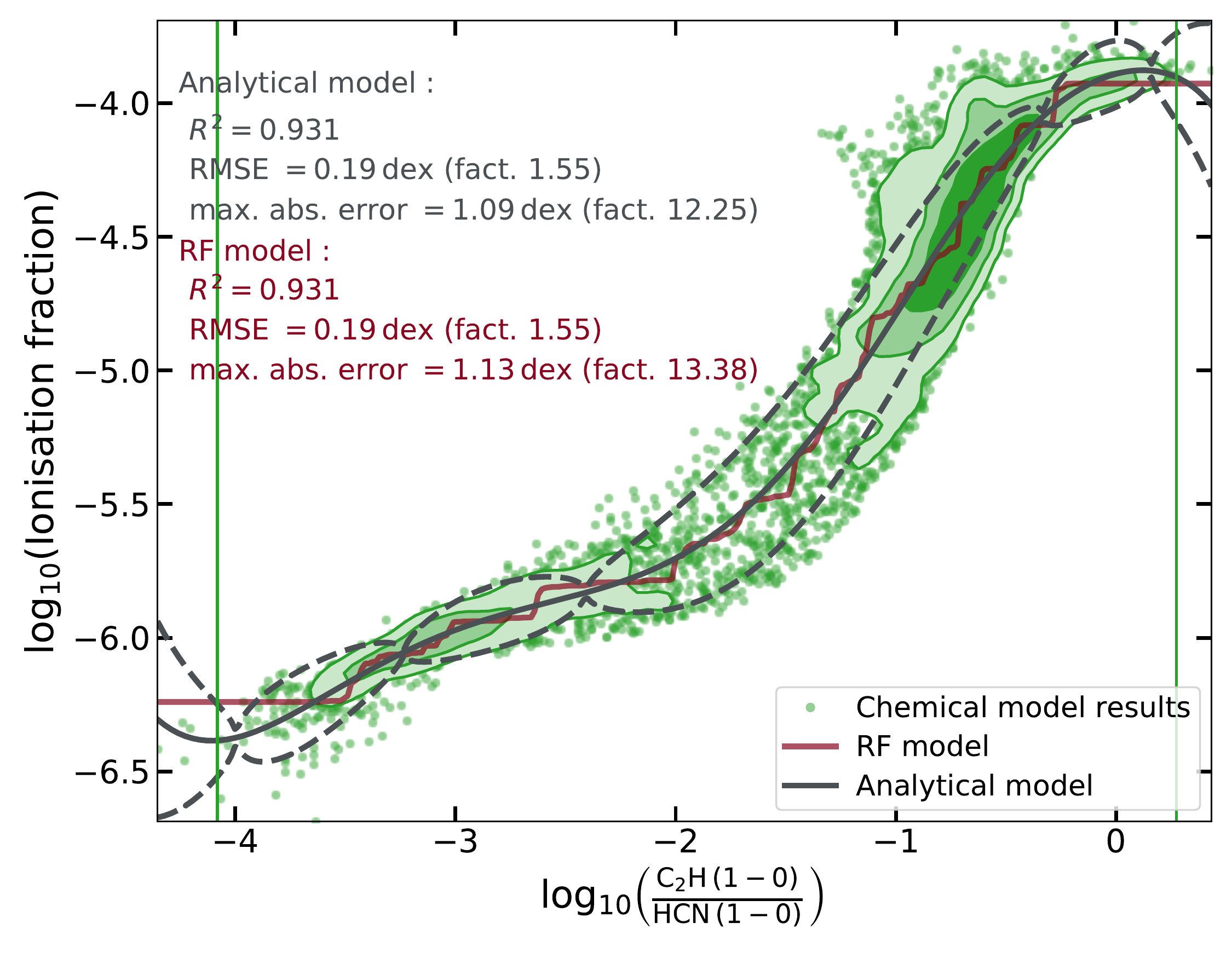}
    \caption{Ionization fraction versus the best column density ratio, C$_2$H/HCN (left panel), and the best line intensity ratio, C$_2$H (1-0) / HCN (1-0) (right panel) for tracing the ionization fraction in translucent medium conditions. 
The model grid is shown as a scatter plot, with the central crowded regions replaced by PDF isocontours containing 25\%, 50\%, and 75\% of the points. 
Superimposed are the RF model (red line), the analytical fit (solid black line, presented in Sect.~\ref{sect:AnalyticalFits}), the analytical fit of the 1$\sigma$ uncertainty (dashed black lines, presented in Sect.~\ref{sect:AnalyticalFits}), and the bounds of the validity range of the analytical fit (vertical lines, presented in Sect.~\ref{sect:AnalyticalFits}).
The quality estimates of the two models are indicated on the figure.}
    \label{fig:RFmodelBestRatiosTranslucent}
  \end{figure*}
}

\newcommand{\FigRankingSingleRatiosColdDense}{
  \begin{figure*}
    \includegraphics[width=0.5\linewidth,valign=t]{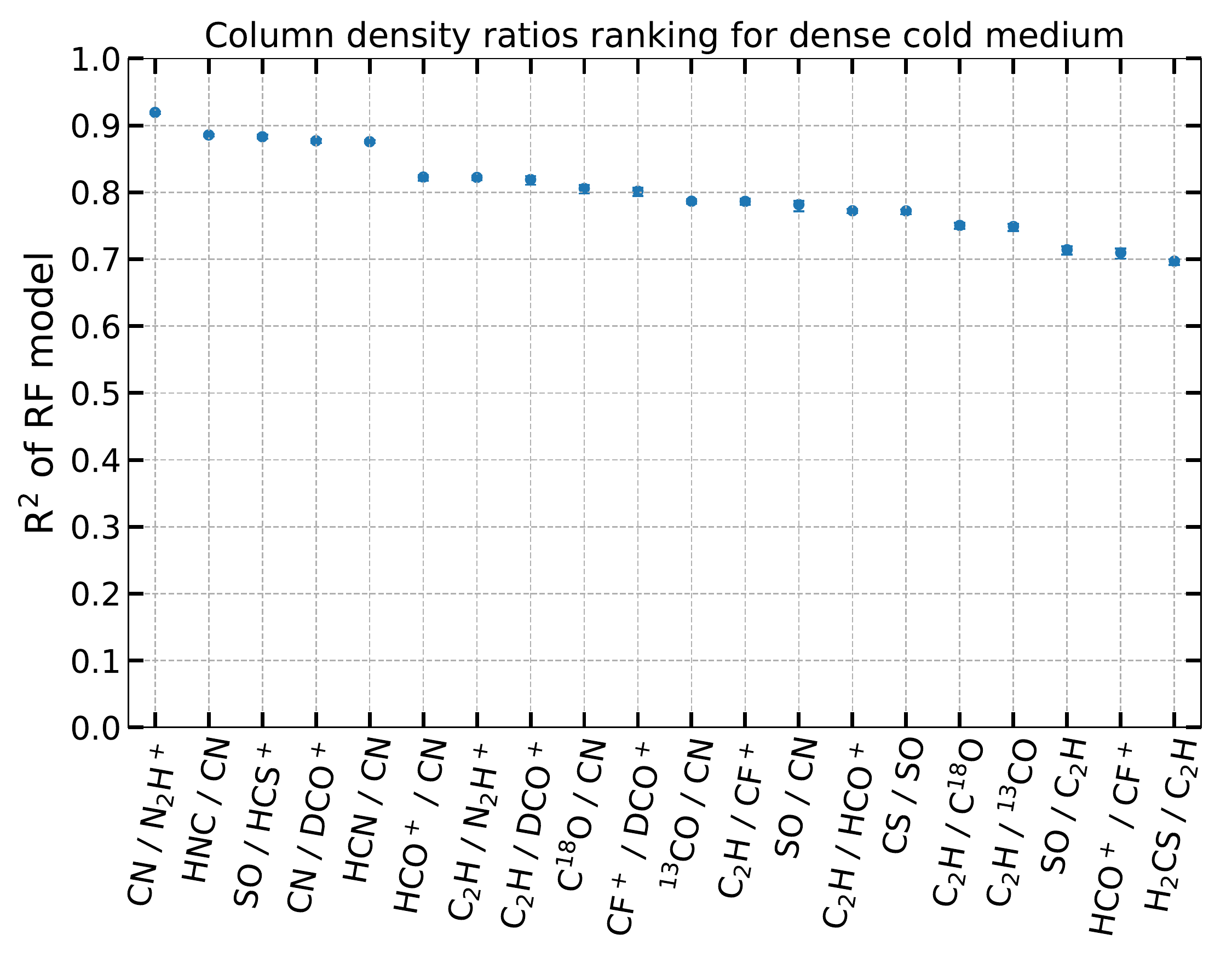}
    \includegraphics[width=0.5\linewidth,valign=t]{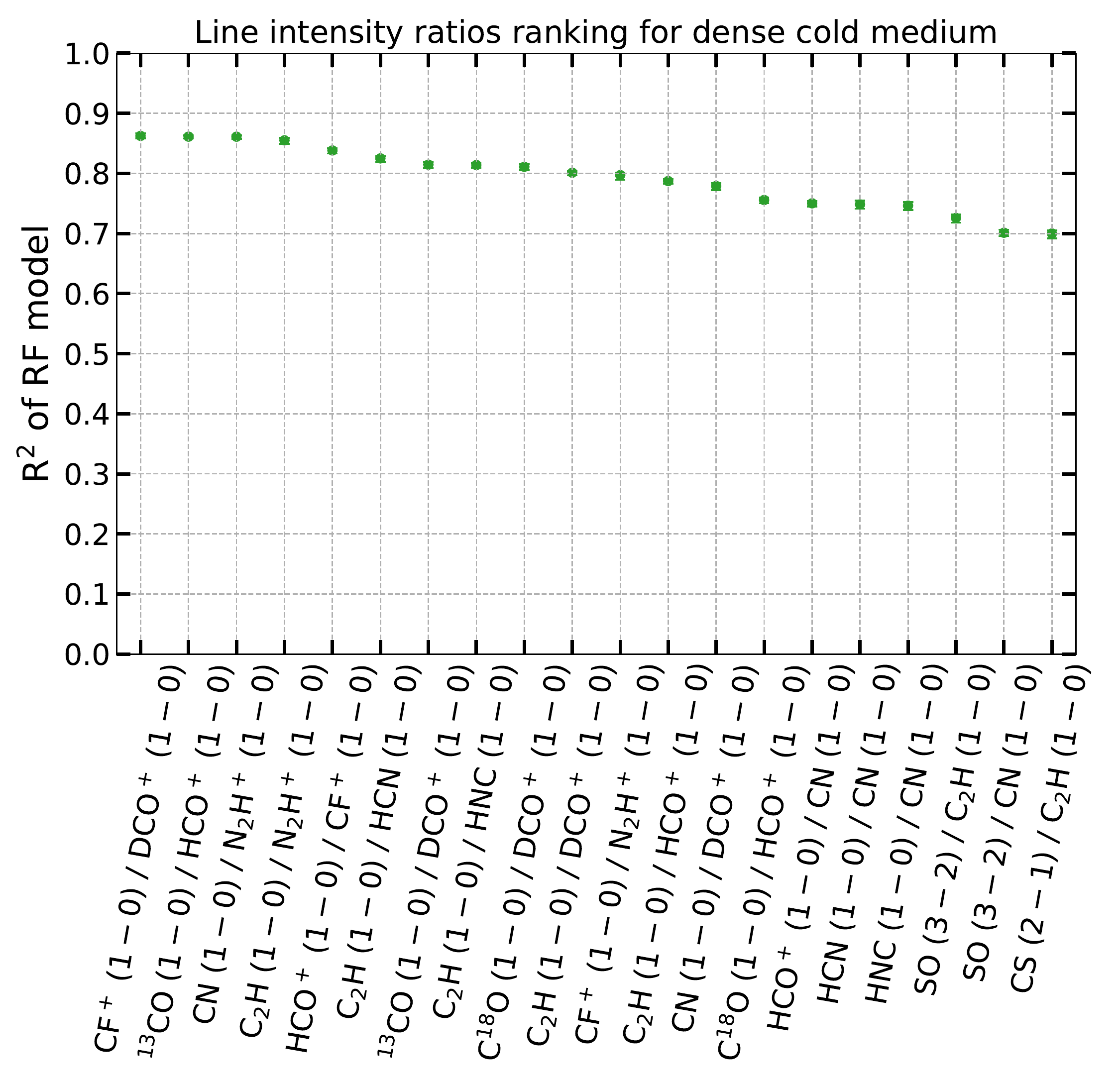}
    \caption{Ranking of column density ratios (left) or line intensity ratios (right) of observable tracers by order of the predictive power for predicting the ionization fraction (measured by the $R^2$ coefficient), in the case of dense cold medium conditions (showing only the first 20). 
Errorbars of the $R^2$ estimates are computed by cross-validation (see text for explanations).}
    \label{fig:RankingSingleColdenRatiosColdDense}
  \end{figure*}
}

\newcommand{\FigRFmodelBestRatiosDense}{
  \begin{figure*}
    \includegraphics[width=0.5\linewidth]{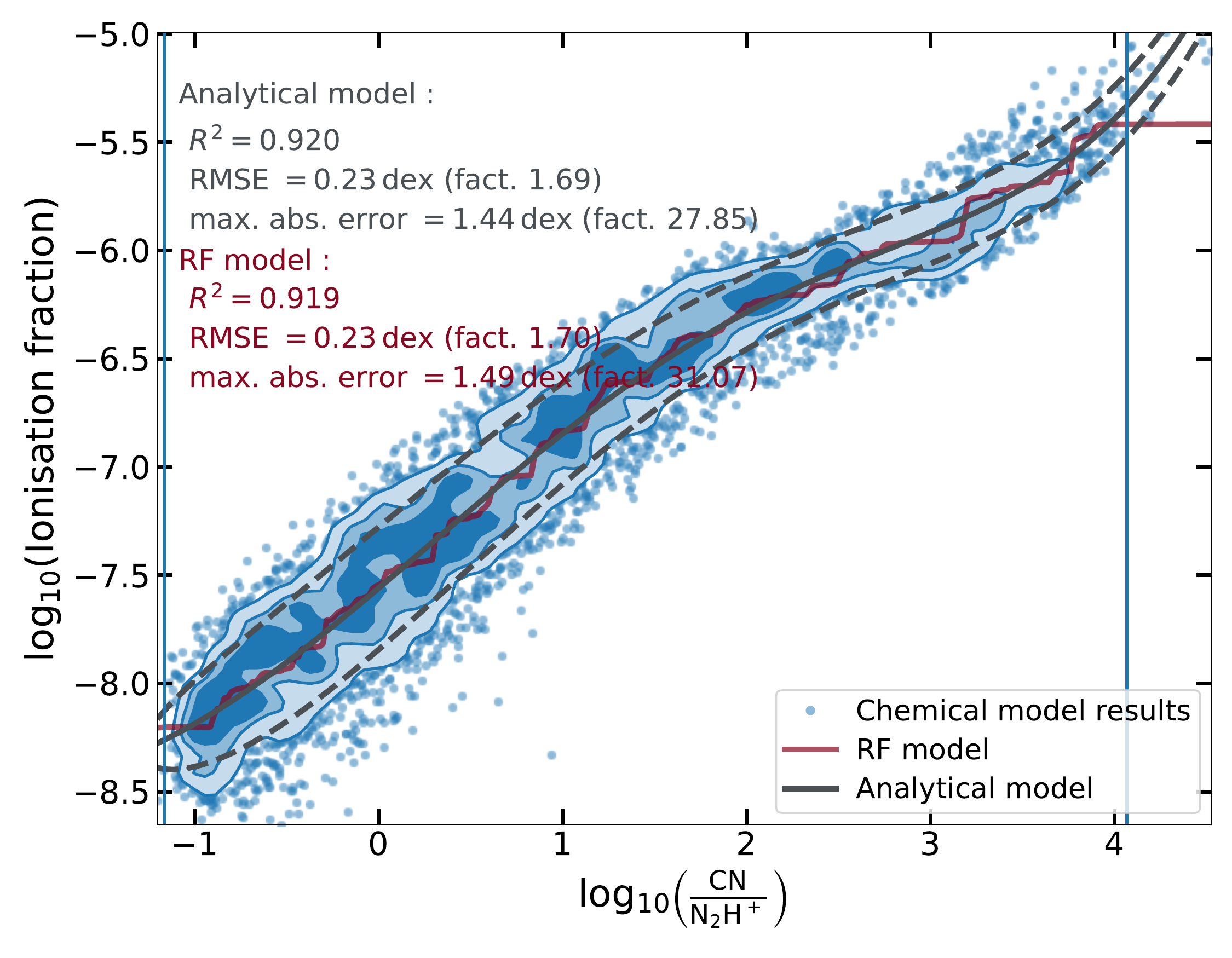}
    \includegraphics[width=0.5\linewidth]{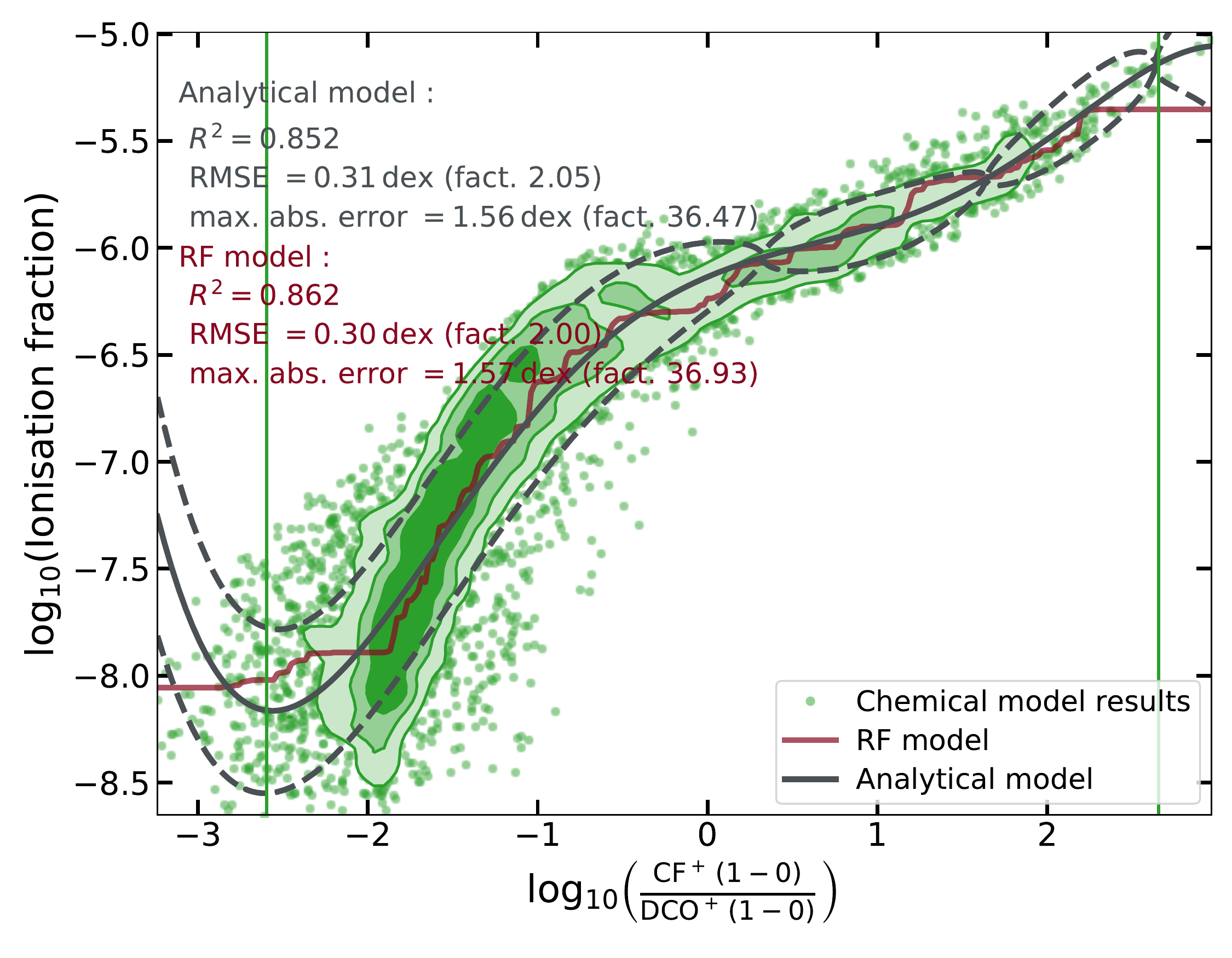}
    \caption{Same as Fig.~\ref{fig:RFmodelBestRatiosTranslucent} for cold dense medium conditions.}
    \label{fig:RFmodelBestRatiosDense}
  \end{figure*}
}

\newcommand{\FigAnalyticalmodelUncertainties}{
  \begin{figure*}
    \includegraphics[width=0.5\linewidth]{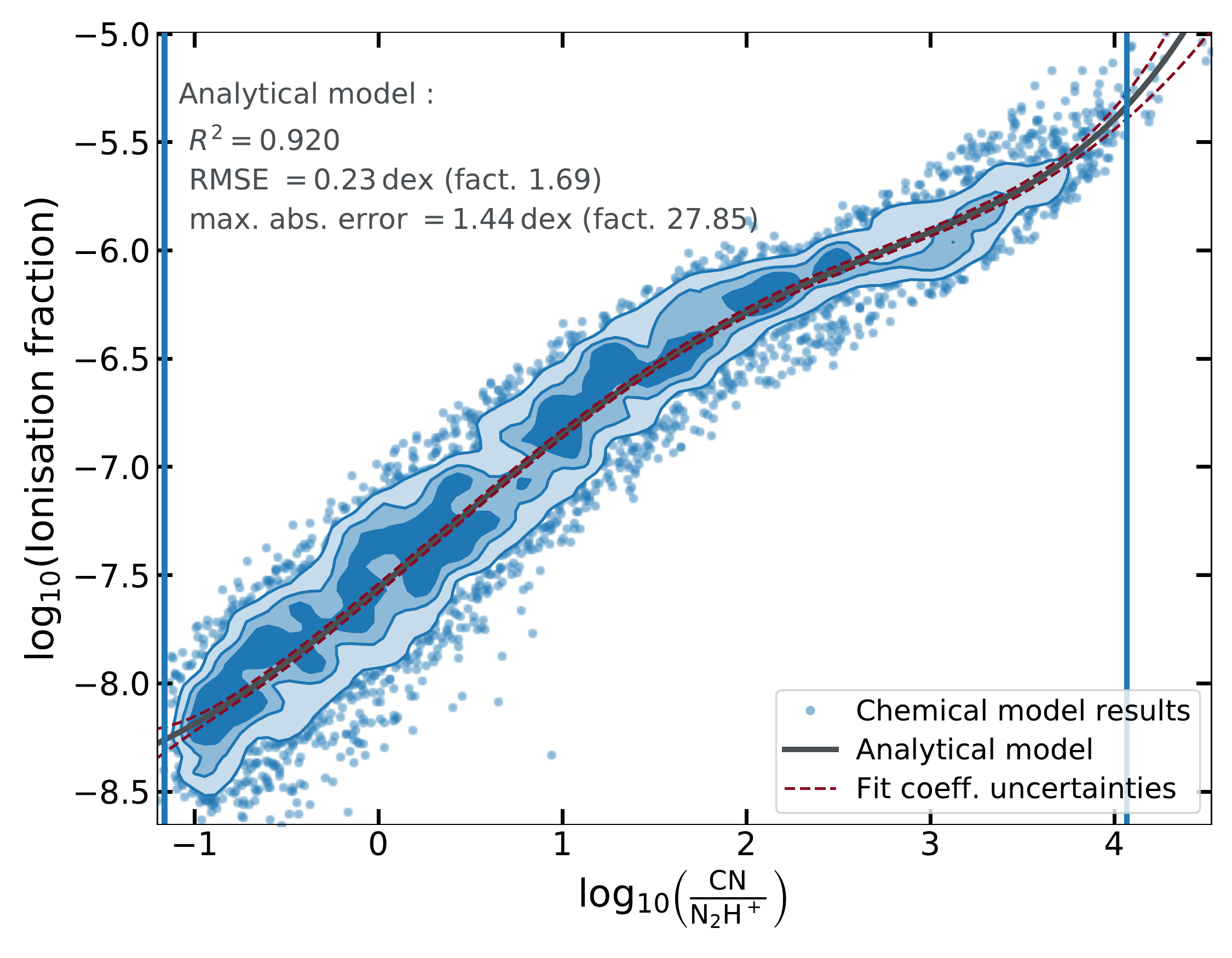}
    \includegraphics[width=0.5\linewidth]{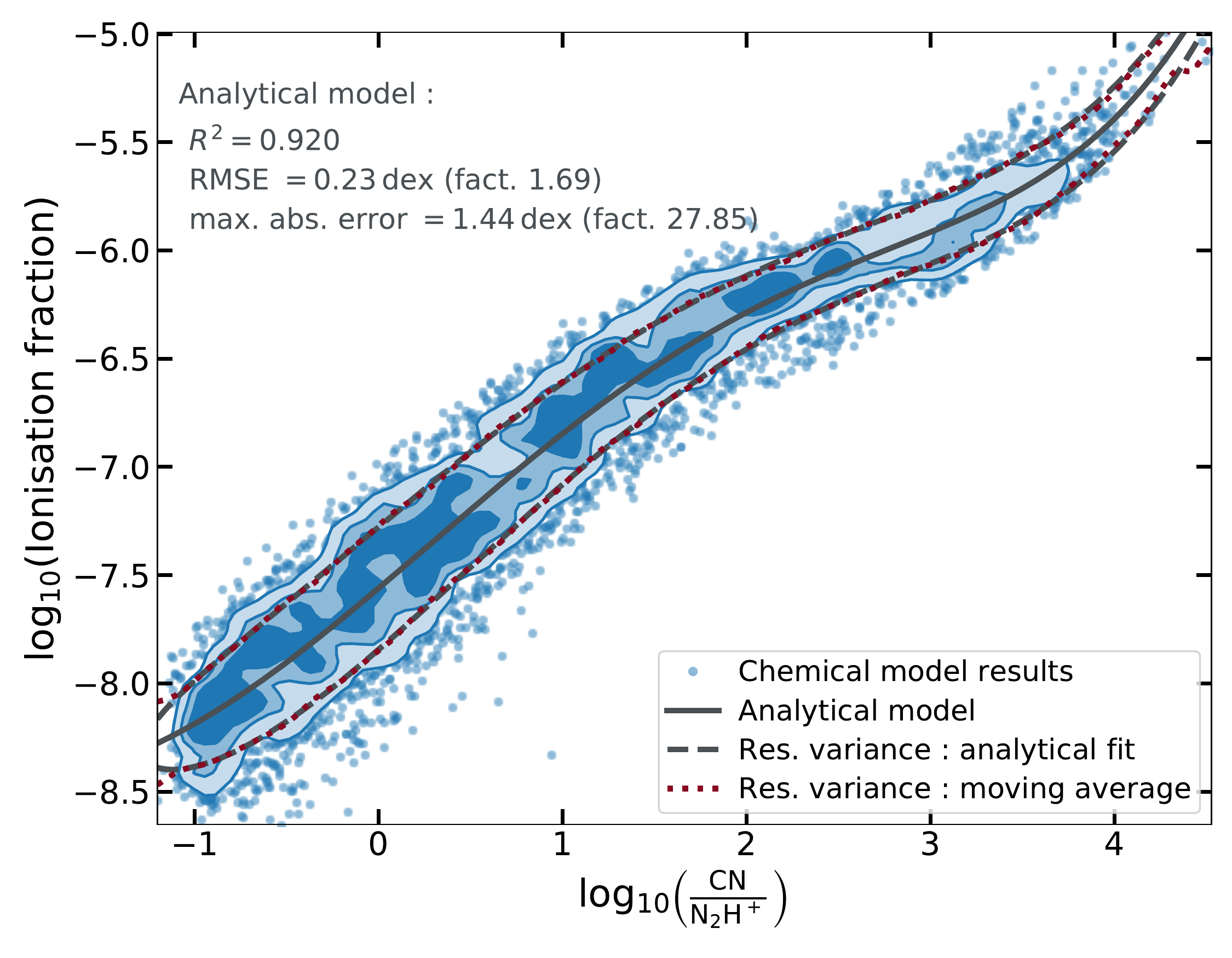}
    \caption{Illustration of the two sources of uncertainties for the best column density ratio for tracing the ionization fraction in dense cold medium, CN/N$_2$H$^+$. The left panel shows the analytical fit (solid black line), the standard deviation around this curve corresponding to the uncertainties on the fit coefficients (thin dashed red lines representing the $3\sigma$ level), and the bounds of the validity range defined in the text (vertical blue lines). The right panel shows the analytical fit (solid black line) and two estimates of the standard deviation corresponding to the residual scatter of the data points around the curve: the red dotted line presents a moving-average estimate of the local standard deviation, and the dashed black lines shows our analytical fit of the residual standard deviation.
On both panels, the chemical model grid is shown as a scatter plot, with the central crowded regions replaced by PDF isocontours containing 25\%, 50\%, and 75\% of the points.}
    \label{fig:AnalyticalmodelUncertainties}
  \end{figure*}
}

\newcommand{\FigAnalyticalmodelBestSixTranslucentColDenRatios}{
  \begin{figure*}
    \includegraphics[width=0.5\linewidth]{figures/Translucent2_oldH2_abratios_1ratio_best0_c2h_over_hcn_residualfit_paper.pdf}
    \includegraphics[width=0.5\linewidth]{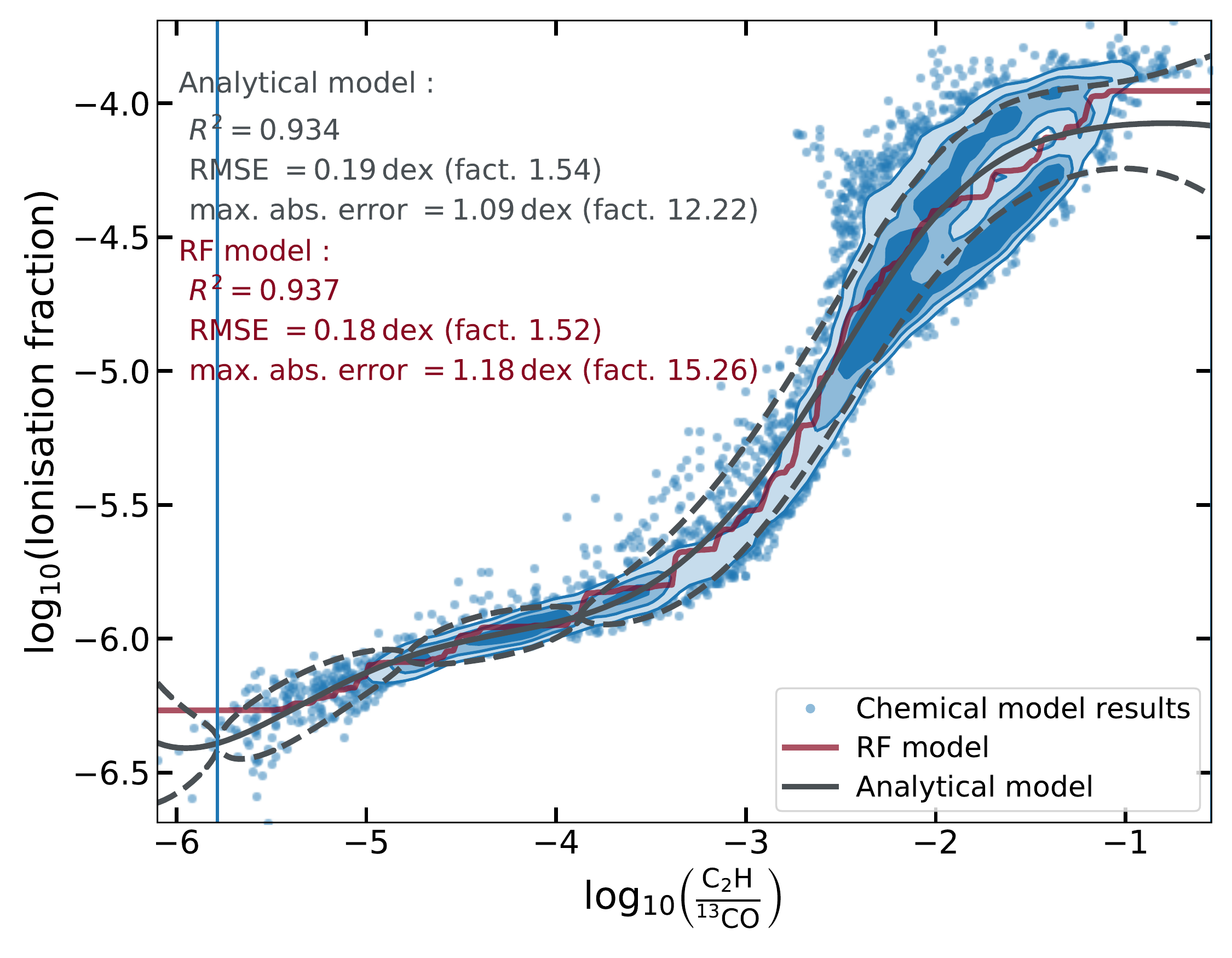}\\
    \includegraphics[width=0.5\linewidth]{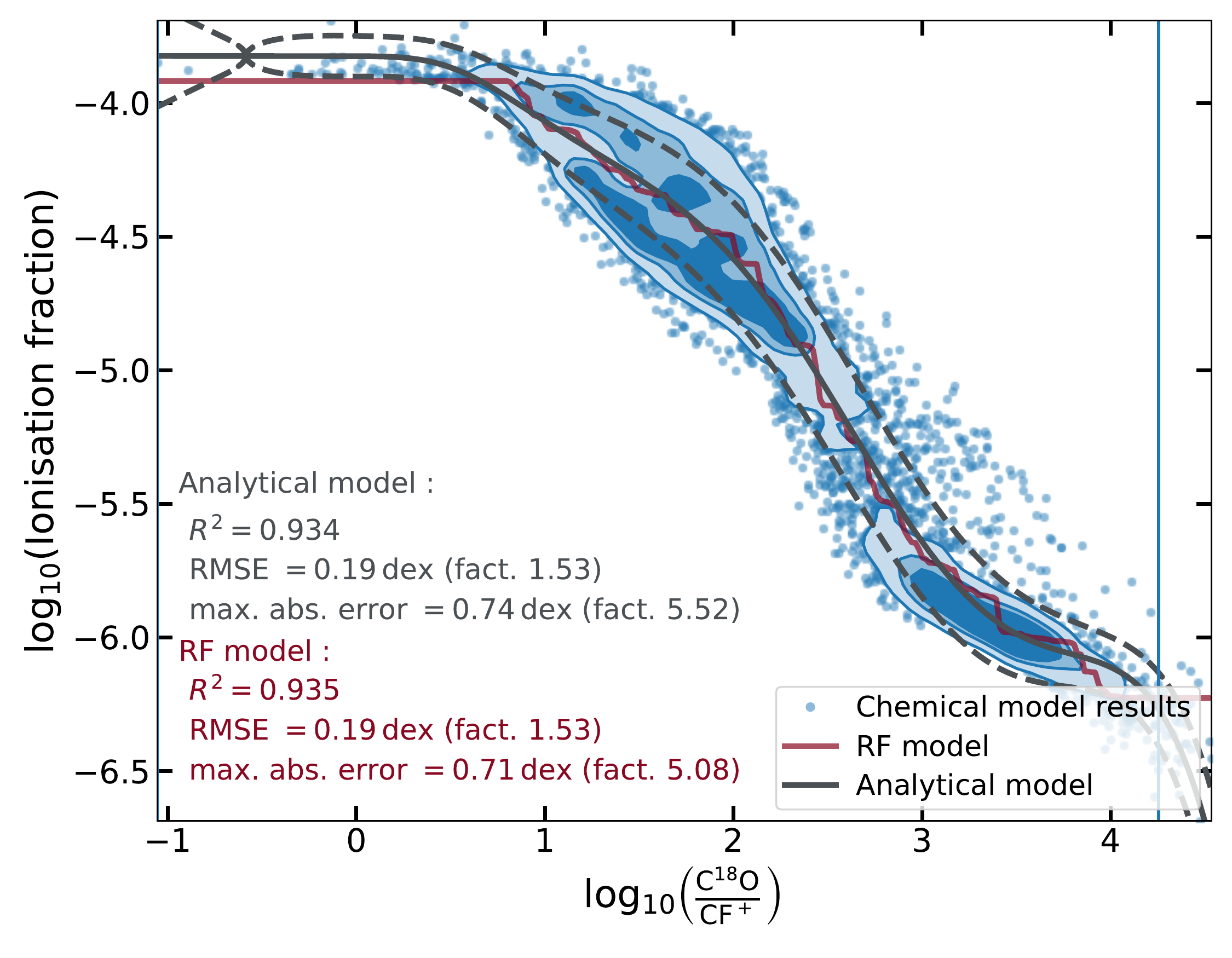}
    \includegraphics[width=0.5\linewidth]{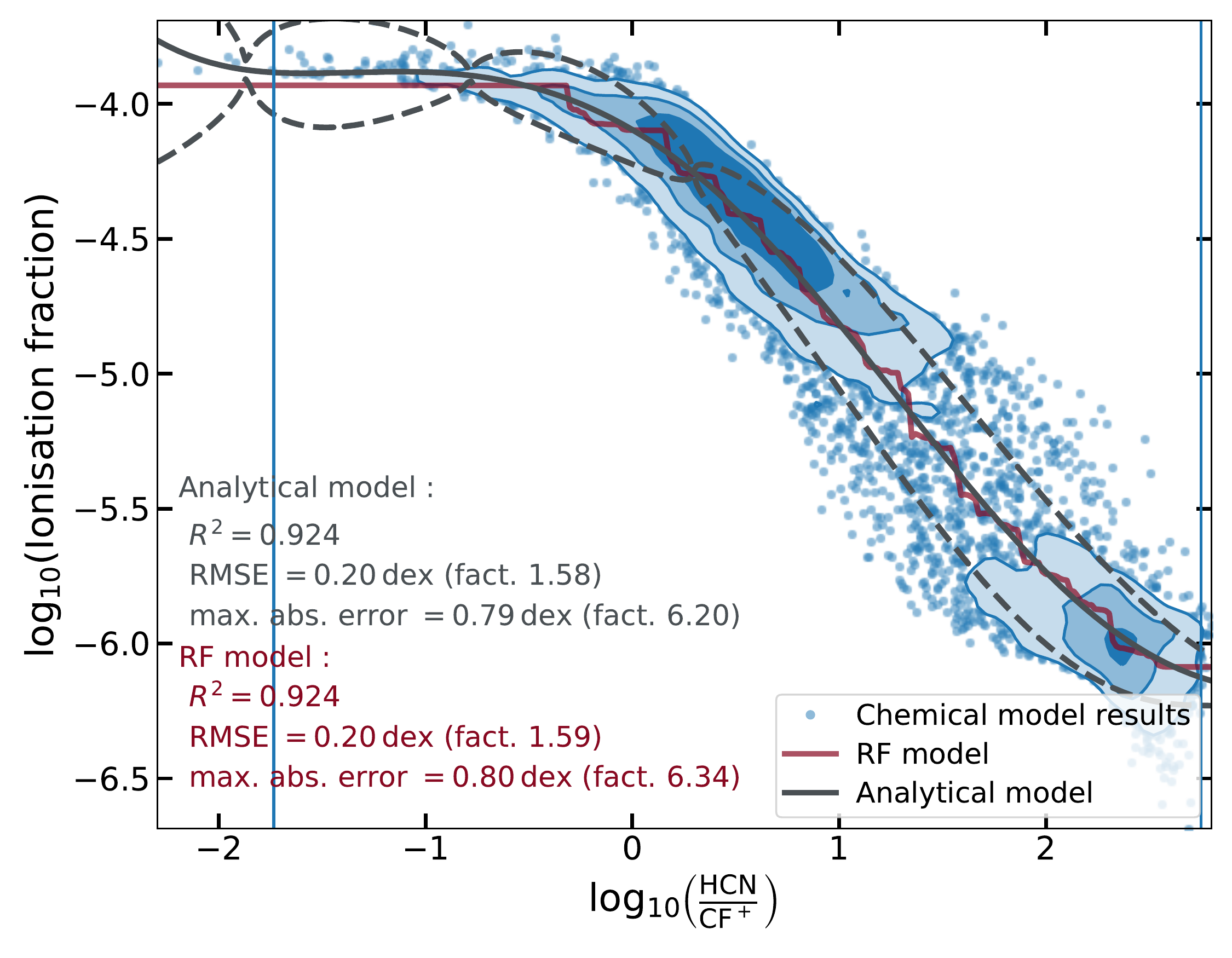}\\
    \includegraphics[width=0.5\linewidth]{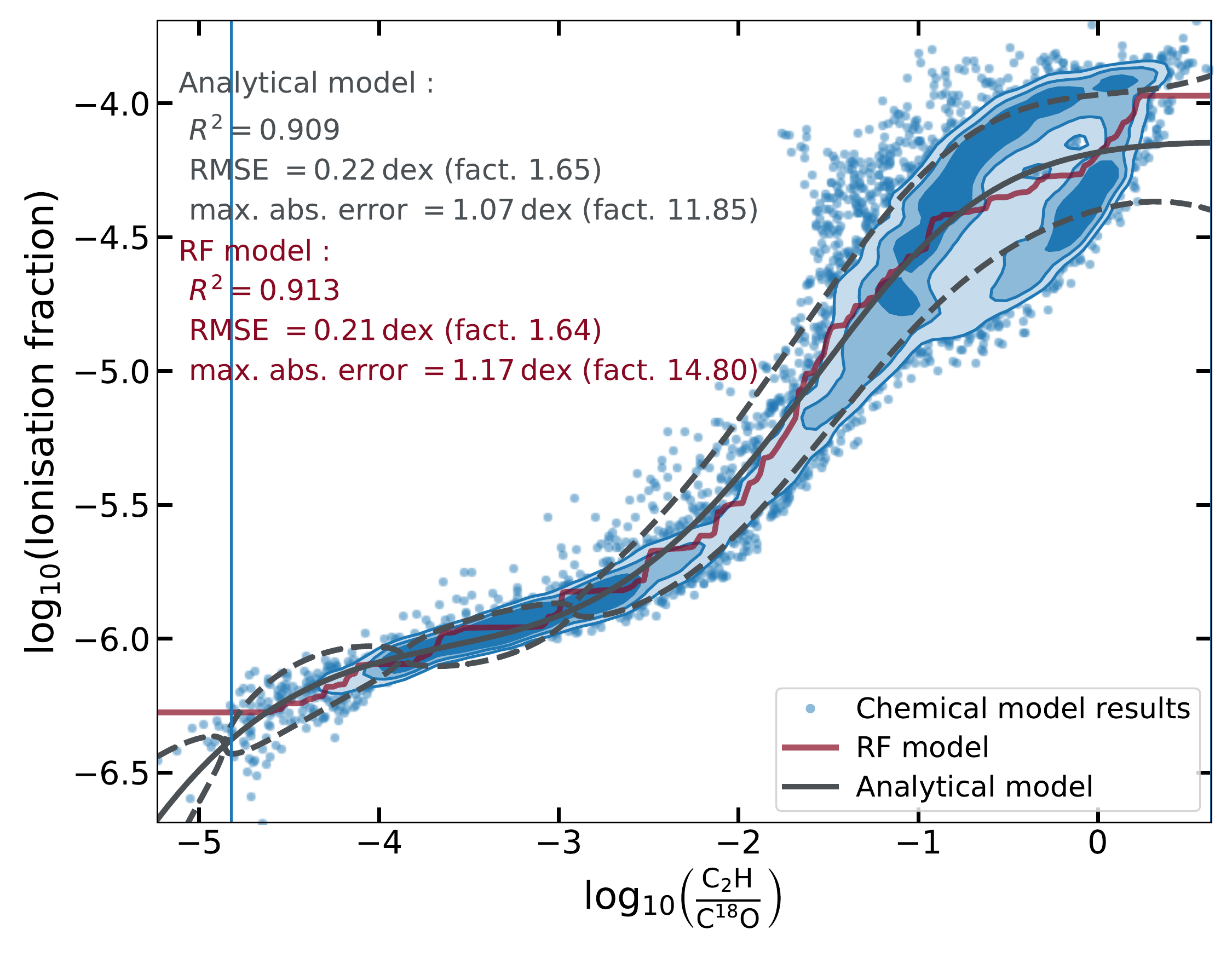}
    \includegraphics[width=0.5\linewidth]{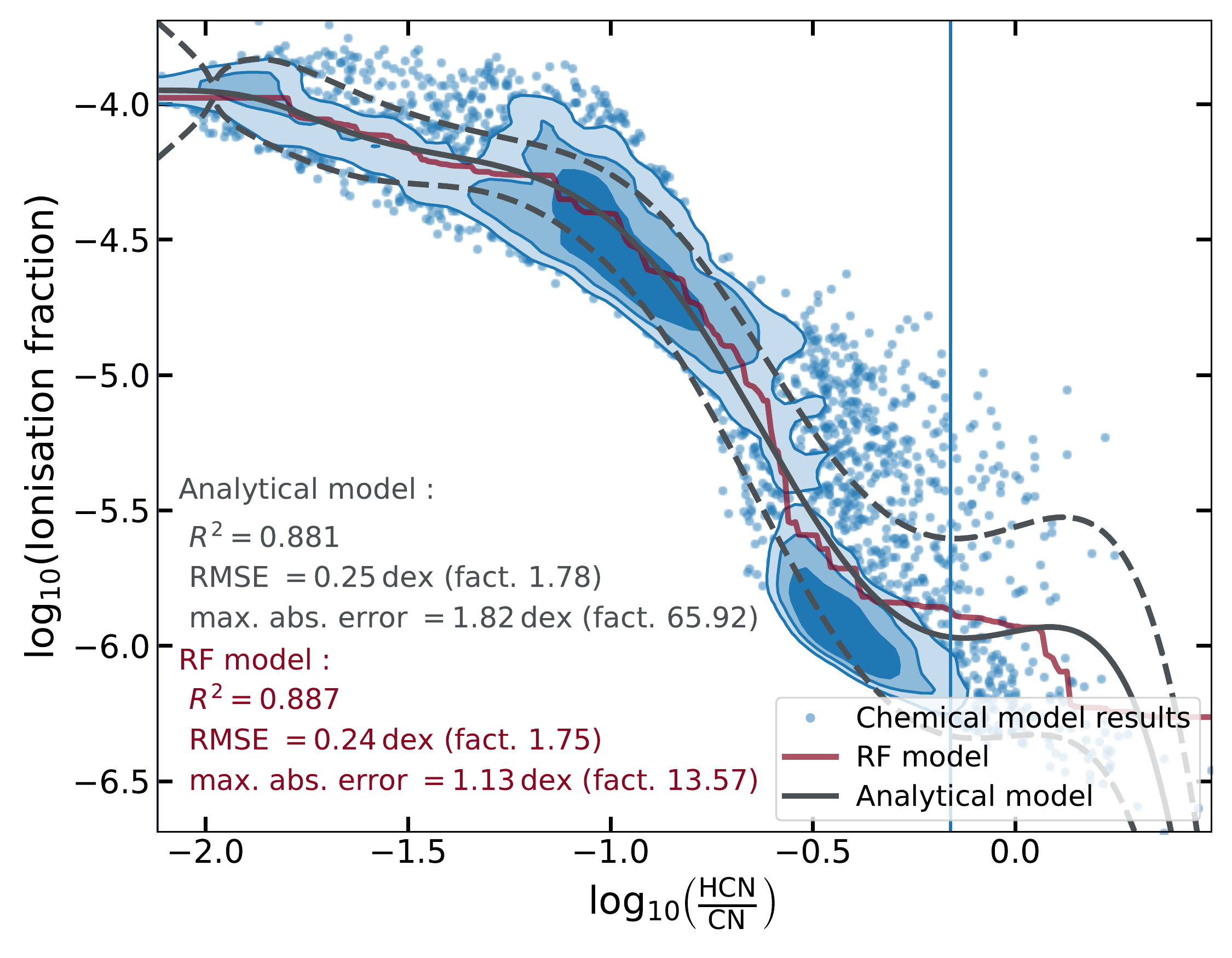}
    \caption{Ionization fraction versus column density ratio for the best six ratios found in Sect.\ref{sect:Tracers} for translucent medium conditions. 
The chemical model grid is shown as a scatter plot, with the central crowded regions replaced by PDF isocontours containing 25\%, 50\%, and 75\% of the points.
Superimposed are the RF model (red line), the analytical fit (solid black line), and the analytical fit of the 1$\sigma$ uncertainty (dashed black lines). 
The quality estimates of the two models are indicated on the figure.}
    \label{fig:AnalyticalmodelBestSixTranslucentColDenRatios}
  \end{figure*}
}

\newcommand{\FigAnalyticalmodelBestSixTranslucentIntensityRatios}{
  \begin{figure*}
    \includegraphics[width=0.5\linewidth]{figures/Translucent2_oldH2_intratios_new_1ratio_best0_c2h_1-0_over_hcn_1-0_residualfit_paper.pdf}
    \includegraphics[width=0.5\linewidth]{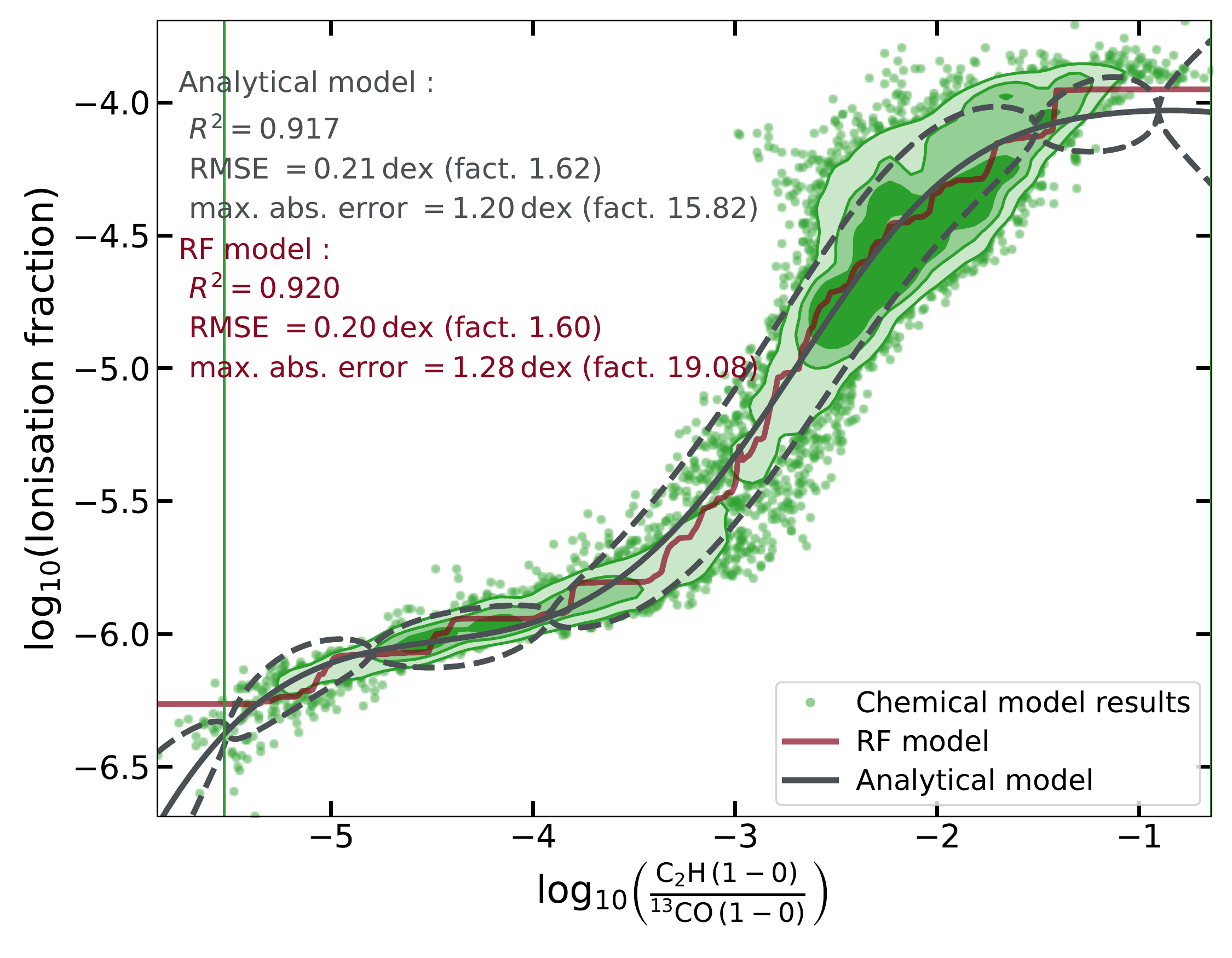}\\
    \includegraphics[width=0.5\linewidth]{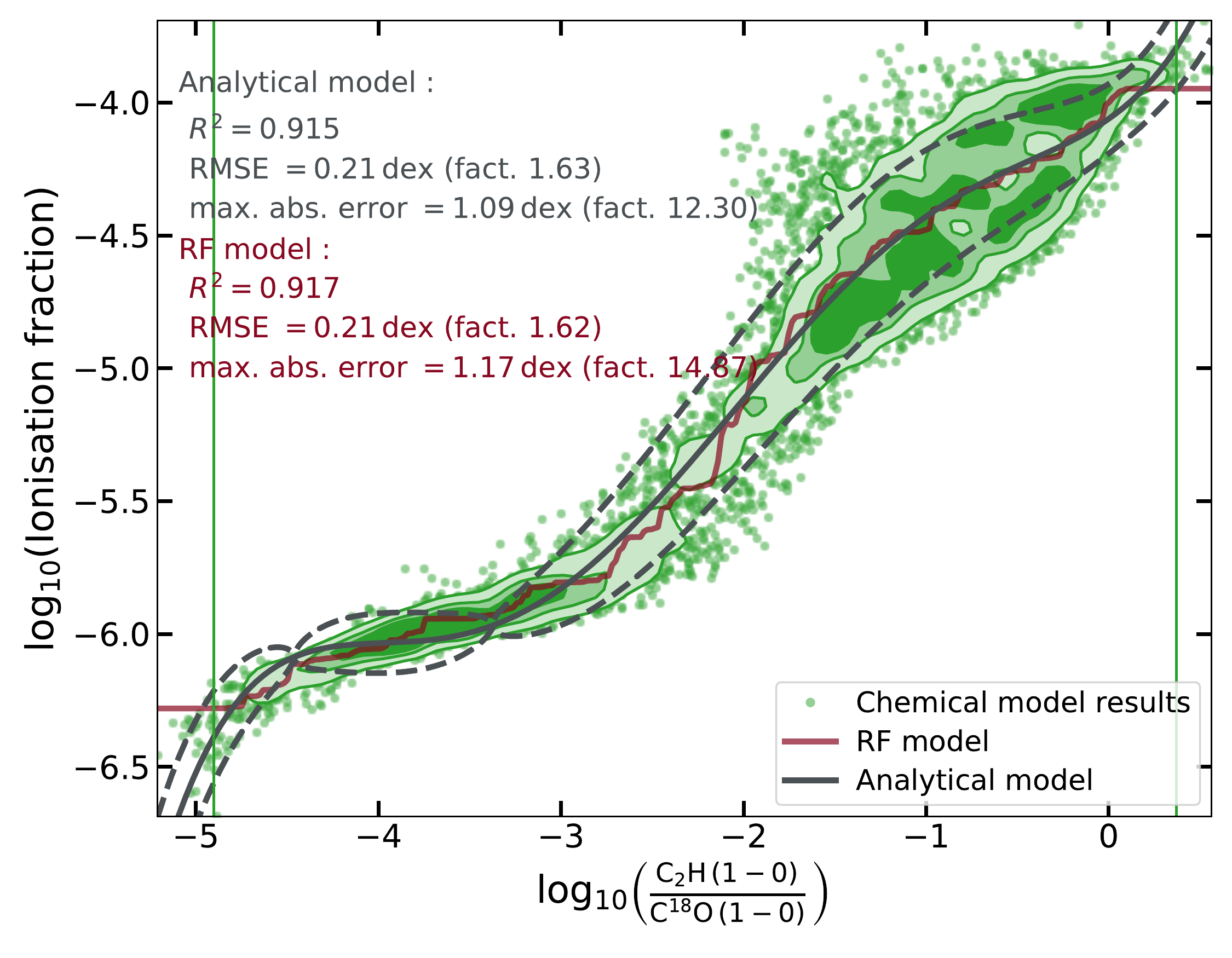}
    \includegraphics[width=0.5\linewidth]{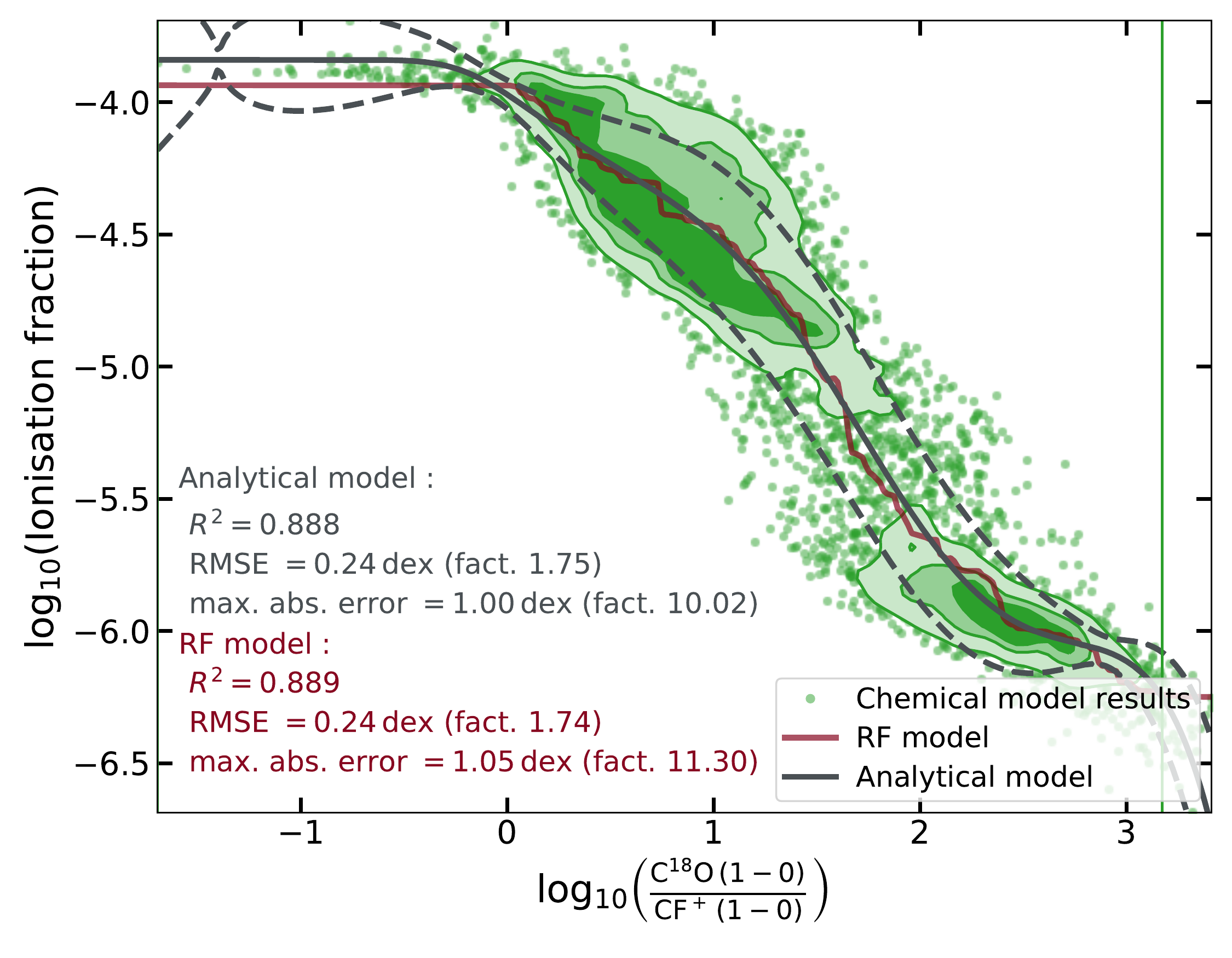}\\
    \includegraphics[width=0.5\linewidth]{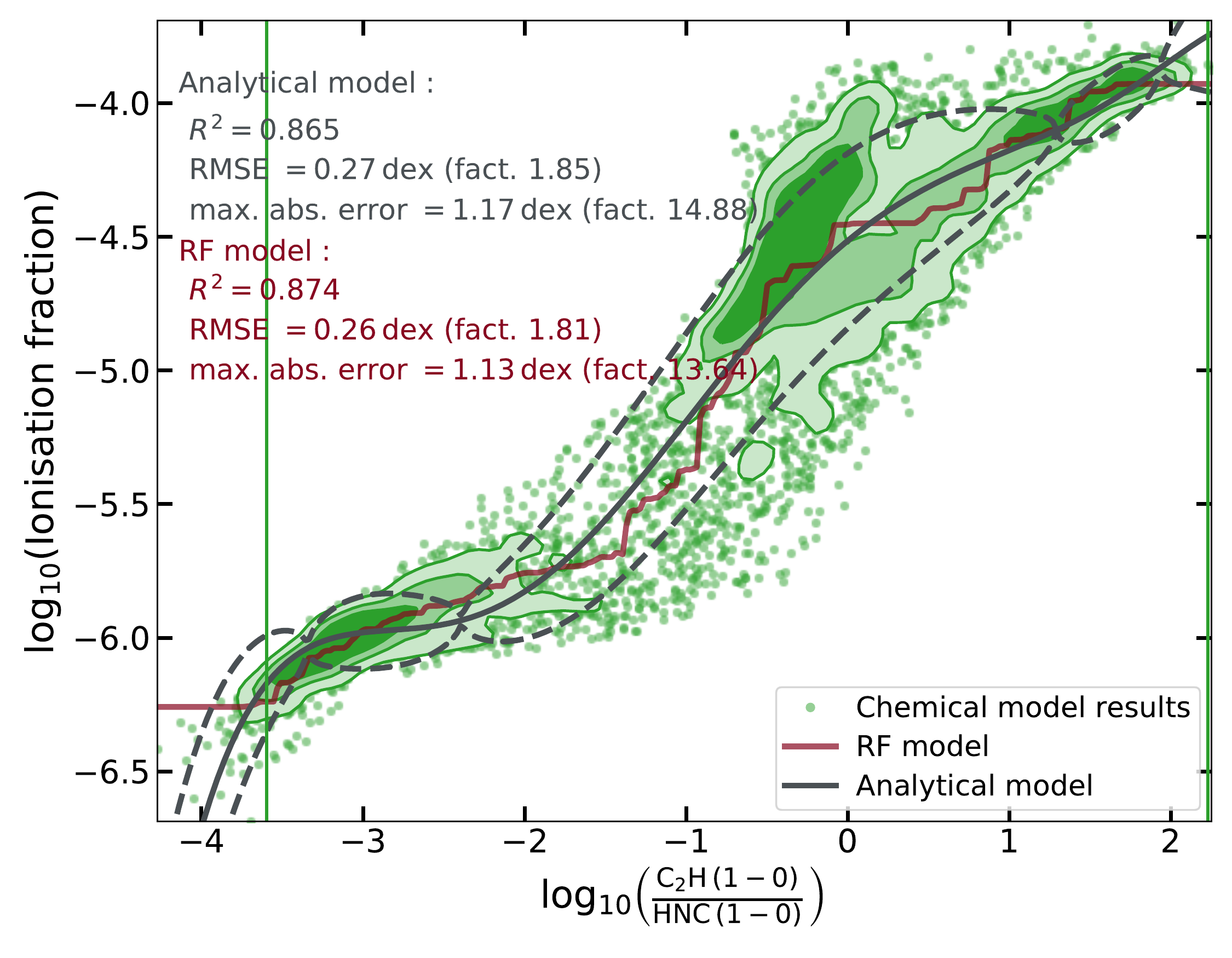}
    \includegraphics[width=0.5\linewidth]{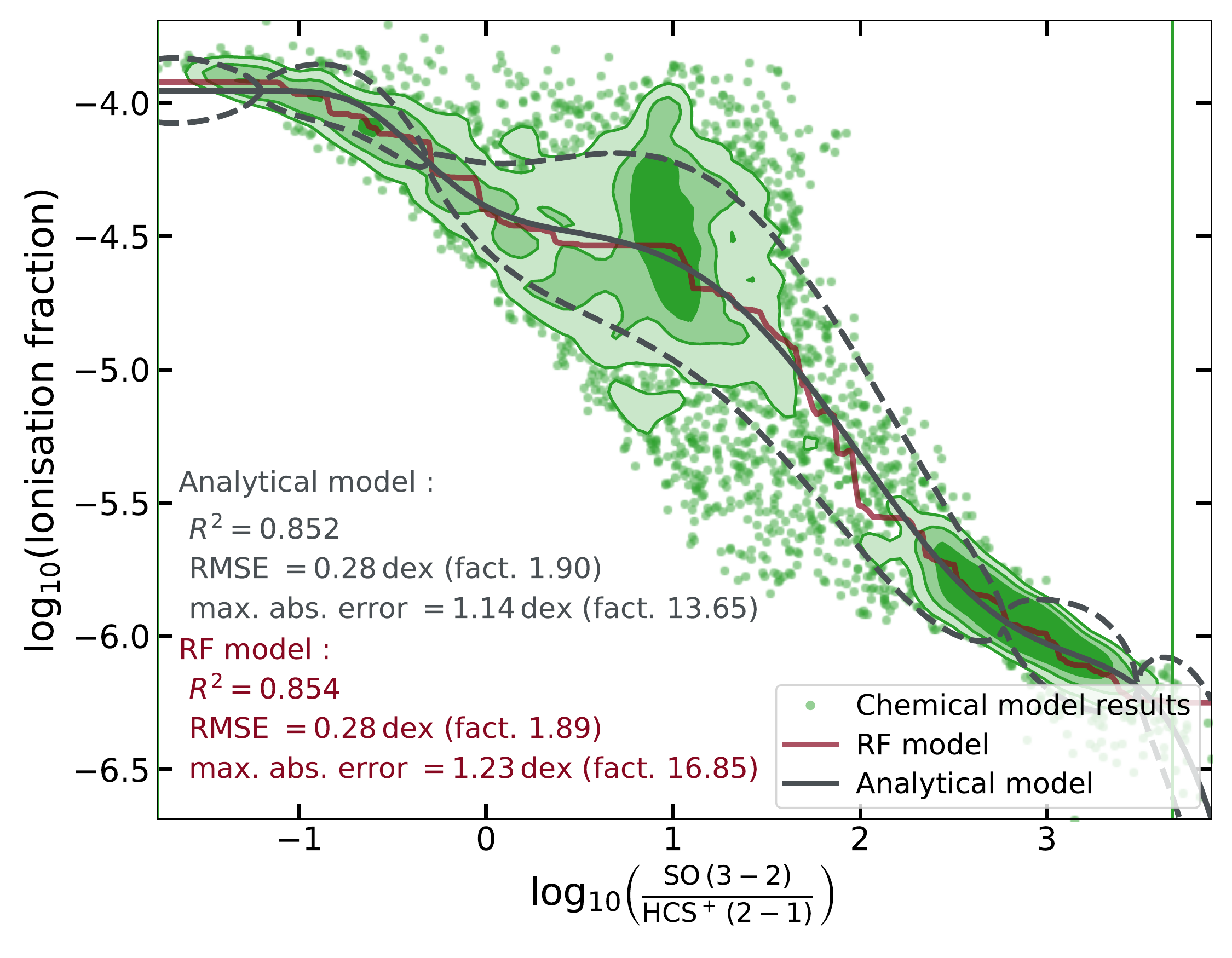}
    \caption{Ionization fraction versus integrated line intensity ratio for the best six ratios found in Sect.\ref{sect:Tracers} for translucent medium conditions. 
The chemical model grid is shown as a scatter plot, with the central crowded regions replaced by PDF isocontours containing 25\%, 50\%, and 75\% of the points.
Superimposed are the RF model (red line), the analytical fit (solid black line), and the analytical fit of the 1$\sigma$ uncertainty (dashed black lines). 
The quality estimates of the two models are indicated on the figure.}
    \label{fig:AnalyticalmodelBestSixTranslucentIntensityRatios}
  \end{figure*}
}

\newcommand{\FigAnalyticalmodelBestSixDenseColDenRatios}{
  \begin{figure*}
    \includegraphics[width=0.5\linewidth]{figures/Dense_cold2_oldH2_abratios_1ratio_best0_cn_over_n2hp_residualfit_paper.pdf}
    \includegraphics[width=0.5\linewidth]{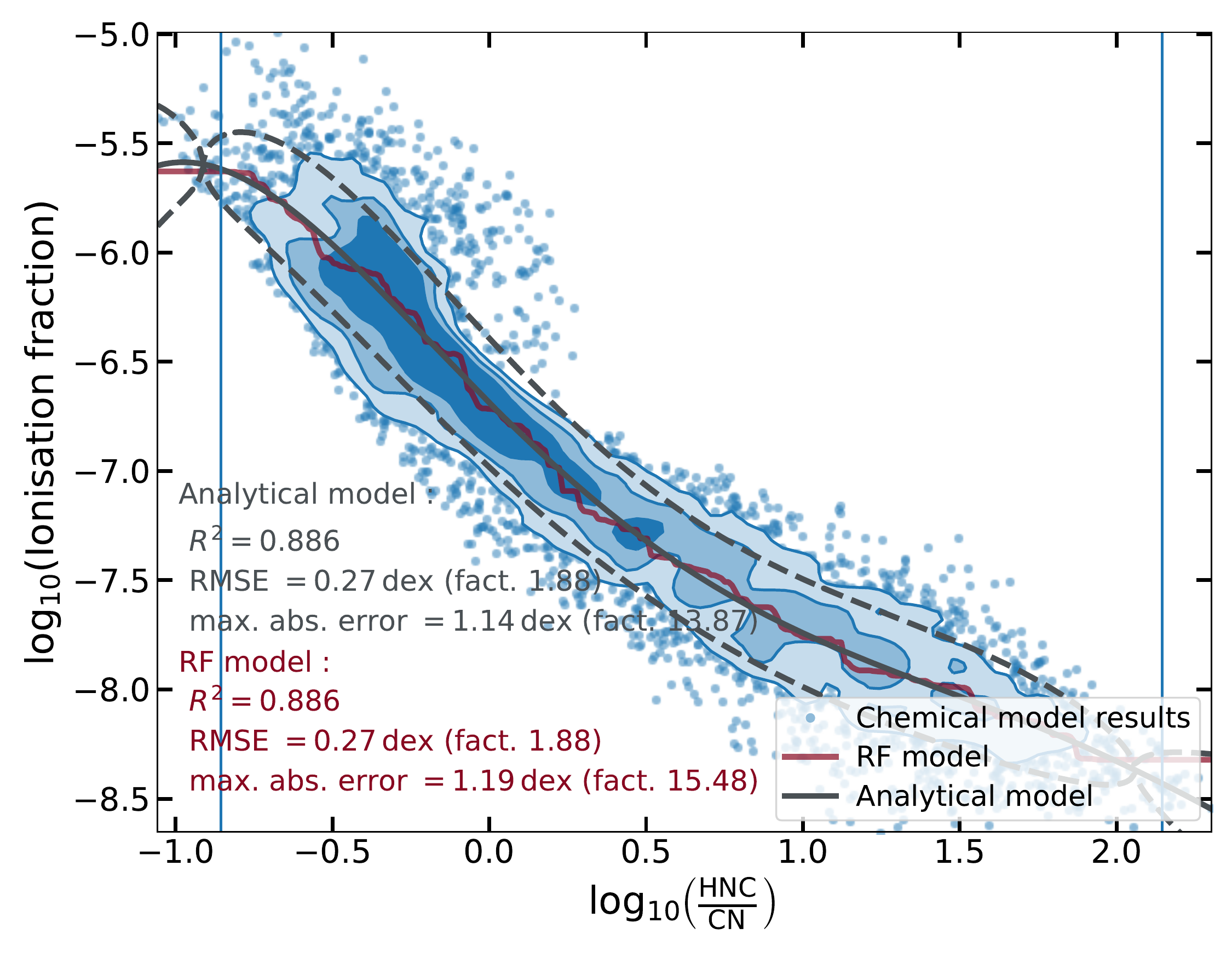}\\
    \includegraphics[width=0.5\linewidth]{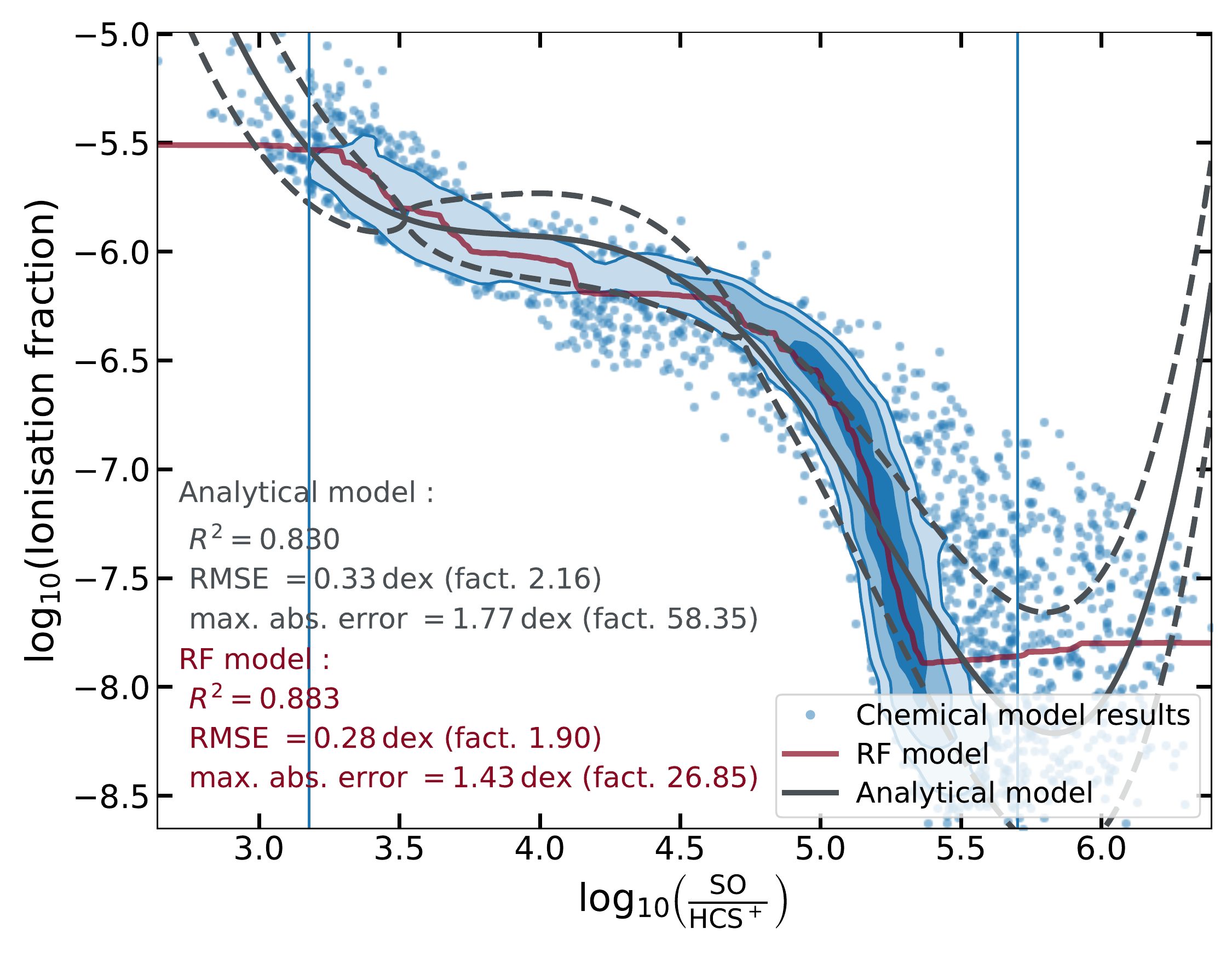}
    \includegraphics[width=0.5\linewidth]{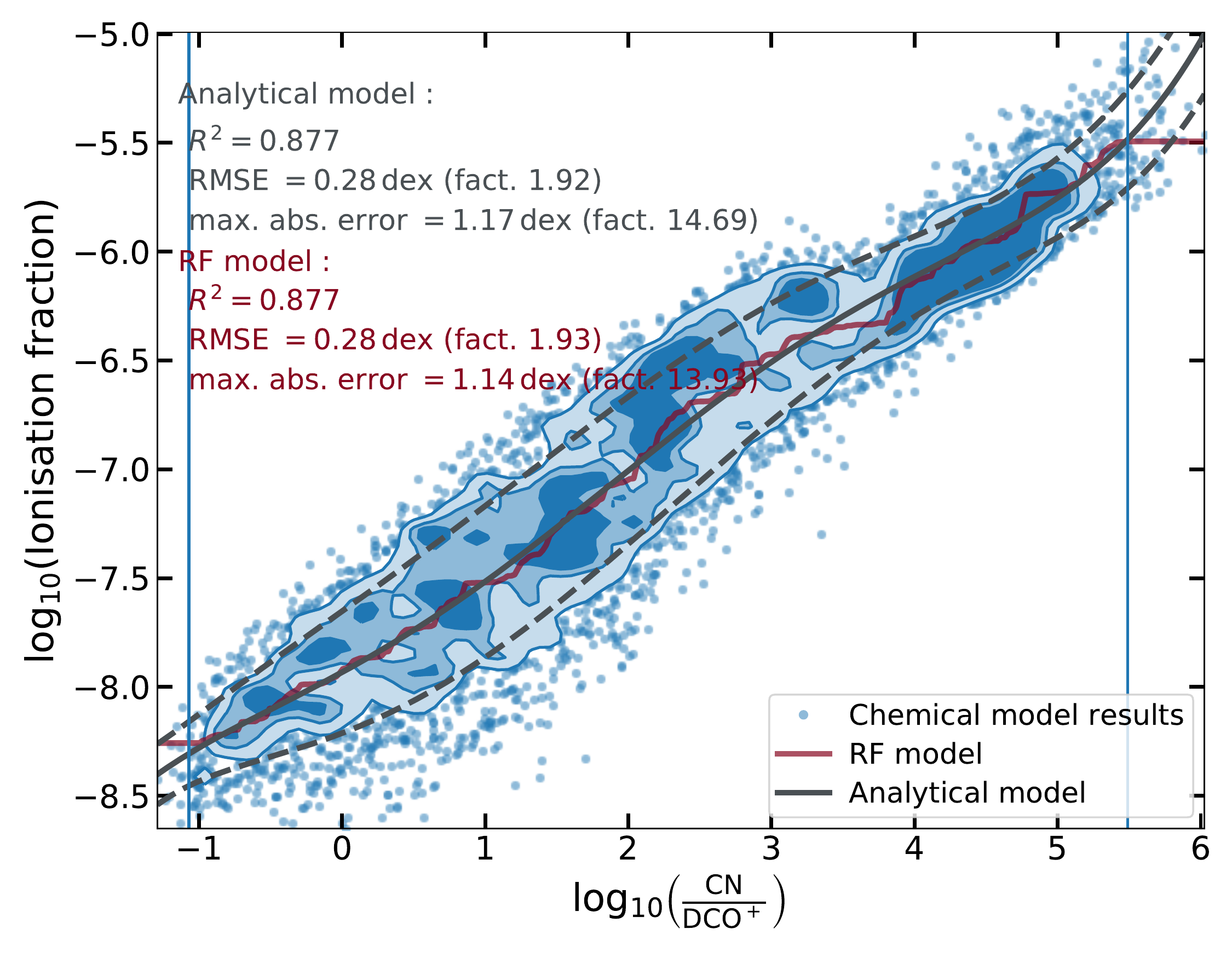}\\
    \includegraphics[width=0.5\linewidth]{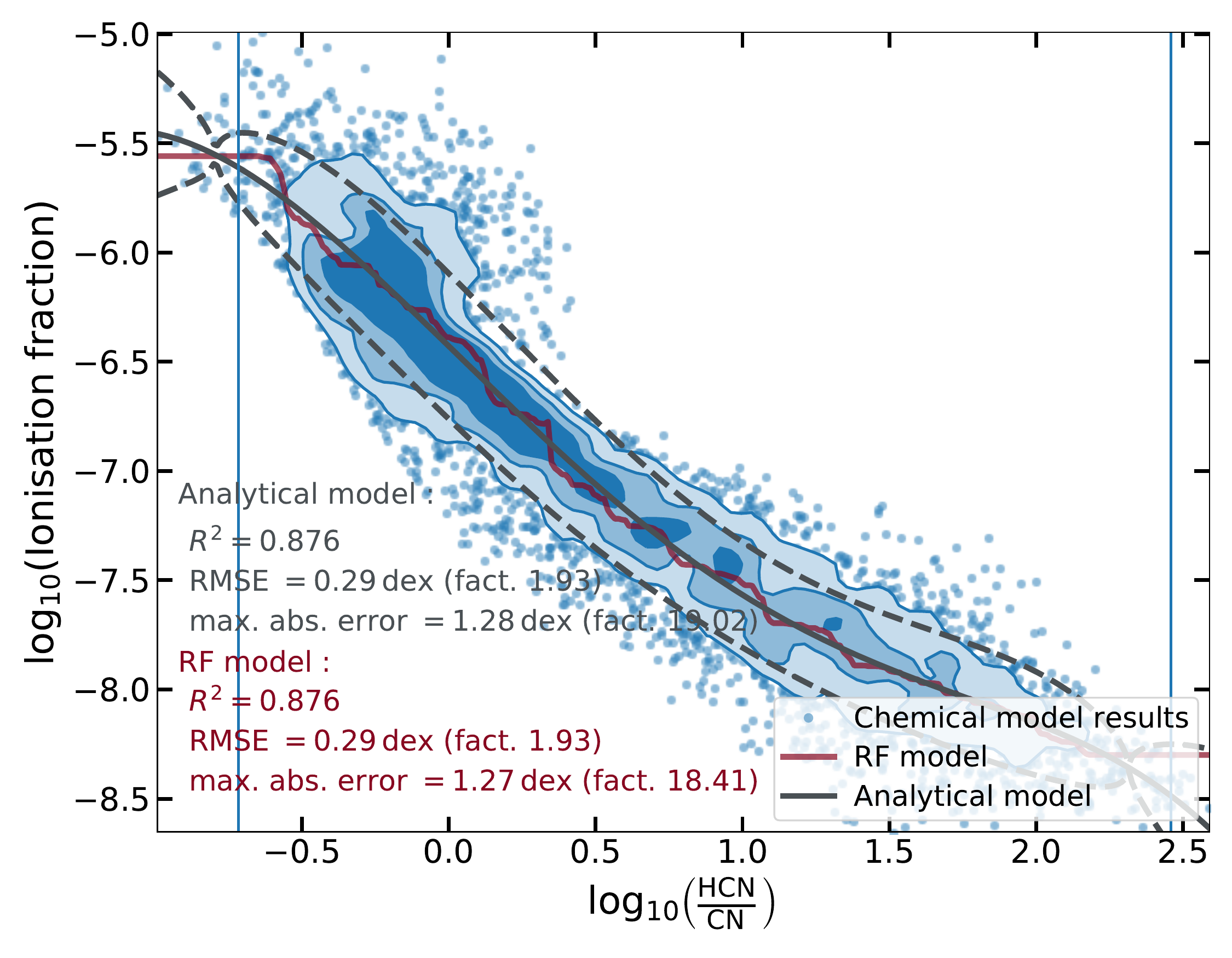}
    \includegraphics[width=0.5\linewidth]{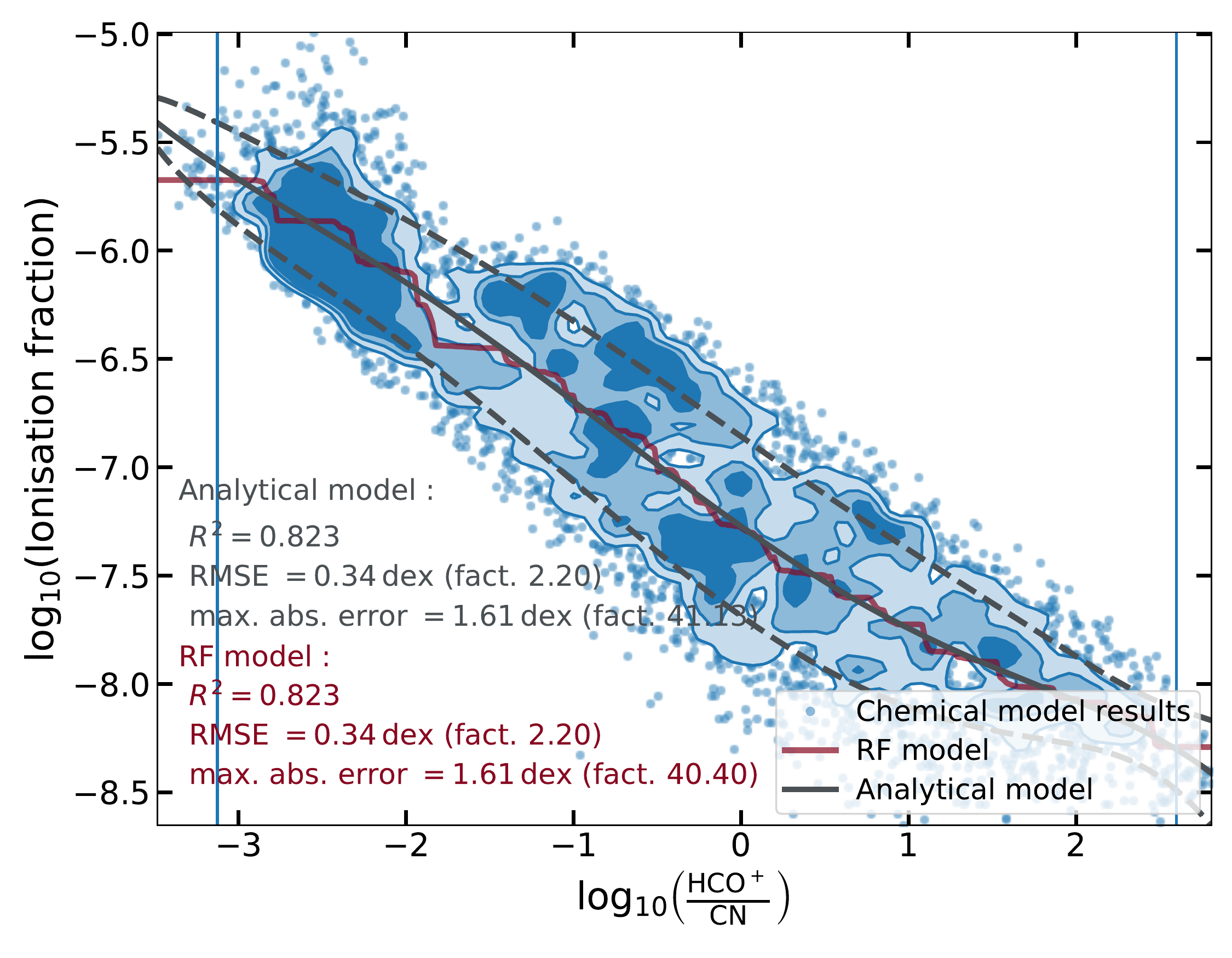}
    \caption{Ionization fraction versus column density ratio for the best six ratios found in Sect.\ref{sect:Tracers} for cold dense medium conditions. 
The chemical model grid is shown as a scatter plot, with the central crowded regions replaced by PDF isocontours containing 25\%, 50\%, and 75\% of the points.
Superimposed are the RF model (red line), the analytical fit (solid black line), and the analytical fit of the 1$\sigma$ uncertainty (dashed black lines). 
The quality estimates of the two models are indicated on the figure.}
    \label{fig:AnalyticalmodelBestSixDenseColDenRatios}
  \end{figure*}
}

\newcommand{\FigAnalyticalmodelBestSixDenseIntensityRatios}{
  \begin{figure*}
    \includegraphics[width=0.5\linewidth]{figures/Dense_cold2_oldH2_intratios_new_1ratio_best0_cfp_1-0_over_dcop_1-0_residualfit_paper.pdf}
    \includegraphics[width=0.5\linewidth]{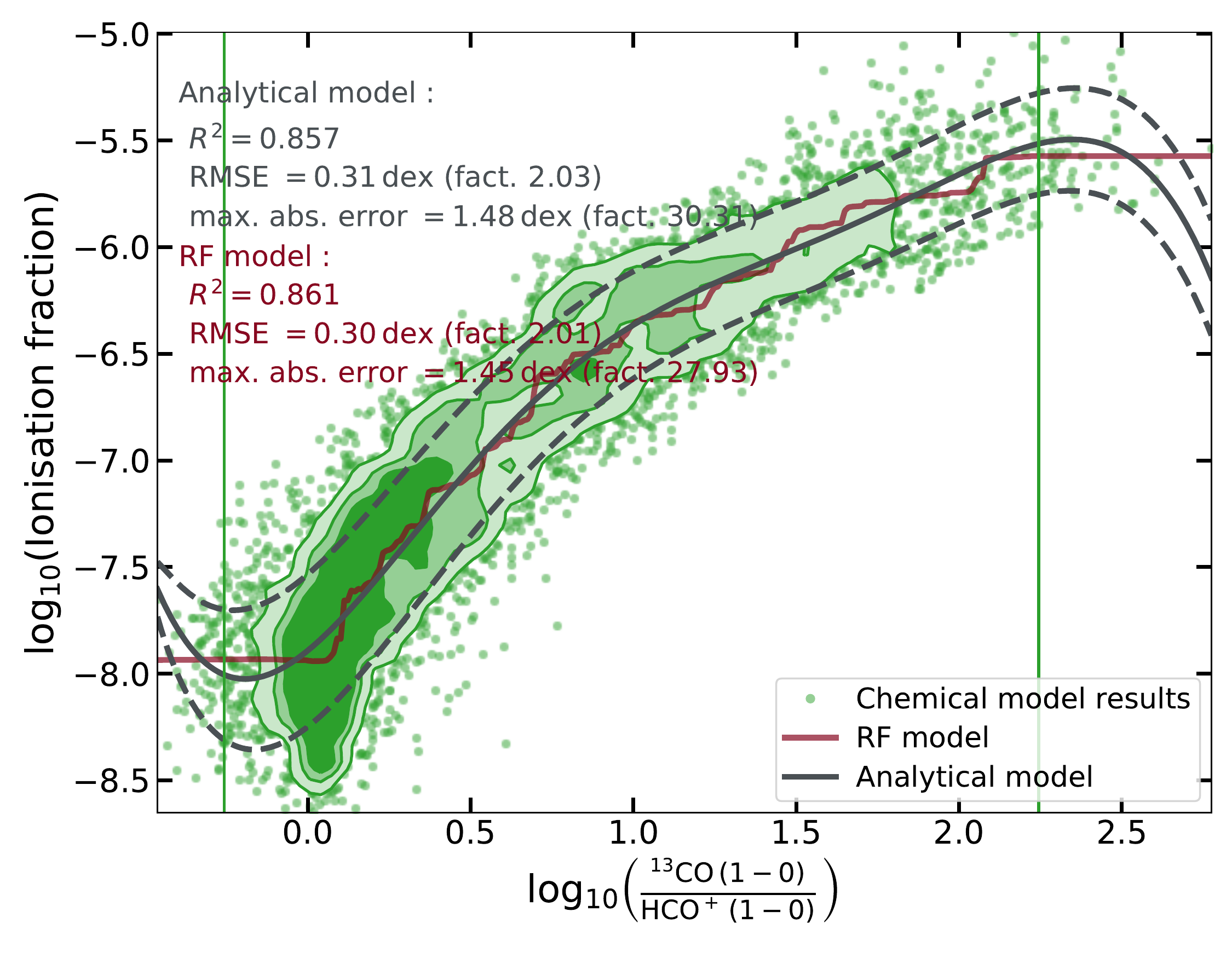}\\
    \includegraphics[width=0.5\linewidth]{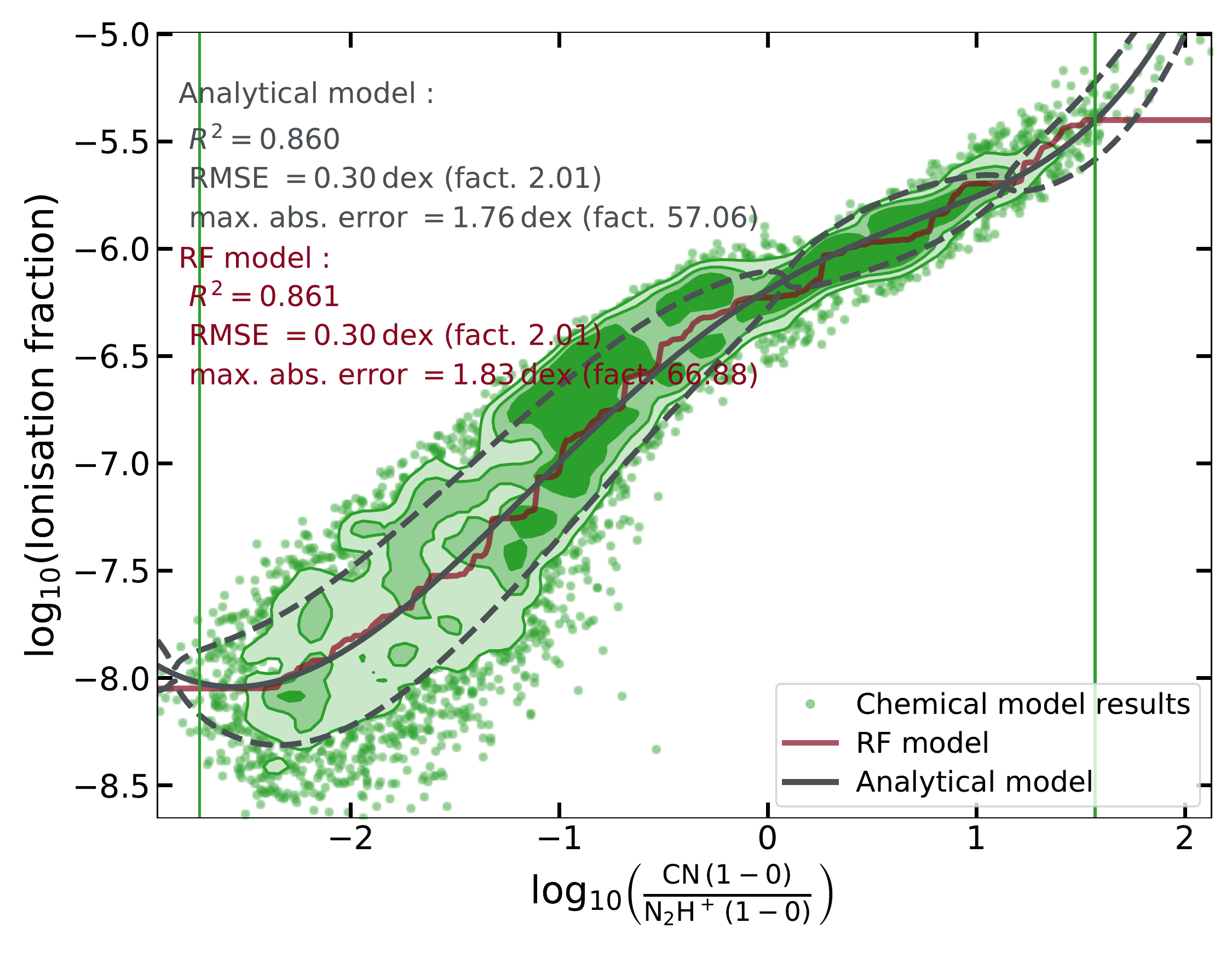}
    \includegraphics[width=0.5\linewidth]{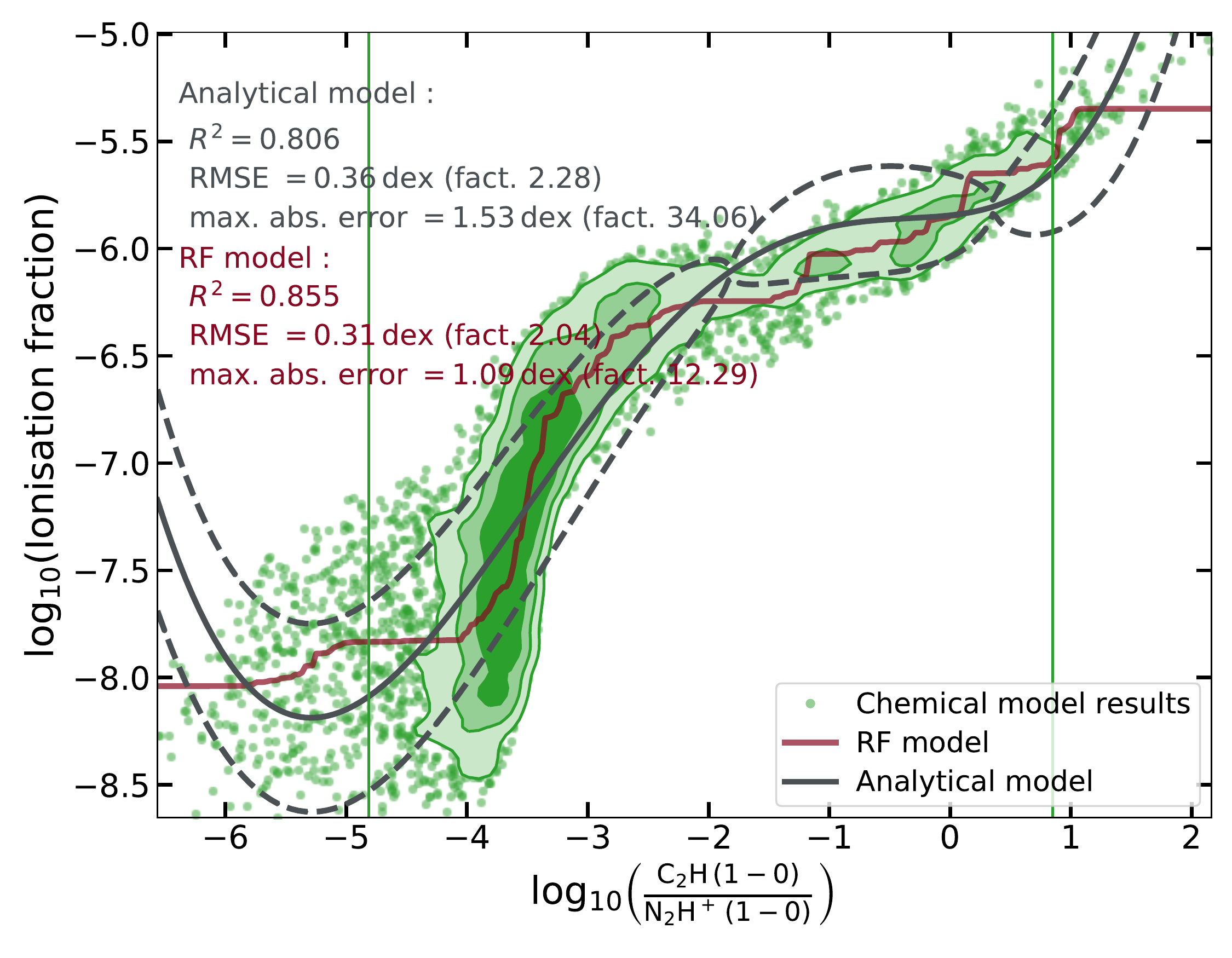}\\
    \includegraphics[width=0.5\linewidth]{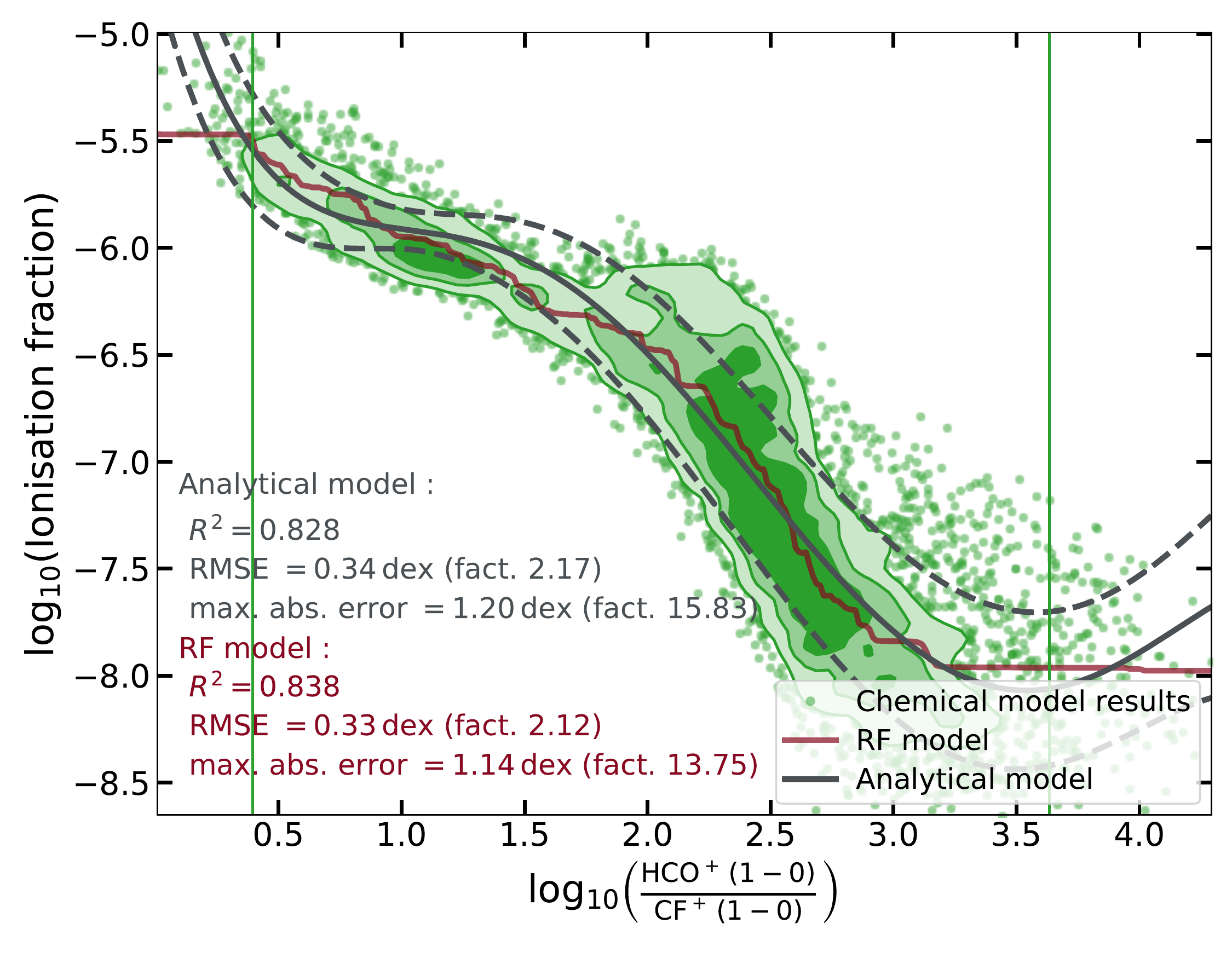}
    \includegraphics[width=0.5\linewidth]{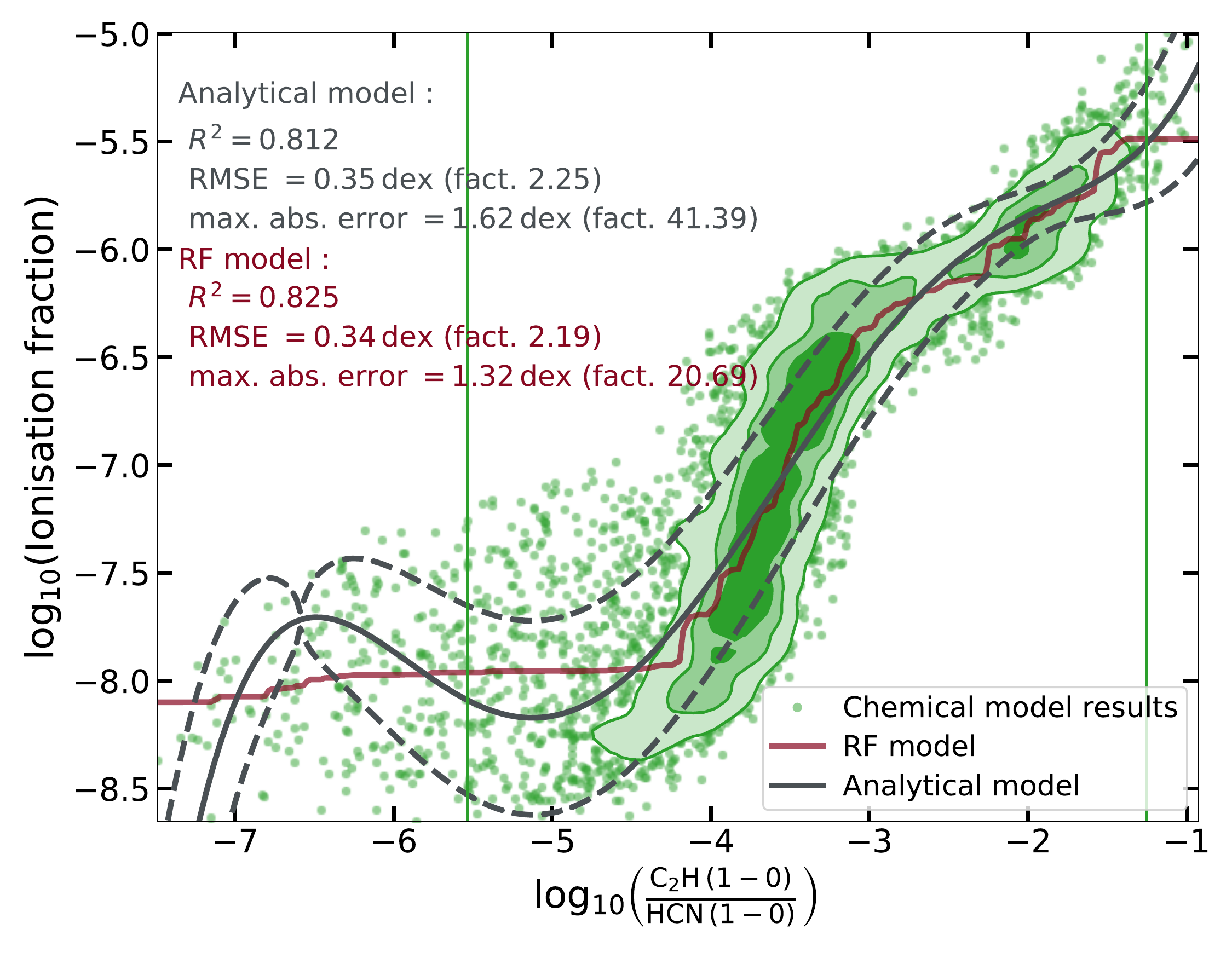}
    \caption{Ionization fraction versus integrated line intensity ratio for the best six ratios found in Sect.\ref{sect:Tracers} for cold dense medium conditions. 
The chemical model grid is shown as a scatter plot, with the central crowded regions replaced by PDF isocontours containing 25\%, 50\%, and 75\% of the points.
Superimposed are the RF model (red line), the analytical fit (solid black line), and the analytical fit of the 1$\sigma$ uncertainty (dashed black lines). 
The quality estimates of the two models are indicated on the figure.}
    \label{fig:AnalyticalmodelBestSixDenseIntensityRatios}
  \end{figure*}
}

\newcommand{\FigDCOpoverHCOp}{
  \begin{figure}
    \includegraphics[width=1\linewidth]{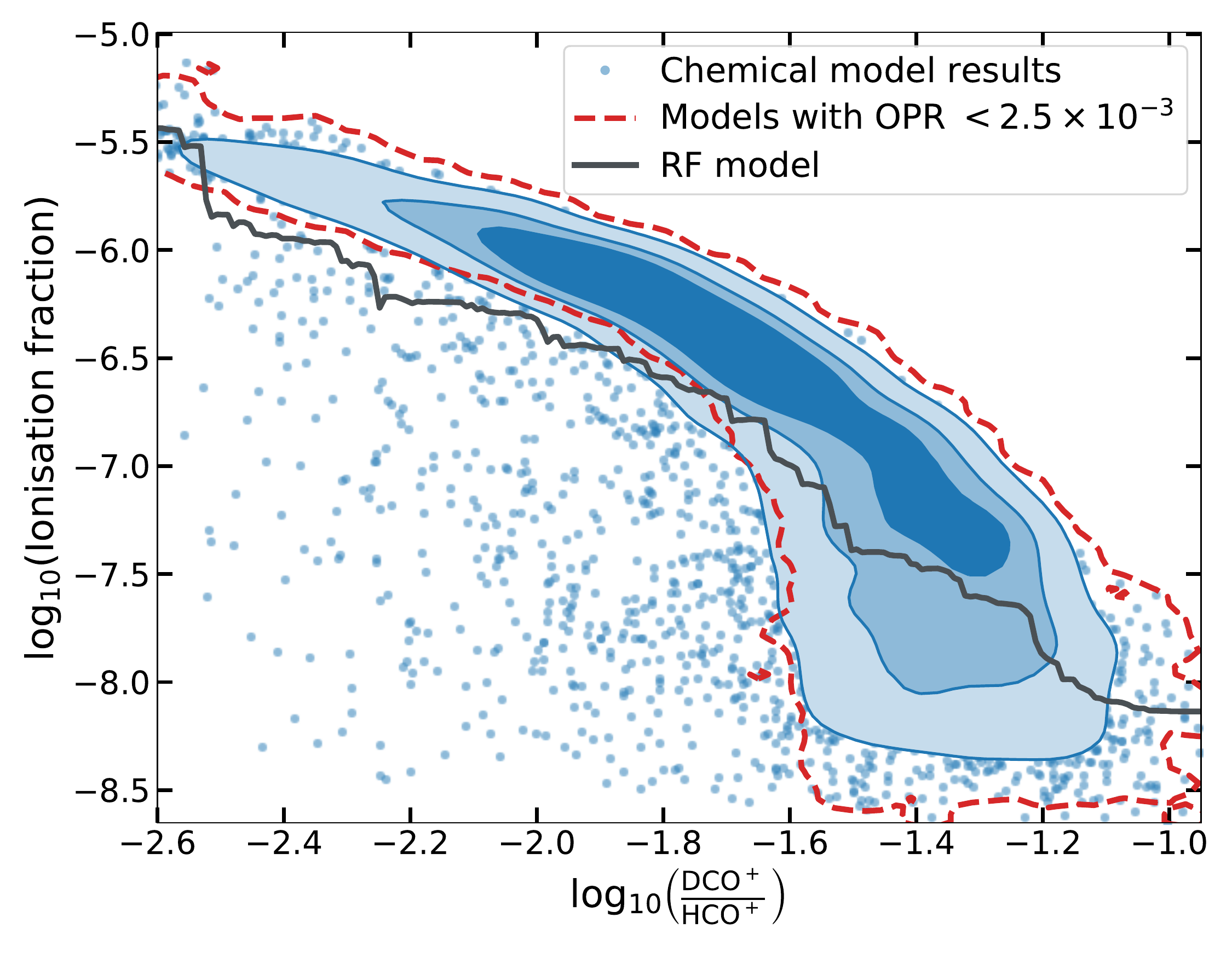}
    \caption{DCO$^+$ over HCO$^+$ column density ratio in our grid of dense cold gas models (blue points and contours), shown as a scatter plot, with the central crowded regions replaced by PDF isocontours containing 25\%, 50\%, and 75\% of the points. 
Our best fit Random Forest model is shown as a black line. The red dashed contours shows the distributions of the models with OPR$_{\mathrm{H}_2} < 2.5\times10^{-3}$.}
    \label{fig:DCOpoverHCOp}
  \end{figure}
}

\newcommand{\FigTranslucentNoiseImpactAbs}{
  \begin{figure}
    \includegraphics[width=1\linewidth]{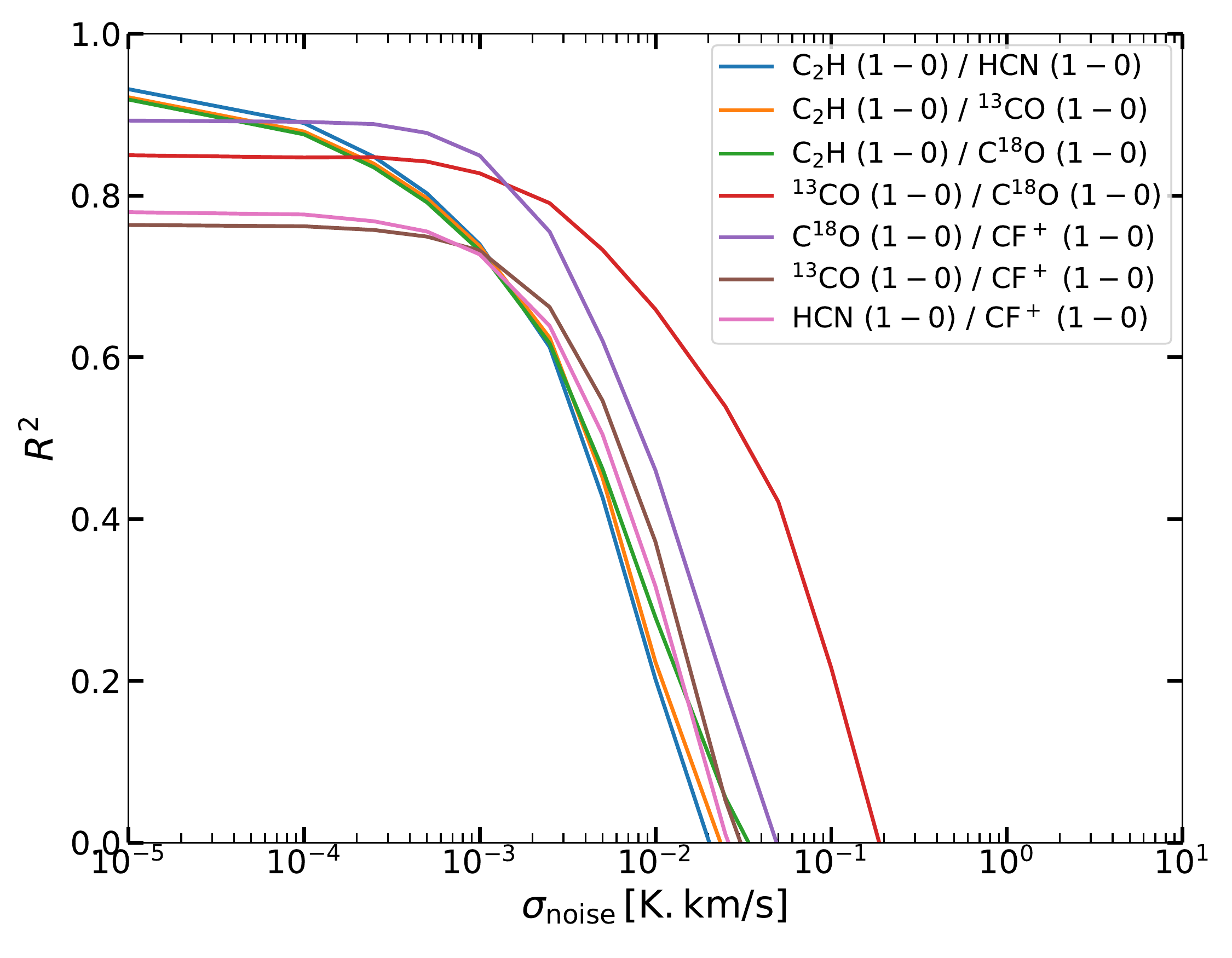}
    \caption{Evolution of the $R^2$ performance of the Random Forest predictors (for translucent gas) when adding noise of constant variance $\sigma_\mathrm{noise}^2$ to all line intensities, as a function of the noise level $\sigma_\mathrm{noise}$, for the 3 best tracers for translucent medium (C$_2$H(1-0)/HCN(1-0), C$_2$H(1-0)/$^{13}$CO(1-0), and C$_2$H(1-0)/C$^{18}$O(1-0)) and for the four tracers least affected by the noise  ($^{13}$CO(1-0)/C$^{18}$O(1-0), C$^{18}$O(1-0)/CF$^+$(1-0), $^{13}$CO(1-0)/CF$^+$(1-0), and HCN/CF$^+$(1-0)).}
    \label{fig:TranslucentNoiseImpactAbs}
  \end{figure}
}

\newcommand{\FigTranslucentNoiseImpactSNR}{
  \begin{figure}
    \includegraphics[width=1\linewidth]{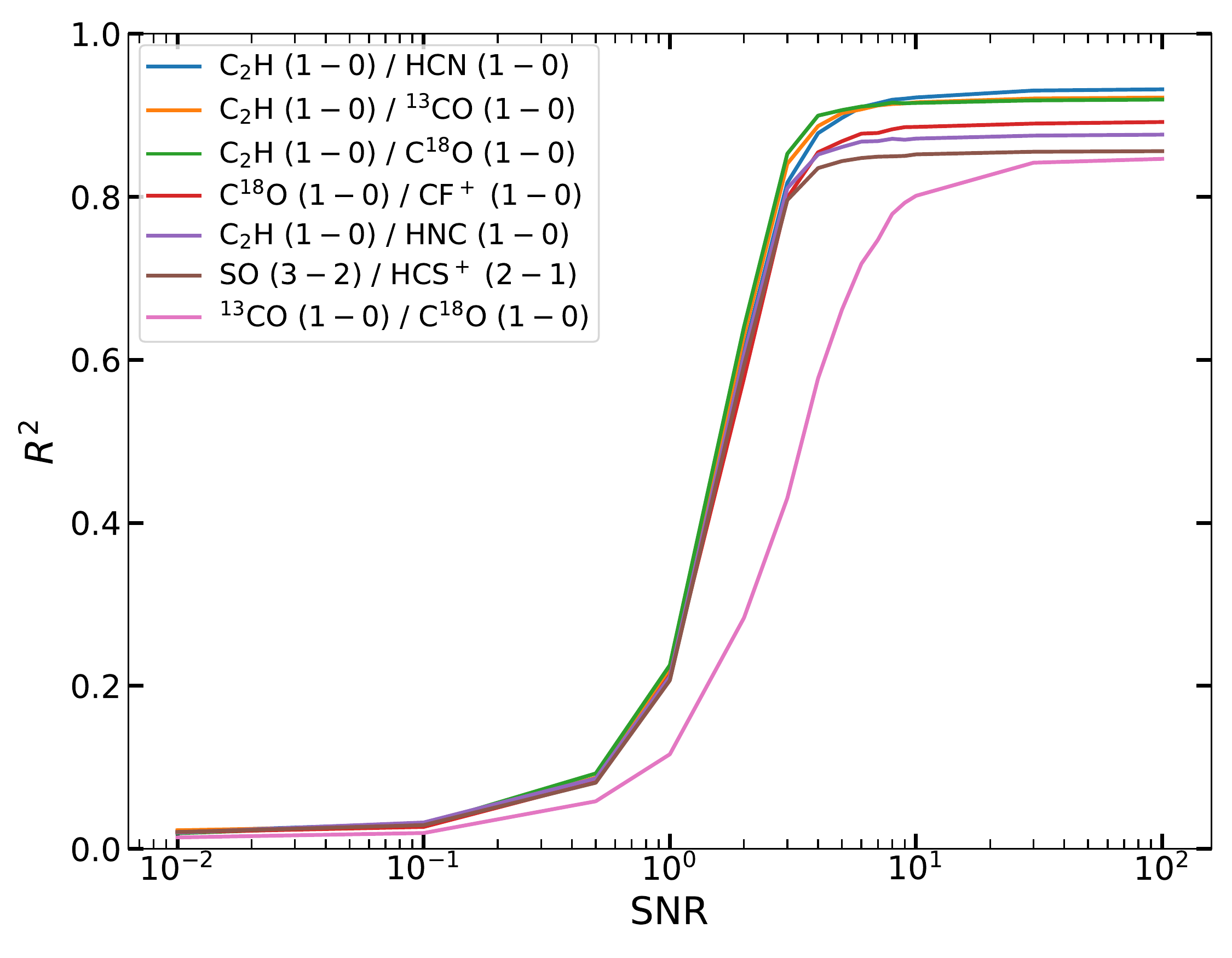}
    \caption{Evolution of the $R^2$ performance of the Random Forest predictors (for translucent gas) when adding noise of constant signal-to-noise ratio to the line intensities, as a function of the SNR, for the 7 best tracers for translucent medium.}
    \label{fig:TranslucentNoiseImpactSNR}
  \end{figure}
}

\newcommand{\FigDenseNoiseImpactAbs}{
  \begin{figure}
    \includegraphics[width=1\linewidth]{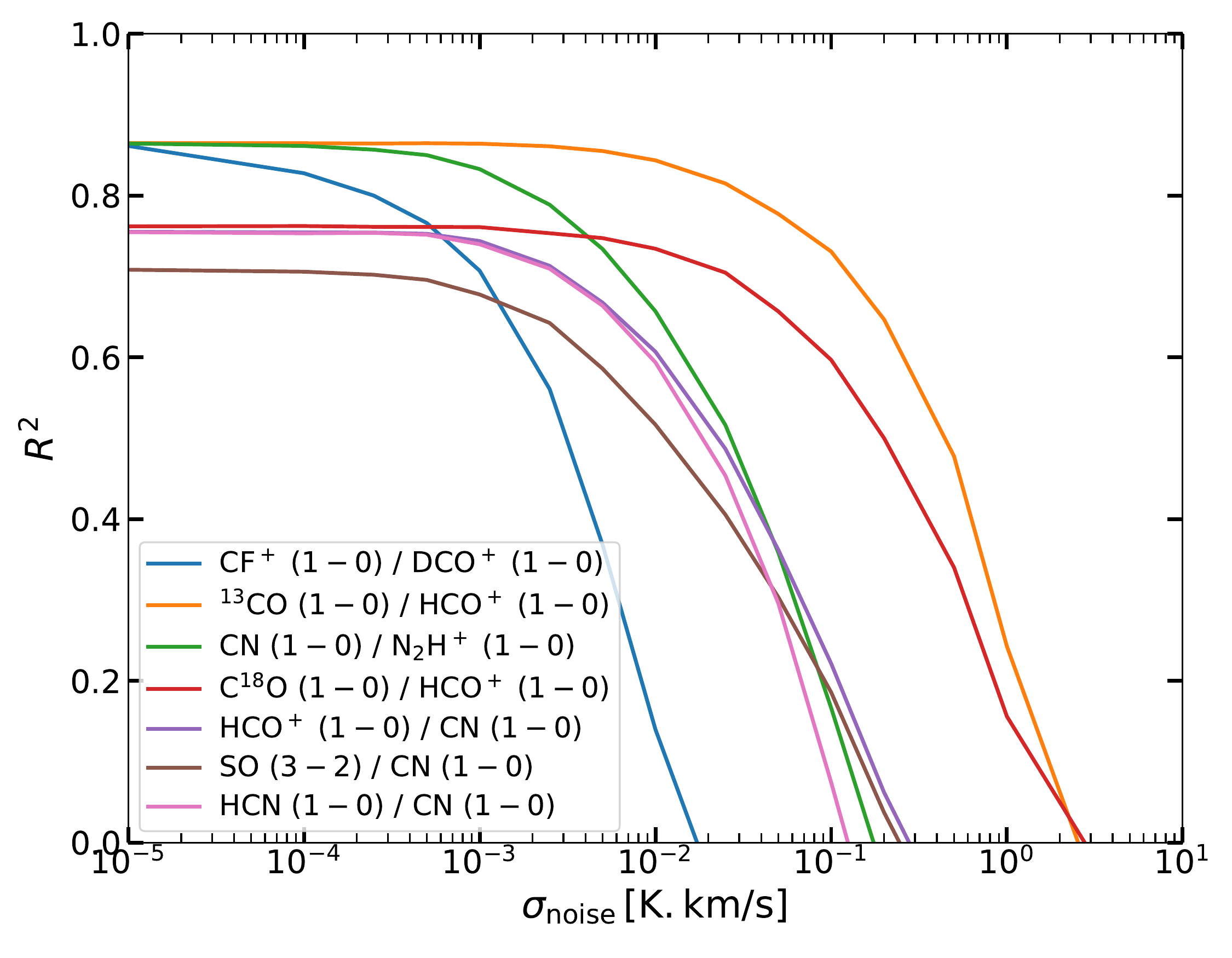}
    \caption{Evolution of the $R^2$ performance of the Random Forest predictors (for cold dense gas) when adding noise of constant variance $\sigma_\mathrm{noise}^2$ to the line intensities, as a function of the noise level $\sigma_\mathrm{noise}$, for the 3 best tracers for cold dense medium (first three curves) and for the four tracers least affected by the noise (next four curves).}
    \label{fig:DenseNoiseImpactAbs}
  \end{figure}
}

\newcommand{\FigDenseNoiseImpactSNR}{
  \begin{figure}
    \includegraphics[width=1\linewidth]{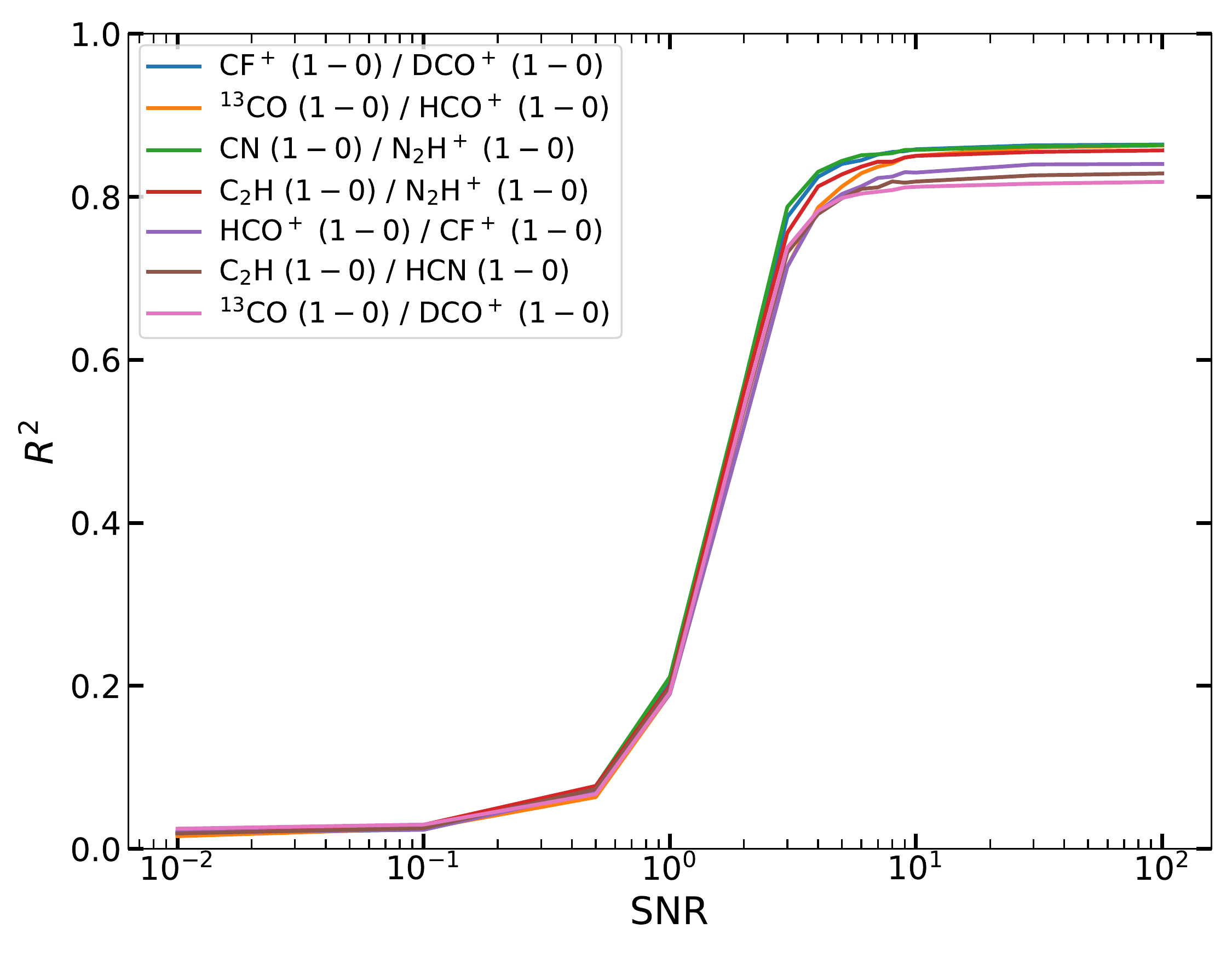}
    \caption{Evolution of the $R^2$ performance of the Random Forest predictors (for cold dense gas) when adding noise of constant signal-to-noise ratio to the line intensities, as a function of the SNR, for the 7 best tracers for cold dense medium.}
    \label{fig:DenseNoiseImpactSNR}
  \end{figure}
}

\newcommand{\FigRFparametersOptimization}{
  \begin{figure}
    \includegraphics[width=1\linewidth]{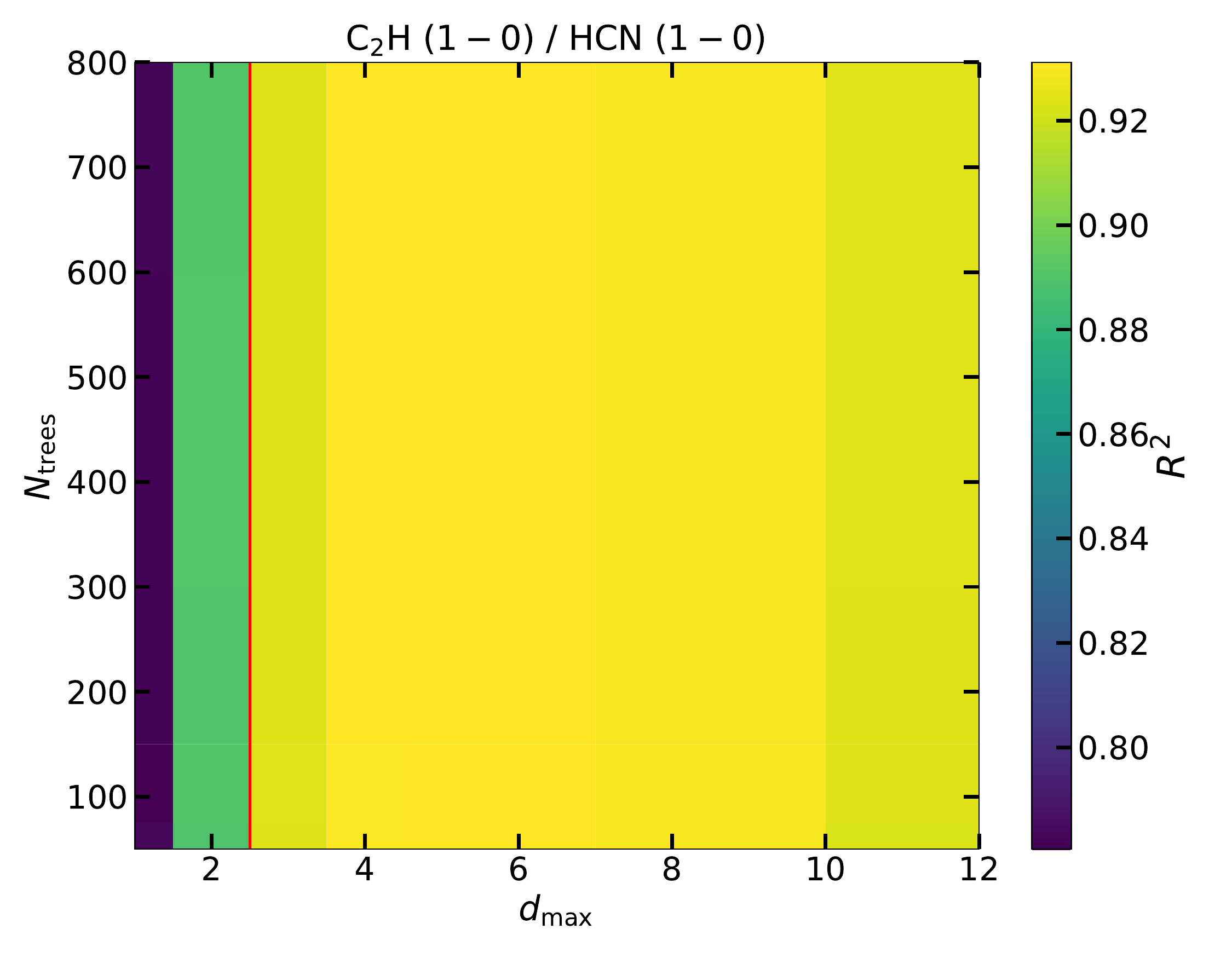}\\
    \includegraphics[width=1\linewidth]{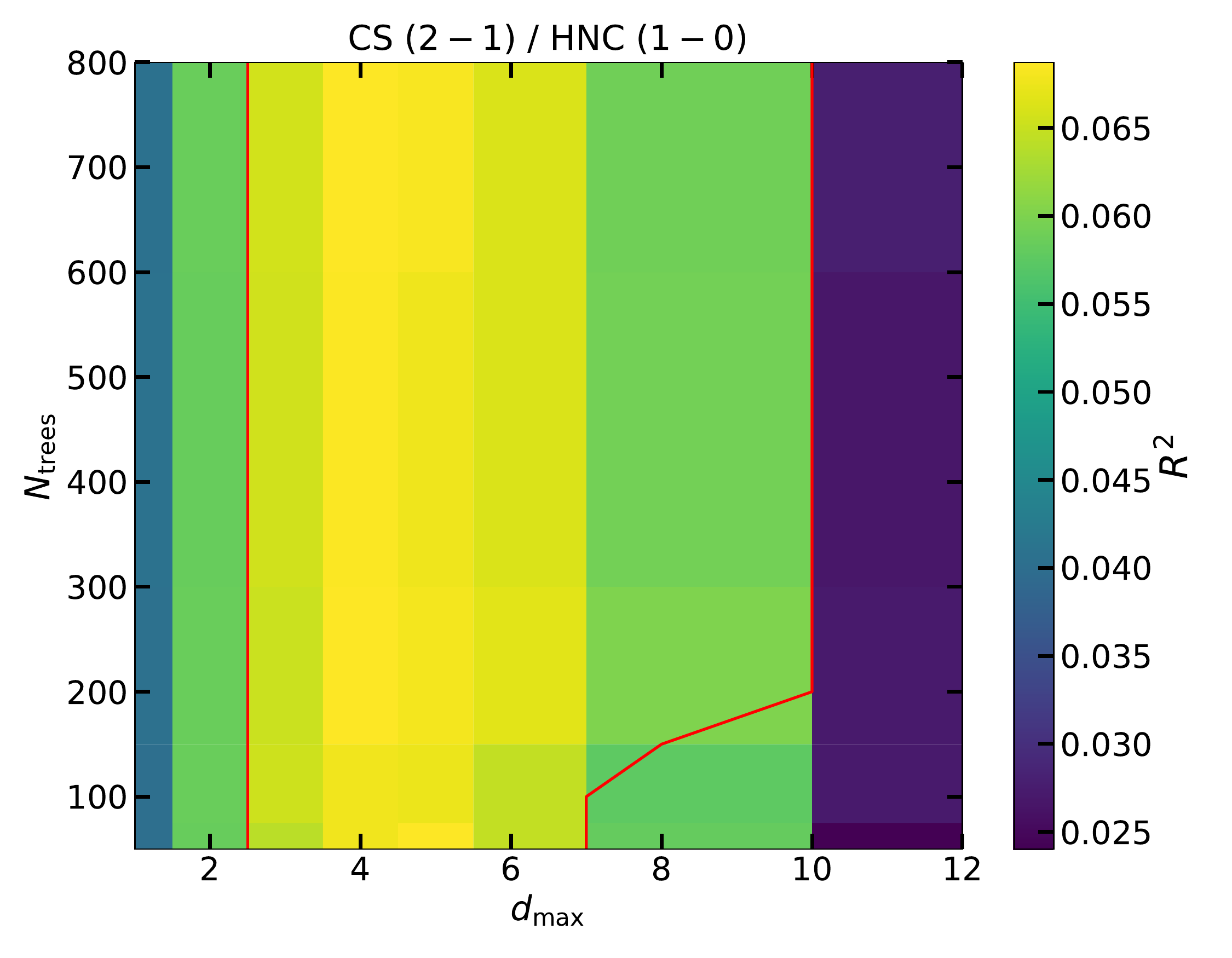}
    \caption{$R^2$ (OOB value) as a function of the tuning parameters of the Random Forest ($N_\mathrm{trees}$ and $d_\mathrm{max}$), for the preliminary best (top panel) and worst (bottom panel) ratios found though a preliminary estimate using defaults values of the parameters, for line integrated intensity ratios and translucent medium conditions. 
The red contours show the region where the $R^2$ value is within 0.01 of the maximum.}
    \label{fig:RFparametersOptimization}
  \end{figure}
}

\newcommand{\TabLineIDs}{%
  \begin{table}
    \caption{We list here the shorthand names, full quantum number designation, and frequency of the molecular lines we considered.}
    \label{tab:LineIDs}
    \begin{center}
    \tiny
      \begin{tabular} {lll}
      \hline 
      Short name & Full quantum numbers & Frequency (GHz) \\
        \hline 
        $^{13}$CO (1-0)    & $J = 1 \rightarrow J = 0$                                                                 & 110.201354 \\
        C$^{18}$O (1-0)    & $J = 1 \rightarrow J = 0$                                                                 & 109.782173 \\
        HCO$^+$ (1-0)      & $J = 1 \rightarrow J = 0$                                                                 & 89.188396 \\
        HCN (1-0)              & $J = 1 \rightarrow J = 0$                                                                 & 88.631602 \\
        HNC (1-0)               & $J = 1 \rightarrow J = 0$                                                                & 90.663568 \\
        CN (1-0)                 & $N=1, J=3/2, F=5/2 \rightarrow N=0, J=1/2, F=3/2$                       & 113.490970 \\
        C$_2$H (1-0)         & $N=1, J=1/2, F=0 \rightarrow N=0,  J=1/2, F=1$                            & 87.40716 \\
        CS (2-1)                 & $J=2\rightarrow J = 1$                                                                    & 97.980953 \\
        SO (3-2)                 & $J=3, N=2 \rightarrow J=2, N=1$                                                    & 99.299870 \\
        HCS$^+$ (2-1)       & $J=2\rightarrow J = 1$                                                                    & 85.347890 \\
        CF$^+$ (1-0)         & $J = 1 \rightarrow J = 0$                                                                 & 102.587533 \\
        H$_2$CS (3-2)             & $ J=3, K_p = 0, K_o = 3 \rightarrow J=2, K_p=0, K_o=3$   & 103.040452 \\
        DCO$^+$ (1-0)      & $J = 1 \rightarrow J = 0$                                                                 &  72.039354 \\
        N$_2$H$^+$ (1-0) & $J=1, F1=2, F=3 \rightarrow J=0, F1=1, F=2$                                & 93.173764 \\
        \hline 
      \end{tabular}
    \end{center}
  \end{table}
}

\begin{document}

\title{Tracers of the ionization fraction in dense and translucent gas:\\
I. Automated exploitation of massive astrochemical model grids}

\titlerunning{Tracers of the ionization fraction in molecular clouds: I.}

 \author{Emeric Bron\inst{\ref{Meudon}} %
   \and  Evelyne Roueff\inst{\ref{Meudon}}%
   \and Maryvonne Gerin\inst{\ref{LERMA}} %
   \and Jérôme Pety\inst{\ref{IRAM},\ref{LERMA}} %
   \and Pierre Gratier \inst{\ref{LAB}} %
   \and Franck Le Petit\inst{\ref{Meudon}} %
   \and Viviana Guzman\inst{\ref{Catholica}} %
   \and Jan H. Orkisz\inst{\ref{Goteborg}} %
   \and Victor de Souza Magalhaes\inst{\ref{IRAM}} %
   \and Mathilde Gaudel\inst{\ref{LERMA}} %
   \and Maxime Vono\inst{\ref{IRIT}} %
   \and S\'ebastien Bardeau\inst{\ref{IRAM}} %
   \and Pierre Chainais\inst{\ref{Lille}} %
   \and Javier R. Goicoechea\inst{\ref{Madrid}} %
   \and Annie Hughes\inst{\ref{IRAP}} %
   \and Jouni Kainulainen\inst{\ref{Goteborg}} %
   \and David Languignon\inst{\ref{Meudon}}
   \and Jacques Le Bourlot\inst{\ref{Meudon}} %
   \and François Levrier\inst{\ref{LPENS}} %
   \and Harvey Liszt\inst{\ref{NRAO}} %
   \and Karin \"Oberg\inst{\ref{CFA}} %
   \and Nicolas Peretto\inst{\ref{UC}} %
   \and Antoine Roueff\inst{\ref{Marseille}} %
   \and Albrecht Sievers\inst{\ref{IRAM}} %
   }

 \institute{%
   LERMA, Observatoire de Paris, PSL Research University, CNRS,
   Sorbonne Universit\'es, 92190 Meudon, France. \label{Meudon}
   \and LERMA, Observatoire de Paris, PSL Research University, CNRS,
   Sorbonne Universit\'es, 75014 Paris, France. \label{LERMA} %
   \and IRAM, 300 rue de la Piscine, 38406 Saint Martin d'H\`eres,
   France. \label{IRAM} %
   \and Laboratoire d'Astrophysique de Bordeaux, Univ. Bordeaux, CNRS, B18N,
   Allee Geoffroy Saint-Hilaire,33615 Pessac, France.\label{LAB} %
   \and Instituto de Astrof\'isica, Pontificia Universidad Cat\'olica de Chile, Av. Vicuña Mackenna 4860, 7820436 Macul, Santiago, Chile \label{Catholica}%
   \and Chalmers University of Technology, Department of Space, Earth and Environment, 412 93 Gothenburg, Sweden \label{Goteborg}%
   \and University of Toulouse,  IRIT/INP-ENSEEIHT,  CNRS,  2  rue  Charles Camichel, BP 7122, 31071 Toulouse cedex 7, France \label{IRIT} %
   \and Univ.  Lille, CNRS, Centrale Lille, UMR 9189 - CRIStAL, 59651 Villeneuve d’Ascq, France \label{Lille} %
   \and Instituto de F\'isica Fundamental (CSIC). Calle Serrano 121, 28006, Madrid, Spain \label{Madrid} %
   \and Institut de Recherche en Astrophysique et Planétologie (IRAP), Université Paul Sabatier, Toulouse cedex 4, France \label{IRAP}%
   \and Laboratoire de Physique de l’Ecole normale supérieure, ENS, Université PSL, CNRS, Sorbonne Université,Université Paris-Diderot, Sorbonne Paris Cité, Paris, France \label{LPENS}%
   \and National Radio Astronomy Observatory, 520 Edgemont Road,
   Charlottesville, VA, 22903, USA. \label{NRAO} %
   \and Harvard-Smithsonian Center for Astrophysics, 60 Garden Street,
   Cambridge, MA, 02138, USA. \label{CFA} %
   \and School of Physics and Astronomy, Cardiff University, Queen's
   buildings, Cardiff CF24 3AA, UK. \label{UC} %
   \and Aix Marseille Univ., CNRS, Centrale Marseille, Institut Fresnel, Marseille \label{Marseille} %
 } %

\offprints{E. Bron, \email{emeric.bron@obspm.fr}}

\date{Received 27 March 2020 / Accepted 6 July 2020}

\abstract%
{The ionization fraction in the neutral interstellar medium (ISM) plays a key role in the physics and chemistry of the ISM, from controlling the  coupling of the gas to the magnetic field to allowing fast ion-neutral reactions that drive interstellar chemistry. 
Most estimations of the ionization fraction have relied on deuterated species such as DCO$^+$, 
whose detection is limited to
dense cores representing an extremely small fraction of the volume of the giant molecular clouds (GMC) they are part of. 
As large field-of-view hyperspectral maps become available, new tracers may be found. 
The growth of observational datasets is paralleled by the growth of massive modeling datasets, and new methods need to be devised to exploit the wealth of information they contain.} 
{We search for the best observable tracers of the ionization fraction based on a grid of astrochemical models, with the broader aim of finding a general automated method applicable to the search of tracers of any unobservable quantity based on grids of models.} 
{We build grids of models that sample randomly a large space of physical conditions (unobservable quantities such as gas density, temperature, elemental abundances, etc.) and compute the corresponding observables (line intensities, column densities) and the ionization fraction. 
We estimate the predictive power of each potential tracer by training a Random Forest model to predict the ionization fraction from that tracer, based on these model grids.} 
{In both translucent medium and cold dense medium conditions, several observable tracers with very good predictive power for the ionization fraction are found. 
Many tracers in cold dense medium conditions are found to be better and more widely applicable than the traditional DCO$^+$/HCO$^+$ ratio. 
We also provide simpler analytical fits for estimating the ionization fraction from the best tracers, and for estimating the associated uncertainties. 
We discuss the limitations of the present study and select a few recommended tracers in both types of conditions.} 
{The method presented here is very general and can be applied to the measurement of any other quantity of interest (cosmic ray flux, elemental abundances, etc.) from any type of model (PDR models, time-dependent chemical models, etc.).} 

\maketitle 

\section{Introduction}

The so-called neutral component of the interstellar medium, despite being shielded from EUV (13.6 to 124 eV) stellar photons able to ionize hydrogen, retains a small ionization fraction ($x(\mathrm{e}^-)=n(\mathrm{e}^-)/n_\mathrm{H}$). 
The ionization mechanism depends on the type of region: FUV (6 to 13.6 eV) photons ionizing C and S in the low $A_\mathrm{V}$ surface layer of clouds, or cosmic rays, X rays, shocks, etc., in the densest parts.
As a result, ionization fractions range between $\sim10^{-4}$ in low $A_\mathrm{V}$ cloud surfaces and down to $\sim10^{-9}$ in dense cores \citep[e.g.][]{Goicoechea2009,Draine2011}.

This ionization fraction controls several key aspects of neutral interstellar clouds. 
It determines the degree of coupling of the gas to the magnetic field: the neutrals, accounting for most of the mass of the fluid, are only indirectly sensitive to the presence of a magnetic field through their friction with the ions that remain coupled to the field, a process called ion-neutral friction. 
This coupling can provide a significant magnetic support against gravitational collapse of dense cores despite the low ionization fractions values found there, between 10$^{-9}$ and 10$^{-7}$ \citep{Mestel1956,Mouschovias1976,Basu1994}. The ionization fraction also controls the onset of the magneto-rotational instability \citep{Balbus1991}, the main mechanism of angular momentum transport in accretion disks.
Moreover, the gas phase chemistry in dense molecular clouds is to a large extent driven by fast ion-neutral reactions \citep{Herbst1973,Oppenheimer1974}. 
The build-up of chemical complexity thus depends on the ionization fraction of the medium.
Finally, some common molecular tracers with high dipole moments, such as HCN and HCO$^+$, have high inelastic collision cross sections with electrons, and their excitation can be significantly affected by electron collisions for ionization fractions $\gtrsim 10^{-5}$ \citep{Black1991,Liszt2012,Liszt2016,Goldsmith2017}. This makes the interpretation of their emission (e.g. to estimate gas density) sensitive to our knowledge on the local ionization fraction .

Direct observational estimation of the ionization fraction in neutral clouds is difficult, except in very specific regions (e.g. \citealt{Goicoechea2009,Cuadrado2019}, at the dissociation front in a photodissociation region). 
Direct estimation of the total charge accounted for by observable molecular ions in molecular clouds only yields a loose lower limit (e.g. \citealt{Miettinen2011}). 
Indirect methods based on tracers that are chemically sensitive to the ionization fraction have thus been commonly used. 
These methods have mostly involved measuring the deuterium fractionation through abundance ratios involving simple molecular ions like DCO$^+$/HCO$^+$ \citep{Guelin1977,Guelin1982,Dalgarno1984,Caselli1998} or N$_2$D$^+$/N$_2$H$^+$ which is less affected by depletion \citep{Caselli2002a}. 
The idea is that the deuterium enrichment (defined as H$_2$D$^+$/H$_3^+$), initiated by the exchange reaction 
\begin{equation}
  \mathrm{H}_3^+ + \mathrm{HD} \rightleftharpoons \mathrm{H}_2\mathrm{D}^+ + \mathrm{H}_2
  \label{eq:deut_reac}
\end{equation}
at low temperature, is limited by electronic dissociative recombination of H$_2$D$^+$, and that the resulting ratio is transmitted (with a known prefactor) to the deuteration fraction of other molecules such as HCO$^+$. 
Using such tracers, the ionization fraction is deduced either by using approximate analytical formulae representing simplified networks \citep{Caselli1998,Miettinen2011,Caselli2002a}, or by adjusting an astrochemical model including a full chemical network to the observations, using stationary chemical models \citep{Williams1998,Bergin1999,Caselli2002b,Fuente2016}, time-dependent ones \citep{Maret2007,Shingledecker2016}, or PDR models \citep{Goicoechea2009}. 
Despite the variety of determination methods, using different deuterated molecules, only very few works have proposed using other tracers than deuterated species. 
For instance, \citet{Flower2007} proposed the C$_6$H$^-$/C$_6$H ratio as a tracer of the ionization fraction, and \citet{Fosse2001} have investigated the relationship between the cyclic-to-linear ratio of C$_3$H$_2$ and the ionization fraction. 
Deuteration-based approaches however suffer from several limitations due to the fact that they depend on other physical or chemical parameters that need to be determined independently.
The initial deuteration reaction (Eq.~\ref{eq:deut_reac}) is sensitive not only to the gas temperature but also to the essentially unmeasurable ortho-to-para ratio of H$_2$  \citep{Pagani1992,Pagani2011,Shingledecker2016}.
Indeed, the endothermicity of the reaction in the backward direction (192K) is very close to the J=1 to J=0 energy difference of H2 (170.5K). Then, even a small fraction of o-H2 (J=1) contributes to H2D+ destruction (with a reduced endothermicity of $\simeq$ 20K) and restricts the deuteration process.
In addition, ratios such as DCO$^+$/HCO$^+$ are linked to H$_2$D$^+$/H$_3^+$ through reactions with neutral species like CO. 
The estimated deuterium fraction is therefore sensitive to the depletion factors of carbon, oxygen and nitrogen that are not easy to evaluate \citep{Caselli2002a}. 
Moreover, deuterated tracers such as DCO$^+$ are typically only detectable in cold dense cores, representing only a tiny fraction of the observable area of a giant molecular cloud (GMC).
Deuteration-based approaches are thus inadequate for an unbiased characterization of the conditions in GMCs as a whole.

Despite the common use of advanced chemical models computing the abundances of hundreds of species, the observed tracers to which these models are compared to estimate $x(\mathrm{e}^-)$ (deuterated molecules such as DCO$^+$) are still those initially proposed based on analytical reasoning using simplified chemical networks.
The wealth of data produced in large chemical model grids remains largely unexploited. 
Their exploration of wide parameter spaces might reveal less intuitive but more efficient tracers. 
Based on this approach, we propose here a general and largely automatic method to identify the best observational predictors of the ionization fraction, when other important parameters such as the gas density, temperature or H$_2$ ortho-to-para ratio are unknown. 
We apply this method to propose new predictors of the ionization fraction as a function of the molecular cloud conditions. 
We use simple stationary chemical models with a complete up-to-date chemical network \citep{Roueff2015}, and use molecular ratios (column density ratios or integrated line intensity ratios) as observable tracers from which we seek to predict the ionization fraction.
We base our investigation on the observed range of physical conditions and detected tracers in the IRAM-30m Large Program ORION-B (Outstanding Radio-Imaging of OrioN B, co-PIs: J. Pety and M. Gerin) \footnote{Informations and data related to the ORION-B program can be found at \url{http://www.iram.fr/~pety/ORION-B/}}.
In this program, we imaged 5 square degrees towards the southern part of the Orion B giant molecular cloud over most of the 3 mm atmospheric window~\citep{Pety2017,Gratier2017,Orkisz2017,Bron2018,Orkisz2019,Roueff2020,Gratier2020}.

In the context of dense cores, the ionization fraction is linked with the cosmic ray ionization rate (CRIR) and both are often studied from the same molecular ratios (although direct tracers of the cosmic ray ionization rate can also be used in more diffuse medium, in particular H$_3^+$, e.g. \citealt{Indriolo2012,LePetit2016}). In our context of the Orion B giant molecular cloud, where UV illumination controls the ionization fraction in large parts of the cloud, we focus here on the question of estimating the ionization fraction only, independently of the source of ionization. The task of tracing the cosmic ray ionization rate (which has also been attempted using astrochemical model grids, e.g. \citealt{Barger2020}) will be considered in a future application of the method presented here.

In this first article, we present a generic method to find the best tracers of an unobservable physical parameter and apply it to the search of new tracers of the ionization fraction among the species that are detectable in the ORION-B dataset. The observational application of the tracers found here to study the ionization fraction in the Orion B molecular cloud will be presented in a second paper \citep{Guzman2020}.

In Sect.~\ref{sect:Method}, we describe our general statistical method for determining the best predictors of a given unobservable parameter based on the results of a model grid.
In Sect.~\ref{sect:Models}, we present the models used in this study for the search of ionization fraction tracers. 
We then present the ranking of observable predictors in Sect.~\ref{sect:Tracers}. 
For ease of application of our results, we provide in Sect.~\ref{sect:AnalyticalFits} analytical fit formulae to deduce the ionization fraction from each of the proposed best predictors.
We finally discuss our results in Sect.~\ref{sect:Discussions} and present our conclusions in Sect.~\ref{sect:Conclusions}.

\section{Method}
\label{sect:Method}

Both observable line intensities (or column densities) on the one hand, and the ionization fraction $x(\mathrm{e}^-)$ on the other hand, depend on multiple, unobservable physical parameters (e.g. gas density, elemental abundances, cosmic ray flux, ...). 
Our goal is to find reliable relationships between observable quantities and ionization fraction, despite lacking estimations of these hidden physical parameters. 
To do this, we first run model grids covering the whole possible parameter space. 
We then use a flexible regression method to fit $x(\mathrm{e}^-)$ as a function of one of the potential observational tracers through the whole grid of chemical models.
This means that we treat the effects of the variations of the hidden parameters as sources of noise on the prediction of the ionization fraction. 
Finally, we use a quantitative measurement of the fit quality as an estimate of the predictive power of each potential tracer. 
These estimates are used to rank the tracers and highlight the most powerful predictors of the ionization fraction. 
The fitted models for the best tracers will provide ready-to-use tools to be applied to observations.

\subsection{Predicting the ionization fraction from one ratio of line intensities (or column densities) with Random Forests}

Any a priori information could easily be included in the method by sampling the parameters of the model grid according to a specific prior distribution. 
However, we wish to minimize the amount of a priori information injected in the method and to avoid making assumptions on the shape of the distributions of physical parameters (e.g. gas density, elemental abundances, cosmic ray flux, ...). 
We thus build a model grid that samples uniformly the possible range of values (see Sect.~\ref{sect:Models}).

Our model grids provide us with a dataset comprising ionization fraction values and corresponding values of observable quantities. 
We will consider line intensity ratios or column density ratios as our observable quantities in this paper. 
The hidden physical parameter values introduce a non deterministic aspect to the relationship between $x(\mathrm{e}^-)$ and the observables: models might have identical values of an observable but different $x(\mathrm{e}^-)$ if the underlying physical parameters are different. 
Learning to predict $x(\mathrm{e}^-)$ from a given observable is then a regression problem, with the uncertainty introduced by the hidden physical parameters playing the role of noise. 
Determining the best tracers of $x(\mathrm{e}^-)$ is thus equivalent to finding observables for which the relationship to $x(\mathrm{e}^-)$ is least affected by this noise (i.e. by the hidden physical conditions). 
This means finding the observables for which the most accurate regression model can be found.

For this regression problem, we choose to use Random Forests \citep{Breiman2001} because their flexibility makes it possible to fit general non-linear shapes, while their simplicity provides reasonable computational costs. 
This makes the method presented here very general and applicable to finding tracers of other physical parameters without any assumption on the shape of the relationship between the tracers and the target parameter. 
We will use RF for Random Forest in the rest of the paper.
RF regression models are based on the concept of regression trees \citep{Breiman84}, where a succession of binary decisions are made based on the input variables (e.g. $x_3 < 2$ or $\ge2$) and constant values are predicted in each of the subsets of the partition that the decision tree defines. 
While such decision trees are easily interpretable, they require large tree depths to be flexible but are prone to overfitting if this depth is too large. 
RF tackle this overfitting problem by using the simple idea that multiple overfitted regression models will, when averaged, give a better prediction as long as the errors they individually make are uncorrelated between models. 
In a RF, the individual trees are made as independent as possible by introducing randomness in two aspects: 1) the building of each tree only considers a random subset of the input variables, and 2) each tree is given a bootstrapped sample (i.e. drawn by random sampling with replacement from the original dataset, \citealt{Breiman1996}) instead of the original full sample. 
This way, the datasets seen by the different trees are independent and each tree only sees a subset of the dataset (bootstrapped datasets typically contain only 63\% of the points of the original dataset as repetition is allowed). 
This provides a very flexible regression model, which retains some of the interpretability of decision trees. 
RF have thus quickly become a standard method in Machine Learning (see e.g. \citealt{Hastie2001}). 
In addition, they allow to estimate the generalization error of the fit (i.e., the error made when predicting data not seen during training): as each tree has only seen a random bootstrapped sample from the data, it is possible to estimate for each datapoint a partial prediction using only the trees that have not seen this datapoint during training.
As the sample seen by a given tree is called a bag, these partial predictions are called out-of-bag predictions (OOB). 
\cite{Gratier2020} also use RF in the context of the interstellar medium and introduce the method in detail.

We thus train RF regression models for each observable (using only one observable at a time) and estimate the accuracy of the regression models. 
This accuracy is taken as an estimate of the predictive power of the observable quantity considered, for the purpose of predicting $x(\mathrm{e}^-)$.
The different observables can then be ranked according to this predictive power estimate.
The accuracy of the regression model is estimated with the OOB $R^2$
\[
  R^2 = 1 - \frac{S\!S_\mathrm{res}}{S\!S_\mathrm{tot}}
  \quad \mbox{with the sums of squares}
\]
\[
  S\!S_\mathrm{res} = \sum_i \left( y_i^{pred} - y_i^{true}\right)^2
  \quad \mbox{and} \quad
  S\!S_\mathrm{tot} = \sum_i \left( \overline{y^{true}} - y_i^{true}
    \right)^2,
\]
where the index $i$ runs across data points (individual model results), $y_i^{true}$ is the true value of ionization fraction (computed by the chemical model), $y_i^{pred}$ is the OOB prediction value from the RF, and $\overline{y^{true}}$ is the average of the (true) ionization fraction over the model grid. 
This coefficient $R^2$ gives the fraction of the total ionization fraction variance (across the full model grid) that the RF model can explain from the given observable predictor alone (i.e. it measures the fractional decrease from the initial variance of $x(\mathrm{e}^-)$ to the variance of the residuals).

It is thus <1, with 1 representing perfect prediction (zero residual variance). Note that it can take a negative value when the model performs worse than predicting a constant value set at the average $x(\mathrm{e}^-)$ of the dataset. A value of 0 indicates a performance equivalent to this constant prediction of the average.
This $R^2$ value is used for the ranking of tracers. 
For information, we also provide  below the root mean square error
\[
  \mathrm{RMSE} = \sqrt{ \frac{1}{N} \sum_i \left( y_i^{true} -
      y_i^{pred}\right)^2 },
\]
where $N$ is  the number of chemical models in our grid. 
The RMSE is completely univocally related to the $R^2$ value, but is more interpretable in terms of the amplitude of typical errors. 
We also provide the maximum absolute error
\[
  \mathrm{max.\,\,abs.\,\,err.} = \max _i \left| y_i^{true} -
    y_i^{pred}\right|\jp{.}
\]
This quantity, estimating the maximum error made by our regression model, is not guaranteed to converge when increasing the size of the dataset. 
It should thus not be interpreted further than being the largest error we observed in our limited-size sample.

The RF model depends on a few internal parameters (number of trees, maximum depth of trees, etc...). 
Their values can affect the quality of the model and its tendency to overfit. 
We used a number of trees in the forest $N_\mathrm{trees} = 400$ and a maximum tree depth $d_\mathrm{max} = 4$. 
The procedure used to select these values is described in Appendix~\ref{app:RF_params_optimization}.
Our tests show that this optimization scheme is not critical for our purpose: while the choice of parameter values does affect the quality of the best fit RF model, it does not change significantly the ranking of the predictors that we deduce from it.

The pipeline tool implementing this procedure is available at [link to be added before publication].

\section{Chemical models}
\label{sect:Models}

  \begin{table}
  \small
    \caption{Range of physical parameters explored for each of our two classes of medium: gas density $n_\mathrm{H}$, gas temperature $T_\mathrm{gas}$, incident FUV radiation field intensity $G_0$, line-of-sight visual extinction $A_\mathrm{V}$, cosmic-ray ionization rate $\zeta$, H$_2$ ortho-to-para ratio OPR$_{\mathrm{H}_2}$, depletion factor and sulfur gas-phase elemental abundance [S].}
    \label{tab:grid_ranges}
    \begin{center}
      \begin{tabular} {c|ccc}
        \hline 
         & translucent medium & cold dense medium \\ 
\hline
\nH [cm$^{-3}$] & $3\times10^2-3\times10^3$ & $10^3-10^6$ \\
T$_\mathrm{gas}$ [K] & $15-100$ & $7-20$ \\
G$_0$ & $1-1000$ & 1 \\
\Av & $2-6$ & $5-20$ \\
$\zeta$ [s$^{-1}$] & $10^{-17}-10^{-15}$ & $10^{-17}-10^{-16}$ \\
OPR$_{\mathrm{H}_2}$ & $0.1-3$ & $10^{-4}-10^{-1}$ \\
depletion factor & $1$ & $1-10$ \\
$[$S$]$ & $1.86\times10^{-8}-1.86\times10^{-5}$ & $1.86\times10^{-8}-1.86\times10^{-5}$ \\

        \hline 
      \end{tabular}
    \end{center}
  \end{table}

We use here the chemical code presented in \cite{Roueff2015} to study isotopic fractionation of deuterium, carbon and nitrogen compounds. 
Single zone models with fixed density, temperature, visual extinction, radiation field, cosmic ray ionization rate, ortho-to-para H$_2$ ratio and depletion factors are computed at steady state. 

We consider for the present study a chemical network including deuterium, and isotopic carbon and oxygen species where the deuterium, carbon and oxygen fractionation reactions have been introduced following the recent determinations of exothermicities by \cite{Mladenovic2017}.
We introduce in particular D$^{13}$CO$^+$. 
Apart from these specific fractionation reactions, the chemistry of isotopically substituted species is built automatically from the chemical network of the major components. 
The chemical reactions involving one single carbon-containing reactant and one single carbon-containing product are duplicated with the same reaction rate coefficient. 
Simple statistical assumptions are introduced when two carbon containing molecules are implied in the reaction.
Consider for example the case of the reaction
$$\mathrm{CX} + \mathrm{CY} \rightarrow \mathrm{CX}' + \mathrm{C Y}', $$ taking place with a reaction rate coefficient $k$.
The reactions introduced for the isotopically substituted species are the following: 
\begin{align*}
^{13}\mathrm{CX} + \mathrm{CY} &\rightarrow \vphantom{a}^{13}\mathrm{CX}' + \mathrm{CY}'        \quad \mathrm{with} \,\,k/2 \\
^{13}\mathrm{CX} + \mathrm{CY} &\rightarrow \mathrm{CX}' + \vphantom{a}^{13}\mathrm{CY}'         \quad \mathrm{with} \,\,k/2 \\
\mathrm{CX} + \vphantom{a}^{13}\mathrm{CY} &\rightarrow \mathrm{CX}' + \vphantom{a}^{13}\mathrm{CY}'          \quad \mathrm{with} \,\,k/2 \\
\mathrm{CX} + \vphantom{a}^{13}\mathrm{CY} &\rightarrow \vphantom{a}^{13}\mathrm{CX}' + \mathrm{CY}'          \quad \mathrm{with} \,\,k/2 \\
^{13}\mathrm{CX} + \vphantom{a}^{13}\mathrm{CY} &\rightarrow \vphantom{a}^{13}\mathrm{CX}' + \vphantom{a}^{13}\mathrm{CY}’  \quad \mathrm{with} \,\,k
\end{align*}
Such a procedure leads to an ensemble of 310 species linked through 8711 chemical reactions. 

These models allow us to compute observable column density ratios for the hundreds of species included. 
Although commonly derived by observers, column densities are not the primary observable quantities, and we thus also compute line intensity ratios. 
To do so, we post-process the results of our chemical models using a non-LTE excitation and radiative transfer model (RADEX, \citealt{vanderTak2007}) assuming a typical linewidth of 1 km/s (observed linewidths in the Orion B are typically of a few km/s \citep{Pety2017}).
Results based on column density ratios and based on line intensity ratios will be presented separately in the following sections.

It is unlikely that a single tracer will provide a good estimate of the ionization fraction $x(\mathrm{e}^-)$ in all possible physical conditions. 
Either the tracer will lose its relationship with $x(\mathrm{e}^-)$ in some conditions, or it might be too weak to be observable in other conditions. 
We thus decided to divide the range of possible conditions into subregions corresponding to the different types of environments found in GMCs \citep{Pety2017,Bron2018}. 
We focus on two kinds of environments : the translucent medium and the cold and dense gas, and we derive separate rankings of tracers for these two environments. 
The range of physical conditions explored for each of these environments are chosen based on our previous studies of Orion B and listed in Table~\ref{tab:grid_ranges}.

In both grids, the gas density and gas temperatures were varied covering the typical ranges for translucent medium ($3\times 10^2 - 3\times 10^3$ cm$^{-3}$, 15-100 K) and cold and dense medium ($10^3 - 10^6$ cm$^{-3}$, 7-20 K). 
In the translucent model grid, external FUV photons still play an important role in the chemistry and in controlling the ionization fraction. 
This FUV illumination is controlled through an external FUV field strength $G_0$ (see e.g. \citealt{Hollenbach1999}, p. 177) and an extinction $A_V$ representing the amount of shielding between the FUV source and the gas under consideration (also used as the depth of the slab when computing line intensities). 
We take into account self-shielding of H$_2$ by using the approximate expression of \cite{Draine1996} and introduce also the shielding of CO by H$_2$ from \cite{Heays2017}. 
We consider lower extinctions and higher $G_0$ values in translucent medium ($A_V$ in the range 2-6, $G_0$ of the external field in the range 1 - 1000) than in cold dense medium ($A_V$ in the range 5-20, external $G_0$ set to 1).
We explore average to moderately strong FUV illumination values in the low density translucent grid. 
Regions with both high density high FUV illumination correspond to dense photodissociation regions (PDR), in which strong chemical and physical stratification on small spatial scale is critical. 
These regions would thus require the use of complete PDR models, such as the Meudon PDR Code \citep{LePetit2006}. 
We thus did not explore this type of conditions in the present study. 

Given the uncertainties about the cosmic ray ionization rate in molecular clouds \citep{Lepp1992,McCall2003,Indriolo2007}, we consider the range of value $10^{-17}-10^{-15}$ s$^{-1}$. 
In the cold gas conditions, in order to account for the reduced cosmic ray fluxes \citep{Padovani2009}, we limit this range to $10^{-17}-10^{-16}$ s$^{-1}$. 
As sulfur can be an important contributor of electrons in neutral gas but has a highly uncertain gas-phase elemental abundance \citep{Agundez2013,Goicoechea2006}, we explore in both grids values of [S], the relative sulfur abundance with respect to H, in the range $1.86\times 10^{-8} - 1.86\times 10^{-5}$. 

We also explore ranges of H$_2$ ortho-para ratio (OPR$_{\mathrm{H}_2}$) which impact significantly two reactions: i.e. H$_2$D$^+$ + o-H$_2$ $\rightarrow$ H$_3^+$ + HD \citep{Pagani1992} where the energy endothermicity is reduced to 61.5 K (compared to 232K with p--H$_2$) and N$^+$ + o-H$_2$ $\rightarrow$ NH$^+$ + H which is slightly endothermic ($\sim$ 44 K) whereas N$^+$ + p-H$_2$ $\rightarrow$ NH$^+$ + H is more strongly endothermic ($\sim$ 170 K), as first emphasized by \cite{LeBourlot1991}.
For this reaction, we follow the prescription of \cite{Dislaire2012} which is derived from experimental results.
We use higher values of the OPR in the warmer translucent medium ($0.1-3$) than in cold dense gas ($10^{-4} - 10^{-1}$). 
Finally, cold dense cores offer conditions where molecules can freeze out on dust grains, depleting the gas phase abundances of elements such as C and O. 
In our cold dense medium grid, we thus in addition explore depletion factors going from 1 (no depletion) to 10 (C elemental abundance 10 times lower than the reference values) with a constant C/O elemental ratio value of 0.6 (the elemental abundance of carbon is taken to be [C]$=1.32\times 10^{-4}$ when there is no depletion). 
Other parameters that might have an impact (although second order compared to the parameters considered here), such as variations of the metal elemental abundances or PAH abundance, were not considered in this study. The gas-phase elemental abundances for metals (relative to H) are taken to be [Fe]$=1.5\times 10^{-8}$, [Cl]$=1.8\times 10^{-7}$, [Si]$=8.2 \times 10^{-7}$, [F]$=1.8\times 10^{-8}$, [Ar]$=3.29\times 10^{-6}$. 

For each medium type, a set of 5000 models was run, sampling randomly and uniformly within the chosen parameter space. The adequacy of this number of models for our purpose is ascertained later when estimating the uncertainties on our results.

Given the variation by orders of magnitude both in the parameter values and the computed observables, we choose to work with the logarithm of all quantities.
We sampled uniformly on the logarithm of each parameter within the ranges indicated above. 
The method described in Sect.~\ref{sect:Method} is applied on the logarithm of all quantities (i.e., training RF models using the logarithm of column density ratios or line intensity ratios to predict the logarithm of $x(\mathrm{e}^-)$). 
Note that representative error values such as the RMSE on logarithms are equivalent to representative error \emph{factors} on the actual quantity. 
These corresponding error factors are given in parenthesis in the result tables of the following sections.

To get rid of possible instrumental, calibration and other source geometry effects, we choose to work only on ratios of observable quantities (either column density ratios or line intensity ratios). 
In the following, we use the term tracers for ratios of observable quantities.

Among all the species computed in our chemical model, we made a selection of species that are detected in the radio observations of the ORION-B project and potentially linked to the ionization fraction. 
Our search for the best tracers is made among the ratios of these selected species. 
For the translucent medium condition, we selected $^{13}$CO, C$^{18}$O, HCO$^+$, HCN, HNC, CN, C$_2$H, CS, SO, H$_2$CS, HCS$^+$, CF$^+$. 
The search for a best ratio was thus done among 66 possible column density ratios (and 66 line intensity ratios). 
For dense cold medium conditions, we considered the same selection with the addition of DCO$^+$ and N$_2$H$^+$. 
We thus had 91 possible column density ratios (and 91 line intensity ratios) in this case.
For line intensities, the exact transition and frequency considered for each species are listed in Table~\ref{tab:LineIDs}.
In the RADEX computations of line intensities, we account for collisional excitation with electrons (using the ionization fraction computed by the chemical model) for species for which collisional data with electrons are available in RADEX (HCO$^+$, HCN and C$_2$H). 
We note that for CN, excitation by electrons was not included as in the current version of RADEX, the collisional data that includes the hyperfine structure of CN does not include collisions with electrons. 
We chose to privilege the fact of accounting for the hyperfine structure here.

\TabLineIDs

\section{Tracers rankings}
\label{sect:Tracers}

We applied the method described in Sect.~\ref{sect:Method} to the two chemical model grids (translucent medium conditions and cold dense medium conditions) presented in Sect.~\ref{sect:Models} in order to obtain a ranking of the selected potential tracers according to their usefulness for predicting the ionization fraction.

\subsection{Translucent medium}

\FigRankingSingleRatiosTranslucent

\FigRFmodelBestRatiosTranslucent

Figure~\ref{fig:RankingSingleRatiosTranslucent} presents the predictive power (estimated as the $R^2$ of a RF fit) of each tracer for the best 20 tracers. 
The left panel shows the result when taking column density ratios as observable quantities, and the right panel the results when considering line intensity ratios. 
We see that the ranking is similar in both cases, suggesting that excitation and radiative transfer only have a moderate effect on the relationship between these tracers and the ionization fraction. 
A more complete ranking (covering all tracers having $R^2>0.5$) is given in Table~\ref{tab:RatiosRankingColumnDensitiesTranslucent} (due to their sizes, the results tables of this Section and of Sect.~\ref{sect:AnalyticalFits} are placed in Appendix~\ref{app:Rankings})\footnote{Datafiles containing the training RF models and tables of the rankings presented in this section are available at [the link will be added in  the published version].}.

In both cases, about ten different ratios are found to be each able to explain more than 80\% of the ionization fraction variance ($R^2 > 0.8$). 
We emphasize that this means that an accurate prediction of the ionization fraction is possible from each of these tracers despite not knowing the values of the 7 physical and chemical parameters that have been varied in our model grid (cf. Table~\ref{tab:grid_ranges}). 
The $R^2$ values are slightly lower when using line intensity ratios rather than column densities ratios, indicating that excitation and radiative transfer effects tend to increase the degeneracy between the ionization fraction and other unknown parameters, but this effect remains moderate. 
To illustrate the performance of the tracers found with this ranking, we show on the left panel of Fig.~\ref{fig:RFmodelBestRatiosTranslucent} the ionization fraction versus the best ranked column density ratio (C$_2$H/HCN) in our grid of models for translucent medium conditions (blue symbols, with contours in shades of blue indicating iso-PDF contours encompassing 25\%, 50\%, and 75\% of the distribution) and the prediction of the fitted RF model (solid red line), which is found in this case to explain 95.7\% of the ionization fraction variance in our grid. 
The remaining scatter around the relationship represents the effect of ignoring all other parameters (gas density, temperature, UV field, H$_2$ OPR,...).
The best ranked line intensity ratio (C$_2$H (1-0) / HCN (1-0)) is similarly shown on the right panel of Fig.~\ref{fig:RFmodelBestRatiosTranslucent} with the corresponding fitted RF model (solid red line).

In translucent gas, the ionization is still dominated by the effect of external FUV photons ionizing carbon (and to a lesser extent sulfur and chlorine), and is slowly decreasing as the total extinction increases. 
In the conditions covered by our translucent grid, we find $x(\mathrm{e}^-)$ ranging from $2\times10^{-4}$ to $2\times10^{-7}$. 
C$_2$H, which we find in several of the best ratios, is known to be enhanced in FUV illuminated environments \citep{Pety2005,Cuadrado2015,Guzman2015,Gratier2017,Pety2017}: as explained in \cite{Beuther2008}, C$_2$H traces the amount of carbon not locked into CO, and is thus sensitive to the FUV flux through CO photodissociation and the presence of C and C$^+$ at a significant abundance level.
In our translucent medium model grid, we indeed find C$^+$ to be the main charge carrier and thus to very strongly correlate with $x(e^-)$. H$^+$ and H$_3^+$ may also contribute to the electronic fraction in environments where the cosmic ionization reaches values above $10^{-16}$ s$^{-1}$ \citep{LePetit2016}. However, C$^+$, an open shell ion, is chemically reactive with various molecules, except H$_2$\footnote{Except in strong PDR environments, where a small fraction of vibrationally excited H$_2$ may overcome the endothermicity barrier. Strong PDR environments are not considered here.}, and is at the origin of a complex chemistry with insertion of carbon atoms. C$^+$ itself is not straightforwardly detectable as its fine structure transition at 158 $\mu$m requires spaceborne or airborne observations. But we may expect that molecules involving C$^+$ in the initial chemical steps allow to probe the electronic fraction.
Our finding of ratios involving C$_2$H (relative to e.g. HCN, HNC or CN) as good proxies of the electronic fraction is a natural consequence of the relevance of C$^+$ as one of the main positive charge carriers. The initial step of C$_2$H formation involves indeed the C$^+$ + CH $\rightarrow$ C$_2^+$ + H reaction, followed by subsequent reactions with H$_2$ up to C$_2$H$_2^+$, which recombines to form C$_2$H.
Molecules such as HCN, HNC or CO and its isotopologues, on the other hand, are saturated stable molecules which scale with column density. As a result, ratios such as C$_2$H/HCN, whose transitions are easily detectable, offer a convenient diagnostic tool of the electronic fraction in translucent medium. The electronic fraction is then, as shown in Fig. \ref{fig:RFmodelBestRatiosTranslucent} (left panel), an increasing function of the C$_2$H/HCN ratio. 

CF$^+$ is another proxy of C$^+$, as described in \cite{Neufeld2006,Guzman2012}, and ratios involving this ion are also found here to be good tracers of the ionization fraction. However, this ion is relatively scarce since it involves fluorine, which has a low relative abundance to H$_2$ and has only been detected in PDR environments so far (detectability issues are investigated in Sect. \ref{sec:noise}).

In order to estimate the reliability of our results and determine if our 5000-model grid is sufficient to explore the chosen parameter space for our purpose, we compute errorbars on the predictive power estimate (the $R^2$ of the RF fit). 
To do so, we use 10-fold cross validation: the model grid is randomly split into 10 parts, and for each of these parts, a RF model is trained on the other 9 parts and tested on the remaining part which it has not seen during training. 
From these 10 estimates of the $R^2$ (all made on samples unseen during training), a reliable estimate of the predictive power on unseen data is made, as well as an estimate of the standard error on this predictive power. 
The corresponding error bars are also shown in Fig.~\ref{fig:RankingSingleRatiosTranslucent}. For most of the points, the errorbars are smaller than the marker, and the inset on the left panel presents a zoom on the first five ratios, showing the magnitude of the errorbars. This shows that the uncertainty induced by the finite size of our model grid in negligibly small and that our 5000-model grid is sufficient for our purpose. 
However, this conclusion should be taken with some caution as it has been shown that there exists no unbiased estimator of the variance of a cross-validation estimate \citep{Bengio2004} and that a naive estimation of this variance tends to underestimate it by a factor of up to 4 \citep{Varoquaux2018}.

\subsection{Cold dense medium}

\FigRankingSingleRatiosColdDense

\FigRFmodelBestRatiosDense

Figure~\ref{fig:RankingSingleColdenRatiosColdDense} presents the ranking of the best tracers in cold dense conditions (cf. Table~\ref{tab:grid_ranges}), for both column density ratios (left panel) and line intensity ratios (right panel). 
Error bars computed by the cross validation procedure described above are also shown, confirming that the size of our model grid is sufficient for estimating the quality of the fits based on each tracer. 
We see that the $R^2$ values of the best tracers are slightly lower than in the translucent medium case, indicating slightly stronger degeneracies with unknown parameters in this case (note that depletion was varied in this cold dense medium, in addition to the parameters varied in the translucent medium grid). 
However, we still find several tracers explaining more than 80\% of the variance in $x(\mathrm{e}^-)$. 
The best column density ratio is here found to be CN/N$_2$H$^+$.
The cold dense environments are essentially ionized through cosmic rays and secondary UV photons induced by cosmic rays. 
Electrons are primarily produced by cosmic ray ionization of H$_2$ and destroyed in the efficient dissociative recombination reactions of the various molecular ions. 
He$^+$ ions, produced by cosmic ray ionization of He, are also particularly efficient in ionizing the molecular reservoirs, CO, HCN, N$_2$, H$_2$O. 
This contributes to forming atomic ions, in addition to the H$_3^+$ molecular ion resulting from H$_2$ ionization and the other stable molecular ions resulting from proton transfer of H$_3^+$ with stable molecules giving ions such as H$_2$D$^+$, HCO$^+$, H$_3$O$^+$, or N$_2$H$^+$.
The ionization carriers are then shared amongst several different species, going from the simple atomic ions that do not react with H$_2$ (i.e. C$^+$, S$^+$, H$^+$) and closed shell molecular ions such as H$_3^+$, HCO$^+$, H$_3$O$^+$, and N$_2$H$^+$. 
Molecular ions are principally destroyed by dissociative recombination reactions whereas atomic ions rather react with the present neutral molecules since radiative recombination is not efficient. One can thus expect that molecular ions are inversely proportional to the electron abundances, as seen with the CN/N$_2$H$^+$ ratio which is found to increase monotonically with $x(e^-)$ (cf. Fig. \ref{fig:RFmodelBestRatiosDense}, left panel).

As in the translucent case, we find slightly lower $R^2$ values for the best line intensity ratios than for the best column density ratio.
However, we find a few ratios that have better scores as intensity ratios than as column density ratios. In particular, while the $^{13}$CO/HCO$^+$ column density ratio is found to be a poor predictor of the ionization fraction, the $^{13}$CO (1-0) / HCO$^+$ (1-0) line intensity ratio appears as one of the best tracers. Contrary to the previous cases, this is entirely an excitation effect. The abundance ratio of $^{13}$CO/HCO$^+$ is found to be mostly uncorrelated with $x(e^-)$ and mostly constant in the cold dense medium model grid (ratio of about $10^3$ with a typical scatter of a factor of 2-3). However, HCO$^+$ and $^{13}$CO have strongly different critical densities: $\sim 2 \times 10^5$ cm$^{-3}$ for HCO$^+$ (in cold dense medium conditions, $x(e^-)$ is too low for electron collisions to play a significant role) in comparison to $\sim 2 \times 10^3$ cm$^{-3}$ for $^{13}$CO. In the range of densities considered in the cold dense medium grid ($10^3 - 10^6$ cm$^{-3}$), $^{13}$CO excitation is thus mostly at local thermodynamic equilibrium (LTE) and its emissivity per molecule is thus constant with gas density, while HCO$^+$ is transitioning from the sub-thermally excited regime to the LTE regime, and its emissivity per molecule thus increases with density. The $^{13}$CO (1-0) / HCO$^+$ (1-0) ratio thus decreases with gas density. On the other hand, we find the gas density to be very strongly anti-correlated with $x(e^-)$ in these conditions (with a typical scatter of a factor of $\sim 3$), as cosmic ray ionization is the dominant source of ionization here and the recombination rates per ion scale with the gas density. These two effects combine to give a $^{13}$CO (1-0) / HCO$^+$ (1-0) ratio that increases with $x(e^-)$ with a relatively tight correlation (cf. Fig. \ref{fig:AnalyticalmodelBestSixDenseIntensityRatios}, top right panel).

Fig.~\ref{fig:RFmodelBestRatiosDense} shows the relation between $x(\mathrm{e}^-)$ and the best column density ratio, CN/N$_2$H (left panel), and the best line intensity ratio, CF$^+$ (1-0) / DCO$^+$ (1-0) (right panel). 
We see a larger scatter than in Fig.~\ref{fig:RFmodelBestRatiosTranslucent}, but a clear relationship is still found.

We note that the classical DCO$^+$/HCO$^+$ ratio does not appear among the best tracers found here for cold dense medium conditions. 
This point is discussed in Sect.~\ref{sec:trad_tracers}.

\section{Analytical fit formulas}
\label{sect:AnalyticalFits}

If possible, we recommend using the RF models described in the previous section when attempting to estimate $x(\mathrm{e}^-)$ from one of the best tracers listed above. 
However, the provided datafiles are dependent on a specific implementation of Random Forests (the \texttt{scikit-learn} module for Python, \citealt{Pedregosa2011}). 
For a simpler and more persistent solution (independent of any external software), we provide in this section simple analytical fit formulae for the best tracers found in Sect.~\ref{sect:Tracers}. 
While the RF models are flexible enough to make the method described in Sect.~\ref{sect:Method} generally applicable to any model grid and any physical quantity we want to find tracers of, the analytical fits provided here use formulas that have been specifically chosen for the application presented here (finding predictors of the ionization fraction from our chemical model grid). 
There is no guarantee that these same formulae would perform adequately to find analytical fits with other model grids and/or another quantity to predict.

\subsection{Prediction formulae}

We use simple polynomial formulae (working as before on the logarithm of both the observable ratios and the ionization fraction) to fit the non-linear relationships between the best tracers found in Sect.~\ref{sect:Tracers} and $x(\mathrm{e}^-)$. 
This is applied to all tracers which where found to have $R^2>0.5$ in the previous RF analysis.

In the cold dense medium conditions,  we thus use a simple polynomial of order 5:
\begin{equation}
  f^\mathrm{dense}(x) = a_0 + a_1 x + a_2 x^2 + a_3 x^3 + a_4 x^4 + a_5 x^5 \label{eq:main_fit_dense}
\end{equation}
where $x$ is the $\log_{10}$ of the column density ratio or line intensity ratio from which we want to predict $\log_{10}(x(\mathrm{e}^-))$, $f^\mathrm{dense}(x)$ is our fitting function to $\log_{10}\left(x(\mathrm{e}^-)\right)$ in cold dense gas conditions\footnote{We use the notation $\log_{10}$ for the logarithm in base 10, and $\log_e$ for the natural logarithm.}, and the parameters $a_0$ to $a_5$ are our fit parameters. 
A fit is made for each of the tracers that had $R^2>0.5$ (more than half of the variance explained) in the rankings of Section~\ref{sect:Tracers}. 
The corresponding fit coefficient values for each column density ratio are given in Table~\ref{tab:RatiosColumnDensitiesDenseFullFitCoeff} and the coefficient values for line intensity ratios are listed in Table~\ref{tab:RatiosIntensitiesDenseFullFitCoeff}.

In the translucent medium conditions, $x(\mathrm{e}^-)$ naturally reaches a plateau at the fractional abundance of carbon ($1.32\times10^{-4}$ in our undepleted models). 
We thus use a modified formula combining a polynomial of order 5 and a saturation:
\begin{equation}
  f^\mathrm{translucent}(x) = f_\mathrm{max} - \log_e \left( 1 + e^{-(a_0 + a_1 x + a_2 x^2 + a_3 x^3 + a_4 x^4 + a_5 x^5)} \right) \label{eq:main_fit_translucent}
\end{equation}
The fit parameters here are $f_\mathrm{max}$ and $a_0$ to $a_5$. 
As in the cold dense medium case, $f^\mathrm{translucent}(x)$ is our fitting function to $\log_{10}\left(x(\mathrm{e}^-)\right)$ in translucent gas conditions.
The corresponding fit coefficient values for each column density ratio are listed in Table~\ref{tab:RatiosColumnDensitiesTranslucentFullFitCoeff} and the coefficient values for line intensity ratios are listed in Table~\ref{tab:RatiosIntensitiesTranslucentFullFitCoeff}.

The order of the polynomial in these functions is chosen so that further increasing it yields only marginal increase in $R^2$ (estimated by cross-validation to avoid overfitting). 
The $R^2$ values found for the best tracers in each case are close to the values initially found with the RF models, indicating that our analytical fits are not significantly worse than the RF models, at least for the high-$R^2$ tracers. 
As an example, Figures \ref{fig:RFmodelBestRatiosTranslucent} and \ref{fig:RFmodelBestRatiosDense} also show the analytical fit (solid red line) in comparison to the RF model (solid black line).
For a few of the lower $R^2$ tracers, the analytical fits perform significantly worse as can be seen on the tables of Appendix~\ref{app:Rankings} by comparing the $R^2$ values of the RF models with the $R^2$ values of the analytical fits. 

\subsection{Uncertainty formulae}

\FigAnalyticalmodelUncertainties

Finally, a key point is to estimate uncertainties on our prediction of the ionization fraction. 
We separate here two sources of uncertainty.

Our best analytical fit is determined on a finite sample of models, so that the fit coefficients are only estimates of the theoretical best fit coefficients. 
Estimating again these fit coefficients from a different sample of models (drawn from the same distribution) would result in slightly different values, and these uncertainties on the fit coefficients in turn imply an uncertainty on the ionization fraction value predicted by the fit formula at any given value of the observable quantity (intensity ratio or column density ratio). 
In order to estimate this uncertainty on the predicted value, we proceed by bootstrapping: we repeat the fitting procedure on 100 bootstrapped samples (drawn from the original model grid) and report the standard deviation of the value of the fit function as its uncertainty. 

The left panel of Fig.~\ref{fig:AnalyticalmodelUncertainties} shows the corresponding uncertainty (showing the $3 \sigma$ level in dashed curves around the main prediction curve) in the case of the best column density ratio in cold dense gas conditions (CN/N$_2$H$^+$). 
We name this uncertainty the fit coefficients uncertainty to distinguish it from the second form of uncertainty below. 
We define the validity domain of our fit as the range of values of the observable ratio where our best fit is sufficiently constrained by our finite grid of models for this fit coefficient uncertainty to be negligible.
In practice, we define it as the range of ratios where the above-defined uncertainty remains lower than 2\% of the predicted value. 
In the following, we will thus assume this uncertainty to be negligible inside of this validity domain and focus on the second form of uncertainties. 
The limits of the corresponding validity range are also shown on the left panel of Fig.~\ref{fig:AnalyticalmodelUncertainties} as blue vertical lines.
Due to the tendency of high order polynomial fits to diverge quickly outside of the domain of the fitted dataset, the analytical fit formulae should not be used outside of the validity range defined here.

The second form of uncertainties comes from the unobservable parameters (density, temperature,...), which induce a scatter in the relationship between any of the line intensity ratios or column density ratios and the ionization fraction. 
Inside of the validity domain of the fit, this scatter in the residuals is much larger than the fit coefficients uncertainty, as can be seen on the left panel of Fig.~\ref{fig:AnalyticalmodelUncertainties}), and is thus the  dominant source of uncertainties. 
When applying the fit formula to real observations, in the ideal case where the chemical model used in this paper would be a perfect model of reality, we would expect the value predicted by the fit formula to commit a mean squared error equal to the variance of this scatter of the residuals, and this error represents our lack of information on the underlying physical conditions. 
As can be seen on the previous figures however, this residual scatter varies as a function of the observable predictor (the vertical scatter is smaller in some regions of the plot than in others). 
This residual variance as a function of the predictor can be estimated by a moving average method (shown in the right  panel of Fig.~\ref{fig:AnalyticalmodelUncertainties}, we used a window of 0.1 dex). 
In order to provide a simpler way of estimating this residual variance function, we fitted the simple analytical function
\begin{equation}
  g(x) = \left| b_0 + b_1 x + b_2 x^2 + b_3 x^3 + b_4 x^4 + b_5 x^5 \right| \label{eq:uncertainty_fit}
\end{equation}
to the squared residuals, thus providing a fit to the local variance. 
Note that this function provides a fit to the variance on the prediction of $\log_{10}(x(\mathrm{e}^-))$, and, as previously, $x$ is the $\log_{10}$ of the observable ratio.
The absolute value in this function was chosen because the residual variance is by definition a positive quantity. 
The best fit coefficients to this residual variance function are given in Tables~\ref{tab:RatiosColumnDensitiesTranslucentFullFitCoeff}, \ref{tab:RatiosIntensitiesTranslucentFullFitCoeff}, \ref{tab:RatiosColumnDensitiesDenseFullFitCoeff} and \ref{tab:RatiosIntensitiesDenseFullFitCoeff} in Appendix \ref{app:Rankings}. 
An example of the corresponding residual standard deviation function is shown in the right panel of Fig.~\ref{fig:AnalyticalmodelUncertainties}, where we compare its moving-average estimate (dotted red curves around the main prediction curve) to its squared polynomial fit (dashed black curve), showing in both cases the $3 \sigma$ level, for the best column density ratio in the cold dense medium case (CN/N$_2$H).

The resulting fits are also presented for the best ratio in the different cases in Fig.~\ref{fig:RFmodelBestRatiosTranslucent} and \ref{fig:RFmodelBestRatiosDense}, showing the RF fit, the analytical fit and the scatter fit.
Similar figures for each of the 6 best tracers for both the translucent medium and the cold dense medium conditions, and for both column density ratios and line intensities ratios, are presented in Appendix~\ref{app:fit_figures}, .







\section{Discussions}
\label{sect:Discussions}

\subsection{Traditional ionization tracers}
\label{sec:trad_tracers}

\FigDCOpoverHCOp

One of the most commonly used ionization fraction tracers is DCO$^+$/HCO$^+$ \citep{Guelin1977,Guelin1982,Dalgarno1984,Caselli1998}. 
It is used mainly in cold dense cores, where the temperature is low enough to allow sufficient deuterium enrichment, and the column density is large enough to make DCO$^+$ detectable. 
However, our results show that even in the cold dense gas regime, and despite including DCO$^+$ in our list of potential tracers, the DCO$^+$/HCO$^+$ column density ratio is not ranked among the best tracers of the ionization fraction.
In fact, it is ranked as the 38\textsuperscript{th} best tracer in dense cold gas conditions, with a $R^2$ of 0.57 only (cf. Table~\ref{tab:RatiosColumnDensitiesDenseFullFitCoeff}). Note that this ratio is often determined using observations of H$^{13}$CO$^+$ as H$^{12}$CO$^+$ can be optically thick in high-column-density lines of sight. We leave this aspect aside in this discussion by showing that the column density ratio DCO$^+$/HCO$^+$ itself (however it might be determined observationally) suffers from several limitations as a tracer of the ionization fraction.

The relationship between the DCO$^+$/HCO$^+$ abundance ratio and the ionization fraction found in our (cold and dense) model grid is shown in Fig.~\ref{fig:DCOpoverHCOp}. 
The blue distribution shows the results of the model grid and the black line our best RF model, with a $R^2$ of 0.57. 
We see that two main problems limit the usability of DCO$^+$/HCO$^+$.

First, a large scatter of the ionization values (by up to 3 orders of magnitude) is present at all values of the ratio, despite having a significant fraction of the distribution tightly located around a clear relationship (the outermost blue contour encloses 75\% of the distribution). 
As a result, the best fit RF model tries to make a compromise between the tightly located part of the distribution, and the scattered points at lower ionization fraction values. 
We found most of this scatter to be related to variations in the ortho-to-para ratio of H$_2$ (OPR$_\mathrm{H_2}$), an unobservable parameter whose value remains difficult to estimate in observations of dense cold cores. 
For instance, selecting only the models having OPR$_\mathrm{H_2} < 2.5\times 10^{-3}$, we see in Fig.~\ref{fig:DCOpoverHCOp} (red dashed contour) that we retain only the unscattered part of the distribution. 
Thus the difficulty of obtaining reliable estimates of the OPR of H$_2$ in dense cores limits the use of DCO$^+$/HCO$^+$ as a tracer of the ionization fraction.

Second, even when selecting the low OPR$_\mathrm{H_2}$ models, we see that the relationship presents a very steep slope at high ratio values (low ionization fraction values). 
As a result, for DCO$^+$/HCO$^+$ ratios above $10^{-1.6}$, a range of ionization fractions of more than two orders of magnitude is possible. 
The ratio would then only be usable for lower ratios, equivalent to ionization fractions larger than $10^{-6.5}$.
Even if the model grid presented no scatter at all, a steep slope implies that small observational uncertainties on the ratio will induce large uncertainties on the predicted ionization fraction. 
Thus, relationships with steep slopes are of limited use.

These different effects combine to make the DCO$^+$/HCO$^+$ ratio a poor predictor of the ionization fraction, compared to the best ranked tracers found by our method.

\subsection{Detectability constraints}\label{sec:noise}

\FigTranslucentNoiseImpactAbs

\FigTranslucentNoiseImpactSNR

So far, detectability constraints have been ignored. Predictive power has been tested from noiseless values of column density or line intensity ratios.
However, the different lines considered here have widely different brightnesses and thus differ in terms of detectability with current instruments.
We explore here the effect of various noise levels on the predictive power of the different line ratios. 
We will consider two noise setups corresponding to two observation scenarios: the case of one constant noise level for all lines, corresponding to the typical case of a line survey where faint lines are detected with a lower signal-to-noise ratio than bright ones, and the case of a fixed signal-to-noise ratio (SNR) for all lines, corresponding to observations being designed to reach a set signal-to-noise ratio for a few desired lines. 
These two cases correspond to the two opposite extremes of possible observation scenarios and will give us a general overview of the possible impact of noise on the performance of the tracers.
In both cases, synthetic noise is added to the line integrated intensity values and we consider the SNR on the integrated intensity and not the peak intensity. Note that all noise and SNR values quoted are for individual lines, not for line ratios.

In both cases, in order to measure how the predictive power (measured as the $R^2$ value) is affected by noise, we perform a modified cross-validation. 
The model grid is randomly split in ten parts. For each of these tenths and for each line ratio :
\begin{enumerate}
\item A RF model is \emph{trained} from the other nine parts of the model grids, \emph{without any noise added}.
\item The trained RF is \emph{tested} on the tenth under consideration, \emph{with added noise} (either with a constant noise variance $\sigma^2_\mathrm{noise}$ for all models and all lines, or with a constant SNR for each model and line). 
Since the RF models take the $\log_{10}$ of the intensity ratio as input and since the addition of noise can produce negative values, we only apply the RF model when both line integrated intensity values are above 1 $\sigma$. Otherwise, we do not use the RF model and simply take the average ionization fraction value of the grid as our prediction.
\end{enumerate}
We finally take the average $R^2$ value obtained over the ten noisy tests as the predictive power of the tracer under noisy conditions.
We thus avoid estimating $R^2$ from datapoints that have been seen during training, and we estimate the predictive power of a model trained on noiseless data when applied to noisy data (as would be the case when applying the results of this article to real observations).

  \begin{table}
    \caption{Impact of adding noise to the line intensities on the predictive power $R^2$ (for translucent medium), measured by the noise level $\sigma_{1/2}$ (in a constant noise level situation) and the signal-to-noise ratio SNR$_{1/2}$ (in a constant SNR situation) for which $R^2$ reaches one half of its value in the absence of noise $R^2_\mathrm{(noiseless)}$.}
    \label{tab:NoiseImpactTranslucent}
    \begin{center}
      \begin{tabular} {c|ccc}
        \hline 
        Line intensity ratio &  $R^2_\mathrm{(noiseless)}$  & $\sigma_{1/2}$  & $\mathrm{SNR}_{1/2} $  \\ 
         &               &  K.km/s  &   \\
\hline
C$_2$H $(1-0)$ / HCN $(1-0)$          & 0.93 & $4.55\times 10^{-3}$ &  1.64 \\
C$_2$H $(1-0)$ / $^{13}$CO $(1-0)$    & 0.92 & $5.00\times 10^{-3}$ &  1.61 \\
C$_2$H $(1-0)$ / C$^{18}$O $(1-0)$    & 0.92 & $5.22\times 10^{-3}$ &  1.57 \\
C$^{18}$O $(1-0)$ / CF$^+$ $(1-0)$    & 0.89 & $1.12\times 10^{-2}$ &  1.61 \\
C$_2$H $(1-0)$ / HNC $(1-0)$          & 0.87 & $2.35\times 10^{-3}$ &  1.58 \\
SO $(3-2)$ / HCS$^+$ $(2-1)$          & 0.85 & $7.53\times 10^{-5}$ &  1.58 \\
$^{13}$CO $(1-0)$ / C$^{18}$O $(1-0)$ & 0.85 & $4.30\times 10^{-2}$ &  2.97 \\
SO $(3-2)$ / C$_2$H $(1-0)$           & 0.80 & $2.18\times 10^{-4}$ &  1.55 \\
C$_2$H $(1-0)$ / CN $(1-0)$           & 0.80 & $2.26\times 10^{-3}$ &  1.60 \\
C$_2$H $(1-0)$ / HCO$^+$ $(1-0)$      & 0.78 & $2.70\times 10^{-3}$ &  1.57 \\
HCN $(1-0)$ / CF$^+$ $(1-0)$          & 0.77 & $8.17\times 10^{-3}$ &  1.79 \\
HCO$^+$ $(1-0)$ / CF$^+$ $(1-0)$      & 0.77 & $7.77\times 10^{-3}$ &  1.62 \\
HNC $(1-0)$ / CF$^+$ $(1-0)$          & 0.77 & $4.67\times 10^{-3}$ &  1.61 \\
$^{13}$CO $(1-0)$ / CF$^+$ $(1-0)$    & 0.76 & $9.95\times 10^{-3}$ &  1.74 \\
CN $(1-0)$ / CF$^+$ $(1-0)$           & 0.75 & $5.32\times 10^{-3}$ &  1.90 \\
H$_2$CS $(3-2)$ / C$_2$H $(1-0)$      & 0.75 & $1.00\times 10^{-5}$ &  1.54 \\
SO $(3-2)$ / CF$^+$ $(1-0)$           & 0.74 & $5.16\times 10^{-4}$ &  1.55 \\
CS $(2-1)$ / SO $(3-2)$               & 0.70 & $4.19\times 10^{-4}$ &  1.64 \\
C$_2$H $(1-0)$ / CF$^+$ $(1-0)$       & 0.69 & $9.78\times 10^{-4}$ &  1.74 \\
CS $(2-1)$ / C$_2$H $(1-0)$           & 0.69 & $4.52\times 10^{-4}$ &  1.57 \\
SO $(3-2)$ / CN $(1-0)$               & 0.67 & $2.06\times 10^{-4}$ &  1.53 \\
SO $(3-2)$ / HCN $(1-0)$              & 0.66 & $1.89\times 10^{-4}$ &  1.57 \\
HCS$^+$ $(2-1)$ / C$_2$H $(1-0)$      & 0.64 & $8.33\times 10^{-5}$ &  1.57 \\
HCN $(1-0)$ / HNC $(1-0)$             & 0.64 & $3.38\times 10^{-3}$ &  2.18 \\
SO $(3-2)$ / $^{13}$CO $(1-0)$        & 0.63 & $1.65\times 10^{-4}$ &  1.54 \\
H$_2$CS $(3-2)$ / CF$^+$ $(1-0)$      & 0.59 & $1.00\times 10^{-5}$ &  1.54 \\
CS $(2-1)$ / HCS$^+$ $(2-1)$          & 0.58 & $3.39\times 10^{-5}$ &  2.00 \\
C$^{18}$O $(1-0)$ / CN $(1-0)$        & 0.56 & $2.09\times 10^{-2}$ &  1.75 \\
SO $(3-2)$ / HNC $(1-0)$              & 0.53 & $8.33\times 10^{-5}$ &  1.58 \\
HCO$^+$ $(1-0)$ / CN $(1-0)$          & 0.53 & $2.48\times 10^{-2}$ &  1.94 \\
SO $(3-2)$ / C$^{18}$O $(1-0)$        & 0.51 & $7.66\times 10^{-5}$ &  1.57 \\
CS $(2-1)$ / CF$^+$ $(1-0)$           & 0.50 & $8.93\times 10^{-4}$ &  1.57 \\
        \hline 
      \end{tabular}
    \end{center}
  \end{table}

The results for the translucent medium grid, for a few possible line ratios, are presented in Fig.~\ref{fig:TranslucentNoiseImpactAbs} and \ref{fig:TranslucentNoiseImpactSNR}. 
Figure~\ref{fig:TranslucentNoiseImpactAbs} presents the scenario of a constant noise level $\sigma_\mathrm{noise}$ for all lines and all models, and shows how the $R^2$ of the prediction from different line ratios decreases when the noise level is increased. 
We show only the three best line ratios (according to the noiseless ranking of Tab.~\ref{tab:RatiosRankingIntensitiesTranslucent}): C$_2$H (1-0) / HCN (1-0), C$_2$H (1-0) / $^{13}$CO (1-0), and C$_2$H (1-0) / C$^{18}$O (1-0), and the four tracers found to be the least sensitive to noise. We define the tracers least sensitive to noise as those having the highest $\sigma_{1/2}$ among tracers with $R_\mathrm{noiseless}\ge 0.7$, thus giving a compromise between a good fit quality (high $R_\mathrm{noiseless}$) and a slow decrease with increasing noise level (high $\sigma_{1/2}$). The four best ratios found according to this definition and shown on the  figure are $^{13}$CO (1-0) / C$^{18}$O (1-0), C$^{18}$O (1-0) / CF$^+$  (1-0), $^{13}$CO (1-0) / CF$^+$ (1-0), and HCN (1-0) / CF$^+$ (1-0).
We see that the best three tracers, all including C$_2$H, have their predictive power decreasing sharply at a relatively low noise level (their $R^2$ decreases by 50\% at $\sigma_\mathrm{noise} \sim 5\times 10^{-3}$ K~km~s$^{-1}$).
This is due to the relatively low brightness of the C$_2$H line in translucent conditions (median integrated intensity of $\sim 7\times 10^{-3}$ K~km~s$^{-1}$ in our translucent medium grid). 
In comparison, some line ratios built from brighter lines, despite a lower $R^2$ on noiseless data, are found to perform better in the presence of noise : the $R^2$ for the $^{13}$CO (1-0) / C$^{18}$O (1-0) ratio decreases by 50\% at $\sigma_\mathrm{noise} \sim 4\times 10^{-2}$ K~km~s$^{-1}$ (C$^{18}$O (1-0) has a median integrated intensity of $\sim 3\times 10^{-1}$ K~km~s$^{-1}$ in this grid), the C$^{18}$O (1-0) / CF$^+$ (1-0) ratio has its $R^2$ decreased by 50\% at $\sigma_\mathrm{noise} \sim 10^{-2}$ K~km~s$^{-1}$ (CF$^+$ (1-0) has a median integrated intensity of  $\sim 1.2 \times 10^{-2}$ K~km~s$^{-1}$ in this grid).
We note, however, that ratios built from CF$^+$ (1-0) actually only perform marginaly better than the best ratios involving C$_2$H at high noise levels (see Fig. \ref{fig:TranslucentNoiseImpactAbs}) due to the low brightness of CF$^+$ (1-0).
For a more exhaustive comparison of the line ratios, Tab.~\ref{tab:NoiseImpactTranslucent} gives for each line ratio the noise level $\sigma_{1/2}$ at which the $R^2$ is decreased by half.

\FigDenseNoiseImpactAbs

\FigDenseNoiseImpactSNR

Figure~\ref{fig:TranslucentNoiseImpactSNR} similarly shows the results for the scenario of a fixed SNR for all lines and all models, still in the case of translucent medium conditions.
The variations of the $R^2$ of the prediction with the SNR are shown for the best seven tracers (according to the noiseless ranking of Table~\ref{tab:RatiosRankingIntensitiesTranslucent}). 
In this scenario, we find as expected that the predictive power drops at a SNR of order unity. 
The only exception is the $^{13}$CO (1-0) / C$^{18}$O (1-0) ratio, which decreases slightly earlier. 
This is due to this ratio spanning a relatively small range of values in our grid of models (approximately one order of magnitude) while the other line ratios span ranges of four to six orders of magnitude. 
As the ionization fraction spans a range of values of 2.5 orders of magnitude in the grid, this implies that the relationship between the ionization fraction and the $^{13}$CO (1-0) / C$^{18}$O (1-0) ratio has a much steeper slope than for the other line ratios. 
A steep slope then implies that small errors in the line ratios result in large errors in the predicted ionization fraction. 
As a result, the $^{13}$CO (1-0) / C$^{18}$O (1-0) ratio requires significantly higher SNRs than the other line ratios. 
Similarly to $\sigma_{1/2}$, we define SNR$_{1/2}$ as the SNR value at which $R^2$ is half of its value on noiseless data. 
The SNR$_{1/2}$ values for all line ratios are also given in Table~\ref{tab:NoiseImpactTranslucent}.

  \begin{table}
    \caption{Impact of adding noise to the line intensities on the predictive power $R^2$ (for cold dense medium), measured by the noise level $\sigma_{1/2}$ (in a constant noise level situation) and the signal-to-noise ratio SNR$_{1/2}$ for which the $R^2$ reaches one half of its value in the absence of noise $R^2_\mathrm{(noiseless)}$.}
    \label{tab:NoiseImpactDense}
    \begin{center}
      \begin{tabular} {c|ccc}
        \hline 
        Line intensity ratio &  $R^2_\mathrm{(noiseless)}$  & $\sigma_{1/2}$  & $\mathrm{SNR}_{1/2} $  \\ 
         &               &  K.km/s  &   \\
\hline
CF$^+$ $(1-0)$ / DCO$^+$ $(1-0)$       & 0.86 & $4.20 \times 10^{-3}$ &  1.64 \\
$^{13}$CO $(1-0)$ / HCO$^+$ $(1-0)$    & 0.86 & $6.07 \times 10^{-1}$ &  1.69 \\
CN $(1-0)$ / N$_2$H$^+$ $(1-0)$        & 0.86 & $3.93 \times 10^{-2}$ &  1.60 \\
C$_2$H $(1-0)$ / N$_2$H$^+$ $(1-0)$    & 0.86 & $3.78 \times 10^{-4}$ &  1.64 \\
HCO$^+$ $(1-0)$ / CF$^+$ $(1-0)$       & 0.84 & $6.31 \times 10^{-3}$ &  1.67 \\
C$_2$H $(1-0)$ / HCN $(1-0)$           & 0.83 & $4.18 \times 10^{-4}$ &  1.68 \\
$^{13}$CO $(1-0)$ / DCO$^+$ $(1-0)$    & 0.81 & $2.36 \times 10^{-2}$ &  1.62 \\
C$_2$H $(1-0)$ / HNC $(1-0)$           & 0.81 & $3.99 \times 10^{-4}$ &  1.68 \\
C$^{18}$O $(1-0)$ / DCO$^+$ $(1-0)$    & 0.81 & $1.93 \times 10^{-2}$ &  1.63 \\
C$_2$H $(1-0)$ / DCO$^+$ $(1-0)$       & 0.80 & $4.00 \times 10^{-4}$ &  1.58 \\
CF$^+$ $(1-0)$ / N$_2$H$^+$ $(1-0)$    & 0.80 & $4.92 \times 10^{-3}$ &  1.70 \\
C$_2$H $(1-0)$ / HCO$^+$ $(1-0)$       & 0.79 & $5.41 \times 10^{-4}$ &  1.58 \\
CN $(1-0)$ / DCO$^+$ $(1-0)$           & 0.78 & $9.57 \times 10^{-3}$ &  1.58 \\
C$^{18}$O $(1-0)$ / HCO$^+$ $(1-0)$    & 0.76 & $4.29 \times 10^{-1}$ &  1.69 \\
HCO$^+$ $(1-0)$ / CN $(1-0)$           & 0.75 & $4.70 \times 10^{-2}$ &  1.59 \\
HCN $(1-0)$ / CN $(1-0)$               & 0.75 & $3.69 \times 10^{-2}$ &  1.78 \\
HNC $(1-0)$ / CN $(1-0)$               & 0.75 & $3.52 \times 10^{-2}$ &  1.83 \\
SO $(3-2)$ / C$_2$H $(1-0)$            & 0.73 & $5.75 \times 10^{-4}$ &  1.58 \\
SO $(3-2)$ / CN $(1-0)$                & 0.70 & $3.87 \times 10^{-2}$ &  1.59 \\
CS $(2-1)$ / C$_2$H $(1-0)$            & 0.70 & $5.78 \times 10^{-4}$ &  1.56 \\
SO $(3-2)$ / $^{13}$CO $(1-0)$         & 0.69 & $6.92 \times 10^{-1}$ &  1.78 \\
H$_2$CS $(3-2)$ / C$_2$H $(1-0)$       & 0.68 & $3.76 \times 10^{-4}$ &  1.55 \\
$^{13}$CO $(1-0)$ / N$_2$H$^+$ $(1-0)$ & 0.65 & $9.98 \times 10^{-2}$ &  1.65 \\
SO $(3-2)$ / CF$^+$ $(1-0)$            & 0.64 & $6.46 \times 10^{-3}$ &  1.64 \\
CS $(2-1)$ / CN $(1-0)$                & 0.64 & $2.39 \times 10^{-2}$ &  1.59 \\
C$_2$H $(1-0)$ / CF$^+$ $(1-0)$        & 0.64 & $3.54 \times 10^{-4}$ &  1.65 \\
HNC $(1-0)$ / DCO$^+$ $(1-0)$          & 0.60 & $5.75 \times 10^{-3}$ &  1.61 \\
SO $(3-2)$ / HNC $(1-0)$               & 0.58 & $2.93 \times 10^{-1}$ &  1.78 \\
SO $(3-2)$ / HCN $(1-0)$               & 0.57 & $2.59 \times 10^{-1}$ &  1.81 \\
H$_2$CS $(3-2)$ / CN $(1-0)$           & 0.56 & $5.26 \times 10^{-3}$ &  1.58 \\
HCN $(1-0)$ / DCO$^+$ $(1-0)$          & 0.56 & $4.25 \times 10^{-3}$ &  1.64 \\
C$_2$H $(1-0)$ / $^{13}$CO $(1-0)$     & 0.56 & $7.15 \times 10^{-4}$ &  1.60 \\
SO $(3-2)$ / C$^{18}$O $(1-0)$         & 0.56 & $3.61 \times 10^{-1}$ &  1.68 \\
C$^{18}$O $(1-0)$ / N$_2$H$^+$ $(1-0)$ & 0.56 & $6.34 \times 10^{-2}$ &  1.68 \\
HNC $(1-0)$ / N$_2$H$^+$ $(1-0)$       & 0.55 & $6.29 \times 10^{-2}$ &  1.71 \\
HCS$^+$ $(2-1)$ / C$_2$H $(1-0)$       & 0.54 & $1.95 \times 10^{-4}$ &  1.57 \\
HCO$^+$ $(1-0)$ / HNC $(1-0)$          & 0.54 & $1.97 \times 10^{-1}$ &  1.67 \\
C$_2$H $(1-0)$ / C$^{18}$O $(1-0)$     & 0.53 & $7.32 \times 10^{-4}$ &  1.63 \\
CS $(2-1)$ / CF$^+$ $(1-0)$            & 0.52 & $5.83 \times 10^{-3}$ &  1.59 \\
HCO$^+$ $(1-0)$ / DCO$^+$ $(1-0)$      & 0.52 & $1.63 \times 10^{-3}$ &  1.99 \\
HCO$^+$ $(1-0)$ / HCN $(1-0)$          & 0.51 & $1.75 \times 10^{-1}$ &  1.70 \\
CS $(2-1)$ / HCS$^+$ $(2-1)$           & 0.51 & $4.99 \times 10^{-4}$ &  1.97 \\
        \hline 
      \end{tabular}
    \end{center}
  \end{table}

The results for dense and cold medium conditions are similarly shown in Fig.~\ref{fig:DenseNoiseImpactAbs} and \ref{fig:DenseNoiseImpactSNR} and Table~\ref{tab:NoiseImpactDense}. 
In the case of a constant noise level, we find that the $R^2$ drop generally occurs at higher noise levels than in the translucent medium case, as expected because the lines are brighter in the cold dense medium due to higher column densities. 
Figure~\ref{fig:DenseNoiseImpactAbs} shows the decrease of $R^2$ with $\sigma_\mathrm{noise}$ for the best three dense gas tracers (according to the noiseless ranking of Tab.~\ref{tab:RatiosRankingIntensitiesTranslucent}), and for the four ratios least sensitive  two noise (according to the same definition as previously). We note that $^{13}$CO (1-0) / HCO$^+$ (1-0), the second best ratio on noiseless data (therefore already  shown on the figure), has also the highest $\sigma_{1/2}$ value among ratios with $R_\mathrm{noiseless}\ge 0.7$, so that we show the next four ratios least sensitive to noise on the figure.
These four ratios  are found to be C$^{18}$O (1-0) / HCO$^+$ (1-0), HCO$^+$ (1-0) / CN (1-0), SO (3-2) / CN (1-0), and HCN (1-0) / CN (1-0).
The $\sigma_{1/2}$ values for all line ratios are listed in Table~\ref{tab:NoiseImpactDense}.
When considering a constant SNR scenario for the dense cold medium case, we again find that the $R^2$ drop occurs at a SNR value of order unity, as shown in Fig.~\ref{fig:DenseNoiseImpactSNR}. 
The SNR$_{1/2}$ values for all line ratios are listed in Table~\ref{tab:NoiseImpactDense}.

When interpreting the results of this study, one must keep in mind that the decrease in $R^2$ in our two scenarios (constant $\sigma_\mathrm{noise}$ or constant SNR) does not come from the same effect. 
In the constant $\sigma_\mathrm{noise}$ scenario, at a given $\sigma_\mathrm{noise}$ level one part of the grid has undetected or very low SNR values for the ratio under consideration, while the other part has high SNR. 
The $R^2$ decrease is indicative of the growing fraction of the parameter space with undetected/low SNR ratios. 
As a result, even when finding a low overall $R^2$, there might remain a fraction of the parameter space were the predictor remains very good (usually, high column density, high volume density, high temperature,...), which we did not characterize here. 
In the constant SNR scenario on the other hand, the SNR is by design constant over all models, independent of the physical parameters. 
The decrease in overall $R^2$ is then more representative of the decrease in predictive power at any point in the parameter space.
As a result, a ratio with a low $\sigma_{1/2}$ value might still be usable in real observations with a higher noise level but would be restricted to high brightness regions of GMCs (in the corresponding lines), while a ratio cannot be used at all in observations with SNR significantly lower than its SNR$_{1/2}$ value.

\subsection{Chemical model reliability}\label{sec:chemical_reliability}

Independently of the statistical method that we present in this article, the results obtained rely on the chosen chemical model and its limitations. 
Previous works on ionization fraction tracers have mostly used stationary-state results of single-zone chemical models (although some works have used time-dependent models, e.g., \citealt{Maret2007,Shingledecker2016}). 
As these previous studies have been mostly limited to deuteration-based tracers, the focus of the present article has been on highlighting the non-deuteration-based tracers that can be found for the ionization fraction from similar chemical models. 
We discuss here the impacts of our model's limitations on our results.

Our single-zone model cannot include a detailed treatment of UV radiative transfer through the cloud. 
While most photodissociation rates can be simply estimated based on an assumed optical depth of dust protecting each model from UV photons (parameter $A_\mathrm{V}$ in our models), species such as H$_2$, CO and its isotopologues can be protected from photodissociation by self- or mutual-shielding. 
While self-shielding of H$_2$ and mutual shielding of CO by H$_2$ are included in our model using approximations \citep{Draine1996,Heays2017}, mutual shielding of $^{13}$CO and C$^{18}$O by H$_2$ and $^{12}$CO are not included. 
As a result, in our translucent medium models where photodissociation by external UV photons still plays an important role, we expect the abundances of the rarer CO isotopologues to be less reliable than the other species. 
Note that the observations of $^{13}$CO and C$^{18}$O in our ORION-B dataset indeed present specificities (systematic excitation temperature differences with $^{12}$CO, \citealt{Bron2018,Roueff2020}) that remain unexplained even by more complex 1D PDR models.

The only explicit surface reaction in our chemical model is H$_2$ formation, however we account for the freeze-out of CO through our depletion parameter. 
The list of species that we consider as possible tracers has been restricted to species that are not strongly affected by surface chemistry beyond the depletion effect.
Extension to more complex molecules would require a chemical model including a more complete treatment of surface chemistry.

Another source of uncertainties comes from the experimental or theoretical estimates of the reaction rate coefficients used in our chemical network.
While we did not directly perform a sensitivity analysis of the reaction rate coefficients, our model grids include temperature variations which subsequently impact the reactions rate coefficients through their temperature dependence. 
This is especially true for the important dissociative recombination reactions which display significative temperature dependences.
Temperature is then considered as an unobserved parameter when searching for good tracers of $x(e^-)$. 
The discovery of strong relationships between some of the ratios and $x(e^-)$, despite temperature variations in the grid, thus indicates that these relationships are to some extent robust to the reaction rates.
A careful analysis of the magnitude and correlations of model uncertainties resulting from reaction rate uncertainties in the chemical network would deserve a separate study.

Time-dependent effects are expected to be more important in cold dense medium conditions than in translucent medium conditions as photochemistry has shorter timescales in the latter case. 
Time-dependent effects that result from the time evolution of some physical parameter (e.g. density and temperature during the contraction of a core) while the chemistry follows in a quasi-stationary way are in part accounted for in our models by exploring a large range of the various physical parameters that can be subject to variations (see Table~\ref{tab:grid_ranges}). 
In addition, the progressive freeze out of CO on dust grains is accounted for by considering a range of depletion factors for carbon and oxygen. 
Similarly, the slow evolution of OPR$_\mathrm{H_2}$ in cold gas can keep parts of the chemistry (deuterium chemistry, nitrogen chemistry) in a time-dependent evolution that depends mainly on the evolution of OPR$_\mathrm{H_2}$. 
This is also in part accounted for by exploring a large range of OPR$_\mathrm{H_2}$ values in our models. 
The tracer-finding method presented in this article will be applied to time-dependent chemical models in a future study.

The final and most important limitation of our model is that it does not include a spatial dimension : gas at a single value of density, temperature, etc. is assumed to be exposed to a given radiation field and protected by a given column density. 
As a result the possibility that different emission lines originate in separate layers of gas on the line of sight is completely neglected. Variations of the physical conditions along the line of sight can indeed have an important impact on the observables (e.g. \citealt{Levrier2012}).
This limitation could be important both in the translucent medium where physical and chemical gradients are present due to the progressive extinction of the external UV field, and in dense cores with density and temperature gradients.
As a result some caution must be exercised when choosing the line ratios to consider, ratios involving two species expected to emit in completely different regions should be avoided. 
For instance, the C$_2$H (1-0) / N$_2$H$^+$ (1-0) that is found as the fourth best line ratio in dense cold medium should be avoided : C$_2$H is known to be a tracer of UV illumination and thus more likely to be emitted at the external surface of a given clump, while N$_2$H$^+$ is abundant in the inner regions of the core where CO is already significantly depleted. 
An application of our method to 1D PDR models to better account for this effect in translucent medium conditions will be carried out in a future work.

\subsection{Parameter PDF in the model grids}

In the model grids used in this study, we sampled uniformly (in logarithm) for the values of the unobservable physical conditions (gas density, temperature, UV field, etc.) in an hypercube defined by lower and upper bounds for each of the parameters. The results of our ranking method will depend on this assumed PDF (probability density function) for the physical conditions in the ISM. We made here the choice of making the minimal assumption: knowing  only reasonable lower and upper bounds on each of the parameters, the uniform distribution is the maximum entropy PDF (i.e. the PDF that best represents our assumed state of knowledge). As a perspective, if more a priori knowledge is available, then more accurate assumptions for the PDF (in particular for the correlations between the different physical parameters) could reveal additional tracers.

We note that this assumption of a uniform PDF over a maximum support (in the sense that any more accurate PDF would have almost all of its weight enclosed in this support) makes it likely that any tracer found to have a very good relationship with the ionization fraction would keep a strong relationship for more accurate PDF choices (if the relationship is strong over the full hypercube, it should with high likelihood stay strong on subregions of this hypercube). In this sense, we expect the tracers found here to remain reliable, but a more accurate PDF choice might strongly increase the performance of some tracers found here to perform poorly and thus reveal additional tracers of the ionization fraction. This argument remains however qualitative as pathological cases of PDF might be constructed that would radically change the rankings of the tracers.

As a result, the precise rankings presented in this article could slightly change, but we expect the good tracers highlighted by these ranking to be reliable for more realistic PDFs of the physical conditions in GMCs.

\subsection{Final recommandation}

Based on the limitations discussed above (detectability and model reliability), we recommend the use of the following integrated line intensity ratios to trace the ionization fraction.
\begin{itemize}
\item In translucent medium conditions, we recommend the use C$_2$H (1-0) / HCN (1-0), C$_2$H (1-0) / HNC (1-0), or C$_2$H (1-0) / CN (1-0). If sensitivity is an issue, HCN (1-0) / CF$^+$ (1-0) can sometimes perform as well as the previously listed ratios.
\item In cold dense gas conditions, we recommend the use of CF$^+$ (1-0) / DCO$^+$ (1-0), $^{13}$CO (1-0) / HCO$^+$ (1-0) or CN (1-0) / N$_2$H$^+$ (1-0) if detectability is not an issue for these species, and of $^{13}$CO (1-0) / HCO$^+$ (1-0) or C$^{18}$O (1-0) / HCO$^+$ (1-0) otherwise.
\end{itemize}
This list is of course not exhaustive, and other ratios can give satisfactory predictions (see Tables \ref{tab:NoiseImpactTranslucent} and \ref{tab:NoiseImpactDense}) if the species listed above are not available.

In translucent gas conditions, this recommendation is based on the following points. 
After eliminating rarer CO isotopologues based on our discussion of mutual-shielding effects on selective photodissociation in low/moderate $A_\mathrm{V}$ regions (cf Sect.~\ref{sec:chemical_reliability}), and eliminating ratios involving sulfur species that were found to require unreasonably low noise levels of $10^{-4}-10^{-5}$ K km s$^{-1}$, the three remaining best line intensity ratios are C$_2$H (1-0) / HCN (1-0) and C$_2$H (1-0) / HNC (1-0) and C$_2$H (1-0) / CN (1-0). 
If noise sensitivity is critical, the tracer found to have the best predictive power at high noise levels is found in Sect.~\ref{sec:noise} to be HCN (1-0) / CF$^+$ (1-0) but is negligibly better than the three previously mentioned at high noise levels.

In cold dense gas conditions, the three best ratios are CF$^+$ (1-0) / DCO$^+$ (1-0), $^{13}$CO (1-0) / HCO$^+$,and CN (1-0) / N$_2$H$^+$ (1-0). 
If noise sensitivity is critical, we found in Sect.~\ref{sec:noise} that the ratios with the best predictive power at high noise levels are $^{13}$CO (1-0) / HCO$^+$ and C$^{18}$O (1-0) / HCO$^+$ (1-0).

\section{Conclusions}
\label{sect:Conclusions}

We have presented a general statistical method to find the best observable tracers of an unobservable parameter based on a grid of models spanning the range of possible values for all the unknown underlying physical parameters (e.g., gas density, temperature, depletion, etc.). 
Our method estimates the predictive power of each potential observable tracer by training a flexible, non-linear regression model (a Random Forest model, making no assumption on the non-linear shape of the relationship to be found) on the task of predicting the target quantity from each of the potential tracers. 
The fit quality on test data, measured as the $R^2$ coefficient by cross-validation and out-of-bag estimation, is used to rank the potential tracers by order of predictive power.

In the context of our recent studies of the Orion B GMC \citep{Pety2017, Gratier2017, Orkisz2017, Bron2018, Orkisz2019}, we have applied this method to the important astrophysical question of tracing the ionization fraction in the neutral ISM, with the goal of being able to probe its variations across a whole GMC, from its translucent enveloppe to its dense cores. 
We considered grids of single-zone, stationary state astrochemical models exploring wide ranges of values in gas density, temperature, external UV field, $A_\mathrm{V}$ on the line of sight, cosmic ray ionization rate, ortho-to-para ratio of H$_2$, depletion factor, and sulfur elemental abundance.
For a finer exploration of the possible conditions, we considered two grids corresponding to translucent medium conditions and cold dense medium conditions respectively, based on the different types of environments found in the Orion B GMC \citep{Pety2017,Bron2018}.

We considered successively column density ratios and line intensity ratios as potential tracers, focusing on species observable in the band at 100 GHz of our observations of Orion B. 
We find that in both cases and in both types of physical conditions, multiple ratios allow accurate predictions of the ionization fraction, with $R^2 > 0.8$ (and up to 0.96).
We investigated the impact of the noise level on the predictive capability of the different ratios
After accounting for detectability and model reliability, we recommend :
\begin{itemize}
\item for translucent medium conditions, C$_2$H (1-0) / HCN (1-0), C$_2$H (1-0) / HNC (1-0) or C$_2$H (1-0) / CN (1-0),
\item for cold dense medium conditions, CF$^+$ (1-0) / DCO$^+$ (1-0), $^{13}$CO (1-0) / HCO$^+$ (1-0) or CN (1-0) / N$_2$H$^+$ (1-0) at low enough noise level, or $^{13}$CO (1-0) / HCO$^+$ (1-0) or C$^{18}$O (1-0) / HCO$^+$ (1-0) if sensitivity is an issue.
\end{itemize}
In order to simplify the use of these predictors, we constructed ad hoc analytical fits (using polynomials or saturated polynomials) of the relationship of each observable tracer to the ionization fraction. Contrary to the Random Forest models, the choice of the analytical form of these fits is specific to the types of relationships observed in this specific application (different analytical forms might be necessary for other applications).
We also provide analytical formulae to estimate the uncertainty on any measurement of the ionization fraction from these tracers. These tracers will be used to study the ionization fraction in the Orion B molecular cloud in a second paper \citep{Guzman2020}.

The method presented here is very general and could be easily applied to finding tracers of other related (cosmic ray ionization rate, absolute electron abundance) or unrelated (gas density, temperature, OPR$_{\mathrm{H}_2}$,...) unobservable quantities. This method can also be extended simply to simultaneously use pairs (or more) of line ratios (by training RF models on the possible combinations of line ratios), which would likely further increase the quality of the prediction.

\begin{acknowledgements}
We thank the anonymous referee for his comments that helped improve this article.
 We thank the CIAS for their hospitality during the many workshops devoted to the ORION-B project. 
 This work was supported in part by the Programme National "Physique et Chimie du Milieu Interstellaire" (PCMI) of CNRS/INSU with INC/INP, co-funded by CEA and CNES. 
 This project has received financial support from the CNRS through the MITI interdisciplinary programs.
 The authors also acknowledge funding by Paris Observatory through the AF \emph{Astrochimie} program.
 JRG thanks Spanish MICI for funding support under grant AYA2017-85111-P.
\end{acknowledgements}

\bibliographystyle{aa} %
\bibliography{paper_RF_ioniz} %

\begin{thebibliography}{69}
\expandafter\ifx\csname natexlab\endcsname\relax\def\natexlab#1{#1}\fi

\bibitem[{{Ag{\'u}ndez} \& {Wakelam}(2013)}]{Agundez2013}
{Ag{\'u}ndez}, M. \& {Wakelam}, V. 2013, Chemical Reviews, 113, 8710

\bibitem[{{Balbus} \& {Hawley}(1991)}]{Balbus1991}
{Balbus}, S.~A. \& {Hawley}, J.~F. 1991, \apj, 376, 214

\bibitem[{{Barger} \& {Garrod}(2020)}]{Barger2020}
{Barger}, C.~J. \& {Garrod}, R.~T. 2020, \apj, 888, 38

\bibitem[{{Basu} \& {Mouschovias}(1994)}]{Basu1994}
{Basu}, S. \& {Mouschovias}, T.~C. 1994, \apj, 432, 720

\bibitem[{Bengio \& Grandvalet(2004)}]{Bengio2004}
Bengio, Y. \& Grandvalet, Y. 2004, Journal of Machine Learning Research, 5,
  1089

\bibitem[{{Bergin} {et~al.}(1999){Bergin}, {Plume}, {Williams}, \&
  {Myers}}]{Bergin1999}
{Bergin}, E.~A., {Plume}, R., {Williams}, J.~P., \& {Myers}, P.~C. 1999, \apj,
  512, 724

\bibitem[{{Beuther} {et~al.}(2008){Beuther}, {Semenov}, {Henning}, \&
  {Linz}}]{Beuther2008}
{Beuther}, H., {Semenov}, D., {Henning}, T., \& {Linz}, H. 2008, \apjl, 675,
  L33

\bibitem[{{Black} \& {van Dishoeck}(1991)}]{Black1991}
{Black}, J.~H. \& {van Dishoeck}, E.~F. 1991, \apjl, 369, L9

\bibitem[{Breiman(1996)}]{Breiman1996}
Breiman, L. 1996, Machine Learning, 24, 123

\bibitem[{Breiman(2001)}]{Breiman2001}
Breiman, L. 2001, Machine Learning, 45, 5

\bibitem[{{Breiman} {et~al.}(1984){Breiman}, {Friedman}, {Olshen}, \&
  {Stone}}]{Breiman84}
{Breiman}, L., {Friedman}, J.~H., {Olshen}, R.~A., \& {Stone}, C.~J. 1984,
  {Classification and Regression Trees} (New York: Chapman \& Hall), 358

\bibitem[{{Bron} {et~al.}(2018){Bron}, {Daudon}, {Pety}, {Levrier}, {Gerin},
  {Gratier}, {Orkisz}, {Guzman}, {Bardeau}, {Goicoechea}, {Liszt}, {{\"O}berg},
  {Peretto}, {Sievers}, \& {Tremblin}}]{Bron2018}
{Bron}, E., {Daudon}, C., {Pety}, J., {et~al.} 2018, \aap, 610, A12

\bibitem[{{Caselli}(2002)}]{Caselli2002a}
{Caselli}, P. 2002, \planss, 50, 1133

\bibitem[{{Caselli} {et~al.}(1998){Caselli}, {Walmsley}, {Terzieva}, \&
  {Herbst}}]{Caselli1998}
{Caselli}, P., {Walmsley}, C.~M., {Terzieva}, R., \& {Herbst}, E. 1998, \apj,
  499, 234

\bibitem[{{Caselli} {et~al.}(2002){Caselli}, {Walmsley}, {Zucconi}, {Tafalla},
  {Dore}, \& {Myers}}]{Caselli2002b}
{Caselli}, P., {Walmsley}, C.~M., {Zucconi}, A., {et~al.} 2002, \apj, 565, 344

\bibitem[{{Cuadrado} {et~al.}(2015){Cuadrado}, {Goicoechea}, {Pilleri},
  {Cernicharo}, {Fuente}, \& {Joblin}}]{Cuadrado2015}
{Cuadrado}, S., {Goicoechea}, J.~R., {Pilleri}, P., {et~al.} 2015, \aap, 575,
  A82

\bibitem[{{Cuadrado} {et~al.}(2019){Cuadrado}, {Salas}, {Goicoechea},
  {Cernicharo}, {Tielens}, \& {B{\'a}ez-Rubio}}]{Cuadrado2019}
{Cuadrado}, S., {Salas}, P., {Goicoechea}, J.~R., {et~al.} 2019, \aap, 625, L3

\bibitem[{{Dalgarno} \& {Lepp}(1984)}]{Dalgarno1984}
{Dalgarno}, A. \& {Lepp}, S. 1984, \apjl, 287, L47

\bibitem[{{Dislaire} {et~al.}(2012){Dislaire}, {Hily-Blant}, {Faure}, {Maret},
  {Bacmann}, \& {Pineau Des For{\^e}ts}}]{Dislaire2012}
{Dislaire}, V., {Hily-Blant}, P., {Faure}, A., {et~al.} 2012, \aap, 537, A20

\bibitem[{{Draine}(2011)}]{Draine2011}
{Draine}, B.~T. 2011, {Physics of the Interstellar and Intergalactic Medium}

\bibitem[{{Draine} \& {Bertoldi}(1996)}]{Draine1996}
{Draine}, B.~T. \& {Bertoldi}, F. 1996, \apj, 468, 269

\bibitem[{{Flower} {et~al.}(2007){Flower}, {Pineau Des For{\^e}ts}, \&
  {Walmsley}}]{Flower2007}
{Flower}, D.~R., {Pineau Des For{\^e}ts}, G., \& {Walmsley}, C.~M. 2007, \aap,
  474, 923

\bibitem[{{Foss{\'e}} {et~al.}(2001){Foss{\'e}}, {Cernicharo}, {Gerin}, \&
  {Cox}}]{Fosse2001}
{Foss{\'e}}, D., {Cernicharo}, J., {Gerin}, M., \& {Cox}, P. 2001, \apj, 552,
  168

\bibitem[{{Fuente} {et~al.}(2016){Fuente}, {Cernicharo}, {Roueff}, {Gerin},
  {Pety}, {Marcelino}, {Bachiller}, {Lefloch}, {Roncero}, \&
  {Aguado}}]{Fuente2016}
{Fuente}, A., {Cernicharo}, J., {Roueff}, E., {et~al.} 2016, \aap, 593, A94

\bibitem[{{Goicoechea} {et~al.}(2009){Goicoechea}, {Pety}, {Gerin},
  {Hily-Blant}, \& {Le Bourlot}}]{Goicoechea2009}
{Goicoechea}, J.~R., {Pety}, J., {Gerin}, M., {Hily-Blant}, P., \& {Le
  Bourlot}, J. 2009, \aap, 498, 771

\bibitem[{{Goicoechea} {et~al.}(2006){Goicoechea}, {Pety}, {Gerin}, {Teyssier},
  {Roueff}, {Hily-Blant}, \& {Baek}}]{Goicoechea2006}
{Goicoechea}, J.~R., {Pety}, J., {Gerin}, M., {et~al.} 2006, \aap, 456, 565

\bibitem[{{Goldsmith} \& {Kauffmann}(2017)}]{Goldsmith2017}
{Goldsmith}, P.~F. \& {Kauffmann}, J. 2017, \apj, 841, 25

\bibitem[{{Gratier} {et~al.}(2017){Gratier}, {Bron}, {Gerin}, {Pety}, {Guzman},
  {Orkisz}, {Bardeau}, {Goicoechea}, {Le Petit}, {Liszt}, {{\"O}berg},
  {Peretto}, {Roueff}, {Sievers}, \& {Tremblin}}]{Gratier2017}
{Gratier}, P., {Bron}, E., {Gerin}, M., {et~al.} 2017, \aap, 599, A100

\bibitem[{{Gratier} {et~al.}(subm.){Gratier}, {Pety}, {Bron}, {Gerin},
  {Orkisz}, {de Souza Magalhaes}, {Gaudel}, {Vono}, {Bardeau}, {Bourguignon},
  {Chanussot}, {Chainais}, {Goicoechea}, {Guzman}, {Hughes}, {Kainulainen}, {Le
  Bourlot}, {Le Petit}, {Levrier}, {Liszt}, {Öberg}, {Peretto}, {Roueff},
  {Roueff}, {Sievers}, \& {Tremblin}}]{Gratier2020}
{Gratier}, P., {Pety}, J., {Bron}, E., {et~al.} subm., \aap

\bibitem[{{Guelin} {et~al.}(1977){Guelin}, {Langer}, {Snell}, \&
  {Wootten}}]{Guelin1977}
{Guelin}, M., {Langer}, W.~D., {Snell}, R.~L., \& {Wootten}, H.~A. 1977, \apjl,
  217, L165

\bibitem[{{Guelin} {et~al.}(1982){Guelin}, {Langer}, \& {Wilson}}]{Guelin1982}
{Guelin}, M., {Langer}, W.~D., \& {Wilson}, R.~W. 1982, \aap, 107, 107

\bibitem[{{Guzman} {et~al.}(in prep.){Guzman}, {Gerin}, {Pety}, {Bron},
  {Orkisz}, {de Souza Magalhaes}, {Gaudel}, {Vono}, {Bardeau}, {Bourguignon},
  {Chanussot}, {Chainais}, {Goicoechea}, {Gratier}, {Hughes}, {Kainulainen},
  {Le Bourlot}, {Le Petit}, {Levrier}, {Liszt}, {Öberg}, {Peretto}, {Roueff},
  {Roueff}, {Sievers}, \& {Tremblin}}]{Guzman2020}
{Guzman}, V., {Gerin}, M., {Pety}, J., {et~al.} in prep., \aap

\bibitem[{{Guzm{\'a}n} {et~al.}(2012){Guzm{\'a}n}, {Pety}, {Gratier},
  {Goicoechea}, {Gerin}, {Roueff}, \& {Teyssier}}]{Guzman2012}
{Guzm{\'a}n}, V., {Pety}, J., {Gratier}, P., {et~al.} 2012, \aap, 543, L1

\bibitem[{{Guzm{\'a}n} {et~al.}(2015){Guzm{\'a}n}, {Pety}, {Goicoechea},
  {Gerin}, {Roueff}, {Gratier}, \& {{\"O}berg}}]{Guzman2015}
{Guzm{\'a}n}, V.~V., {Pety}, J., {Goicoechea}, J.~R., {et~al.} 2015, \apjl,
  800, L33

\bibitem[{Hastie {et~al.}(2001)Hastie, Tibshirani, \& Friedman}]{Hastie2001}
Hastie, T., Tibshirani, R., \& Friedman, J. 2001, The Elements of Statistical
  Learning, Springer Series in Statistics (New York, NY, USA: Springer New York
  Inc.)

\bibitem[{{Heays} {et~al.}(2017){Heays}, {Bosman}, \& {van
  Dishoeck}}]{Heays2017}
{Heays}, A.~N., {Bosman}, A.~D., \& {van Dishoeck}, E.~F. 2017, \aap, 602, A105

\bibitem[{{Herbst} \& {Klemperer}(1973)}]{Herbst1973}
{Herbst}, E. \& {Klemperer}, W. 1973, \apj, 185, 505

\bibitem[{{Hollenbach} \& {Tielens}(1999)}]{Hollenbach1999}
{Hollenbach}, D.~J. \& {Tielens}, A.~G.~G.~M. 1999, Reviews of Modern Physics,
  71, 173

\bibitem[{{Indriolo} {et~al.}(2007){Indriolo}, {Geballe}, {Oka}, \&
  {McCall}}]{Indriolo2007}
{Indriolo}, N., {Geballe}, T.~R., {Oka}, T., \& {McCall}, B.~J. 2007, \apj,
  671, 1736

\bibitem[{{Indriolo} \& {McCall}(2012)}]{Indriolo2012}
{Indriolo}, N. \& {McCall}, B.~J. 2012, \apj, 745, 91

\bibitem[{{Le Bourlot}(1991)}]{LeBourlot1991}
{Le Bourlot}, J. 1991, \aap, 242, 235

\bibitem[{{Le Petit} {et~al.}(2006){Le Petit}, {Nehm{\'e}}, {Le Bourlot}, \&
  {Roueff}}]{LePetit2006}
{Le Petit}, F., {Nehm{\'e}}, C., {Le Bourlot}, J., \& {Roueff}, E. 2006, The
  Astrophysical Journal Supplement Series, 164, 506

\bibitem[{{Le Petit} {et~al.}(2016){Le Petit}, {Ruaud}, {Bron}, {Godard},
  {Roueff}, {Languignon}, \& {Le Bourlot}}]{LePetit2016}
{Le Petit}, F., {Ruaud}, M., {Bron}, E., {et~al.} 2016, \aap, 585, A105

\bibitem[{{Lepp}(1992)}]{Lepp1992}
{Lepp}, S. 1992, in IAU Symposium, Vol. 150, Astrochemistry of Cosmic
  Phenomena, ed. P.~D. {Singh}, 471

\bibitem[{{Levrier} {et~al.}(2012){Levrier}, {Le Petit}, {Hennebelle},
  {Lesaffre}, {Gerin}, \& {Falgarone}}]{Levrier2012}
{Levrier}, F., {Le Petit}, F., {Hennebelle}, P., {et~al.} 2012, \aap, 544, A22

\bibitem[{{Liszt}(2012)}]{Liszt2012}
{Liszt}, H.~S. 2012, \aap, 538, A27

\bibitem[{{Liszt} \& {Pety}(2016)}]{Liszt2016}
{Liszt}, H.~S. \& {Pety}, J. 2016, \apj, 823, 124

\bibitem[{{Maret} \& {Bergin}(2007)}]{Maret2007}
{Maret}, S. \& {Bergin}, E.~A. 2007, \apj, 664, 956

\bibitem[{{McCall} {et~al.}(2003){McCall}, {Huneycutt}, {Saykally}, {Geballe},
  {Djuric}, {Dunn}, {Semaniak}, {Novotny}, {Al-Khalili}, {Ehlerding},
  {Hellberg}, {Kalhori}, {Neau}, {Thomas}, {{\"O}sterdahl}, \&
  {Larsson}}]{McCall2003}
{McCall}, B.~J., {Huneycutt}, A.~J., {Saykally}, R.~J., {et~al.} 2003, \nat,
  422, 500

\bibitem[{{Mestel} \& {Spitzer}(1956)}]{Mestel1956}
{Mestel}, L. \& {Spitzer}, L., J. 1956, \mnras, 116, 503

\bibitem[{{Miettinen} {et~al.}(2011){Miettinen}, {Hennemann}, \&
  {Linz}}]{Miettinen2011}
{Miettinen}, O., {Hennemann}, M., \& {Linz}, H. 2011, \aap, 534, A134

\bibitem[{{Mladenovi{\'c}} \& {Roueff}(2017)}]{Mladenovic2017}
{Mladenovi{\'c}}, M. \& {Roueff}, E. 2017, \aap, 605, A22

\bibitem[{{Mouschovias}(1976)}]{Mouschovias1976}
{Mouschovias}, T.~C. 1976, \apj, 207, 141

\bibitem[{{Neufeld} {et~al.}(2006){Neufeld}, {Schilke}, {Menten}, {Wolfire},
  {Black}, {Schuller}, {M{\"u}ller}, {Thorwirth}, {G{\"u}sten}, \&
  {Philipp}}]{Neufeld2006}
{Neufeld}, D.~A., {Schilke}, P., {Menten}, K.~M., {et~al.} 2006, \aap, 454, L37

\bibitem[{{Oppenheimer} \& {Dalgarno}(1974)}]{Oppenheimer1974}
{Oppenheimer}, M. \& {Dalgarno}, A. 1974, \apj, 192, 29

\bibitem[{{Orkisz} {et~al.}(2019){Orkisz}, {Peretto}, {Pety}, {Gerin},
  {Levrier}, {Bron}, {Bardeau}, {Goicoechea}, {Gratier}, {Guzm{\'a}n},
  {Hughes}, {Languignon}, {Le Petit}, {Liszt}, {{\"O}berg}, {Roueff},
  {Sievers}, \& {Tremblin}}]{Orkisz2019}
{Orkisz}, J.~H., {Peretto}, N., {Pety}, J., {et~al.} 2019, \aap, 624, A113

\bibitem[{{Orkisz} {et~al.}(2017){Orkisz}, {Pety}, {Gerin}, {Bron},
  {Guzm{\'a}n}, {Bardeau}, {Goicoechea}, {Gratier}, {Le Petit}, {Levrier},
  {Liszt}, {{\"O}berg}, {Peretto}, {Roueff}, {Sievers}, \&
  {Tremblin}}]{Orkisz2017}
{Orkisz}, J.~H., {Pety}, J., {Gerin}, M., {et~al.} 2017, \aap, 599, A99

\bibitem[{{Padovani} {et~al.}(2009){Padovani}, {Galli}, \&
  {Glassgold}}]{Padovani2009}
{Padovani}, M., {Galli}, D., \& {Glassgold}, A.~E. 2009, \aap, 501, 619

\bibitem[{{Pagani} {et~al.}(2011){Pagani}, {Roueff}, \&
  {Lesaffre}}]{Pagani2011}
{Pagani}, L., {Roueff}, E., \& {Lesaffre}, P. 2011, \apjl, 739, L35

\bibitem[{{Pagani} {et~al.}(1992){Pagani}, {Salez}, \& {Wannier}}]{Pagani1992}
{Pagani}, L., {Salez}, M., \& {Wannier}, P.~G. 1992, \aap, 258, 479

\bibitem[{Pedregosa {et~al.}(2011)Pedregosa, Varoquaux, Gramfort, Michel,
  Thirion, Grisel, Blondel, Prettenhofer, Weiss, Dubourg, Vanderplas, Passos,
  Cournapeau, Brucher, Perrot, \& Duchesnay}]{Pedregosa2011}
Pedregosa, F., Varoquaux, G., Gramfort, A., {et~al.} 2011, Journal of Machine
  Learning Research, 12, 2825

\bibitem[{{Pety} {et~al.}(2017){Pety}, {Guzm{\'a}n}, {Orkisz}, {Liszt},
  {Gerin}, {Bron}, {Bardeau}, {Goicoechea}, {Gratier}, {Le Petit}, {Levrier},
  {{\"O}berg}, {Roueff}, \& {Sievers}}]{Pety2017}
{Pety}, J., {Guzm{\'a}n}, V.~V., {Orkisz}, J.~H., {et~al.} 2017, \aap, 599, A98

\bibitem[{{Pety} {et~al.}(2005){Pety}, {Teyssier}, {Foss{\'e}}, {Gerin},
  {Roueff}, {Abergel}, {Habart}, \& {Cernicharo}}]{Pety2005}
{Pety}, J., {Teyssier}, D., {Foss{\'e}}, D., {et~al.} 2005, \aap, 435, 885

\bibitem[{{Roueff} {et~al.}(2020){Roueff}, Gerin, Gratier, Levrier, Pety,
  Gaudel, Goicoechea, Orkisz, de~Souza~Magalhaes, Vono, Bardeau, Bourguignon,
  Bron, Chanussot, Chainais, Guzman, Hughes, Kainulainen, Languignon,
  Le~Bourlot, Le~Petit, Liszt, Marchal, Miville-Deschesnes, Peretto, Roueff, \&
  Sievers}]{Roueff2020}
{Roueff}, A., Gerin, M., Gratier, P., {et~al.} 2020, \aap, accepted

\bibitem[{{Roueff} {et~al.}(2015){Roueff}, {Loison}, \& {Hickson}}]{Roueff2015}
{Roueff}, E., {Loison}, J.~C., \& {Hickson}, K.~M. 2015, \aap, 576, A99

\bibitem[{{Shingledecker} {et~al.}(2016){Shingledecker}, {Bergner}, {Le Gal},
  {{\"O}berg}, {Hincelin}, \& {Herbst}}]{Shingledecker2016}
{Shingledecker}, C.~N., {Bergner}, J.~B., {Le Gal}, R., {et~al.} 2016, \apj,
  830, 151

\bibitem[{{van der Tak} {et~al.}(2007){van der Tak}, {Black}, {Sch{\"o}ier},
  {Jansen}, \& {van Dishoeck}}]{vanderTak2007}
{van der Tak}, F.~F.~S., {Black}, J.~H., {Sch{\"o}ier}, F.~L., {Jansen}, D.~J.,
  \& {van Dishoeck}, E.~F. 2007, \aap, 468, 627

\bibitem[{Varoquaux(2018)}]{Varoquaux2018}
Varoquaux, G. 2018, NeuroImage, 180, 68

\bibitem[{{Williams} {et~al.}(1998){Williams}, {Bergin}, {Caselli}, {Myers}, \&
  {Plume}}]{Williams1998}
{Williams}, J.~P., {Bergin}, E.~A., {Caselli}, P., {Myers}, P.~C., \& {Plume},
  R. 1998, \apj, 503, 689

\end{thebibliography}

\appendix{}

\section{Random Forest parameter optimization}\label{app:RF_params_optimization}

\FigRFparametersOptimization

The Random Forest contains a few tuning parameters for which a value needs to be chosen before training. 
In particular, we will consider only the two most important here : the number of trees in the forest $N_\mathrm{trees}$, and the maximum depth allowed for each tree, $d_\mathrm{max}$. 
Another usual parameter is not considered here: the number of predictors in the random subset considered when choosing along which axis to make a split. 
Indeed, we will only train RF models with a single predictor at a time, so that this number is necessarily 1.

Since optimizing the choice of these parameters on the full model grids (that will then be used for training) and for each potential tracer will lead both to an increased risk of overfitting and a heavy computational time cost, we opt for a very limited optimization, where a single set of parameter values is used for all tracers (i.e. for all RF models trained on a single ratio) and for all model grids. 
The selection of these parameter values is done with a simplified procedure:
\begin{enumerate}
\item For each model grid, RF models are first trained on each ratio using default parameter values (default \texttt{sklearn} values, N$_\mathrm{trees} = 100$ and $d_\mathrm{max} = \infty$), and the OOB $R^2$ value is calculated for each ratio.
\item For the two ratios having the best and worst $R^2$ in the previous step, RF models are trained for a grid of parameter values exploring $N_\mathrm{trees}  = 50 - 800$ and $d_\mathrm{max} = 1 - 12$, and the OOB $R^2$ is again computed for RF models with each possible combination of parameter values (the OOB value is used to limit the risks of overfitting).
\item As the $R^2$ maps obtained show very flat minima, we found a common set of parameter values assuring a $R^2$ within 0.01 of the best value in all cases. 
The parameter values found are $N_\mathrm{trees} = 400$, and $d_\mathrm{max} = 4 $.
\end{enumerate}

In order to illustrate this procedure, Fig.~\ref{fig:RFparametersOptimization} shows for instance the resulting $R^2$ maps as a function of $N_\mathrm{trees}$ and $d_\mathrm{max}$ for the preliminary best (top) and worst (bottom) tracers in the translucent model grid when using line integrated intensity ratios. 
The red contours delimit the region where $R^2$ is within 0.01 of its maximum value.
We see that the variations with $N_\mathrm{trees}$ are limited to a decrease at low values. 
Any large enough value could thus be chosen but larger values will induce higher computation time cost, so that a minimal acceptable value had to be chosen. 
The variations with $d_\mathrm{max}$ show a clear maximum (although rather flat). 
The $R^2$ value decreases for decreasing values of $d_\mathrm{max}$ below the optimum as the RF model then becomes not flexible enough to capture the relationship between the predictor and the target variable. 
Above the optimum, $R^2$ decreases for increasing values of $d_\mathrm{max}$ as overfitting starts to arise. 
The RF model becomes too flexible and starts to learn noise artefacts (we recall that the "noise" here is induced by the random sampling of the unobservable physical parameters of the chemical model).

\section{Tables for tracers ranking, performance, and fit coefficients}
\label{app:Rankings}

\subsection{Tracers ranking and performance}

We provide here the ranking and performance of single ratio RF models, obtained following the method described in Sect.~\ref{sect:Method}.

\subsubsection{Translucent medium}

Tables~\ref{tab:RatiosRankingColumnDensitiesTranslucent} and \ref{tab:RatiosRankingIntensitiesTranslucent} present the ranking we obtain for translucent medium conditions, respectively for column density ratios and integrated line intensity ratios. 
See Sect.~\ref{sect:Method} for a description of the method used, and Sect.~\ref{sect:Tracers} for a discussion of these results. 
For each ratio, we list the performance of the corresponding RF model measured through the (cross-validated) $R^2$, the equivalent root mean square error on $\log_{10}(x(\mathrm{e}^-))$ and the corresponding error factor on $x(\mathrm{e}^-)$, the maximum absolute error on $\log_{10}(x(\mathrm{e}^-))$ and the corresponding error factor on $x(\mathrm{e}^-)$. We also list for comparison the $R^2$ value obtained with the analytical fit described in Sect.~\ref{sect:AnalyticalFits}.

  \begin{table*}
    \caption{Ranking of column density ratios according to their usefulness to predict the ionization fraction in translucent medium conditions (measured through the $R^2$ of a fitted Random Forest model). 
Additional error measures of the Random Forest model (root mean square error and maximum absolute errors) are also given. 
As these errors concern the logarithm of the ionization fraction, we also provide the equivalent error factors on the ionization fraction. 
For comparison, the $R^2$ obtained with the analytical fit described in Sect.~\ref{sect:AnalyticalFits} is also listed in the last column.}
    \label{tab:RatiosRankingColumnDensitiesTranslucent}
    \begin{center}
      \begin{tabular} {c|ccccc|c}
        \hline 
        Column density ratio  & \multicolumn{5}{c|}{Random Forest Model} & Analytical fit \\
                      &  $R^2$  & \multicolumn{2}{c}{Root mean square error}  & \multicolumn{2}{c|}{Maximum absolute error} & $R^2$  \\ 
                      &               &  dex  & (equ. factor)  & dex  & (equ. factor) & \\
\hline
C$_2$H / HCN          & 0.96 & 0.15 & (1.41) & 0.71 & (5.08)   & 0.96 \\
C$_2$H / $^{13}$CO    & 0.94 & 0.18 & (1.52) & 1.19 & (15.43)  & 0.93 \\
C$^{18}$O / CF$^+$    & 0.93 & 0.19 & (1.53) & 0.71 & (5.12)   & 0.93 \\
HCN / CF$^+$          & 0.92 & 0.20 & (1.59) & 0.80 & (6.37)   & 0.92 \\
C$_2$H / C$^{18}$O    & 0.91 & 0.21 & (1.64) & 1.17 & (14.89)  & 0.91 \\
HCN / CN              & 0.89 & 0.24 & (1.75) & 1.11 & (13.01)  & 0.88 \\
$^{13}$CO / CF$^+$    & 0.88 & 0.25 & (1.77) & 1.25 & (17.97)  & 0.88 \\
SO / HCS$^+$          & 0.85 & 0.28 & (1.90) & 1.26 & (18.04)  & 0.85 \\
C$_2$H / HNC          & 0.83 & 0.30 & (1.97) & 1.23 & (17.03)  & 0.82 \\
HCO$^+$ / CF$^+$      & 0.83 & 0.30 & (2.00) & 1.25 & (17.70)  & 0.82 \\
C$_2$H / CN           & 0.82 & 0.30 & (2.01) & 1.28 & (18.92)  & 0.81 \\
$^{13}$CO / C$^{18}$O & 0.81 & 0.31 & (2.05) & 1.03 & (10.66)  & 0.81 \\
C$_2$H / HCO$^+$      & 0.78 & 0.34 & (2.19) & 1.58 & (37.62)  & 0.77 \\
SO / C$_2$H           & 0.78 & 0.34 & (2.20) & 1.24 & (17.57)  & 0.77 \\
HNC / CF$^+$          & 0.77 & 0.35 & (2.23) & 1.32 & (21.07)  & 0.76 \\
H$_2$CS / C$_2$H      & 0.75 & 0.36 & (2.31) & 1.22 & (16.68)  & 0.75 \\
SO / CF$^+$           & 0.74 & 0.37 & (2.33) & 1.27 & (18.42)  & 0.74 \\
CN / CF$^+$           & 0.73 & 0.37 & (2.36) & 1.57 & (37.25)  & 0.73 \\
CS / SO               & 0.71 & 0.39 & (2.47) & 1.91 & (80.46)  & 0.69 \\
C$^{18}$O / CN        & 0.68 & 0.41 & (2.56) & 1.29 & (19.62)  & 0.68 \\
HCO$^+$ / CN          & 0.68 & 0.41 & (2.59) & 2.09 & (123.19) & 0.66 \\
SO / CN               & 0.65 & 0.43 & (2.67) & 1.32 & (20.73)  & 0.65 \\
HNC / CN              & 0.64 & 0.43 & (2.71) & 1.29 & (19.65)  & 0.61 \\
CS / C$_2$H           & 0.64 & 0.44 & (2.74) & 1.36 & (22.89)  & 0.63 \\
CS / HCS$^+$          & 0.63 & 0.44 & (2.75) & 1.49 & (30.96)  & 0.64 \\
H$_2$CS / HCS$^+$     & 0.62 & 0.45 & (2.81) & 1.66 & (45.43)  & 0.62 \\
H$_2$CS / CF$^+$      & 0.62 & 0.45 & (2.82) & 1.34 & (22.12)  & 0.62 \\
HCS$^+$ / C$_2$H      & 0.59 & 0.47 & (2.92) & 1.56 & (36.18)  & 0.57 \\
SO / $^{13}$CO        & 0.56 & 0.48 & (3.02) & 1.44 & (27.56)  & 0.56 \\
C$_2$H / CF$^+$       & 0.51 & 0.51 & (3.22) & 2.37 & (235.58) & 0.51 \\
SO / HNC              & 0.51 & 0.51 & (3.22) & 1.35 & (22.31)  & 0.51 \\
SO / HCN              & 0.50 & 0.51 & (3.24) & 1.78 & (60.42)  & 0.50 \\
CS / CF$^+$           & 0.50 & 0.51 & (3.27) & 1.46 & (28.94)  & 0.50 \\

        \hline 
      \end{tabular}
    \end{center}
  \end{table*}

  \begin{table*}
    \caption{Ranking of line intensity ratios according to their usefulness to predict the ionization fraction in translucent medium conditions (measured through the $R^2$ of a fitted Random Forest model). 
Additional error measures of the Random Forest model (root mean square error and maximum absolute errors) are also given. 
As these errors concern the logarithm of the ionization fraction, we also provide the equivalent error factors on the ionization fraction.
For comparison, the $R^2$ obtained with the analytical fit described in Sect.~\ref{sect:AnalyticalFits} is also listed in the last column.}
    \label{tab:RatiosRankingIntensitiesTranslucent}
    \begin{center}
      \begin{tabular} {c|ccccc|c}
        \hline 
        Line intensity ratio                  & \multicolumn{5}{c|}{Random Forest Model} & Analytical fit \\
                                      &  $R^2$  & \multicolumn{2}{c}{Root mean square error}  & \multicolumn{2}{c|}{Maximum absolute error} & $R^2$ \\ 
         &               &  dex  & (equ. factor)  & dex  & (equ. factor) & \\
\hline
C$_2$H $(1-0)$ / HCN $(1-0)$          & 0.93 & 0.19 & (1.55) & 1.13 & (13.38)  & 0.93 \\
C$_2$H $(1-0)$ / $^{13}$CO $(1-0)$    & 0.92 & 0.20 & (1.60) & 1.28 & (19.08)  & 0.92 \\
C$_2$H $(1-0)$ / C$^{18}$O $(1-0)$    & 0.92 & 0.21 & (1.62) & 1.17 & (14.87)  & 0.91 \\
C$^{18}$O $(1-0)$ / CF$^+$ $(1-0)$    & 0.89 & 0.24 & (1.74) & 1.05 & (11.30)  & 0.89 \\
C$_2$H $(1-0)$ / HNC $(1-0)$          & 0.87 & 0.26 & (1.81) & 1.13 & (13.64)  & 0.87 \\
SO $(3-2)$ / HCS$^+$ $(2-1)$          & 0.85 & 0.28 & (1.89) & 1.23 & (16.85)  & 0.85 \\
$^{13}$CO $(1-0)$ / C$^{18}$O $(1-0)$ & 0.85 & 0.28 & (1.92) & 1.13 & (13.63)  & 0.84 \\
SO $(3-2)$ / C$_2$H $(1-0)$           & 0.80 & 0.33 & (2.11) & 1.25 & (17.68)  & 0.80 \\
C$_2$H $(1-0)$ / CN $(1-0)$           & 0.80 & 0.33 & (2.13) & 1.27 & (18.59)  & 0.78 \\
C$_2$H $(1-0)$ / HCO$^+$ $(1-0)$      & 0.78 & 0.34 & (2.20) & 1.69 & (48.69)  & 0.77 \\
HCN $(1-0)$ / CF$^+$ $(1-0)$          & 0.77 & 0.34 & (2.21) & 1.24 & (17.44)  & 0.77 \\
HCO$^+$ $(1-0)$ / CF$^+$ $(1-0)$      & 0.77 & 0.35 & (2.23) & 1.52 & (32.81)  & 0.76 \\
HNC $(1-0)$ / CF$^+$ $(1-0)$          & 0.77 & 0.35 & (2.24) & 1.34 & (22.03)  & 0.76 \\
$^{13}$CO $(1-0)$ / CF$^+$ $(1-0)$    & 0.76 & 0.36 & (2.27) & 1.57 & (37.42)  & 0.76 \\
CN $(1-0)$ / CF$^+$ $(1-0)$           & 0.75 & 0.36 & (2.29) & 1.37 & (23.24)  & 0.75 \\
H$_2$CS $(3-2)$ / C$_2$H $(1-0)$      & 0.75 & 0.36 & (2.32) & 1.25 & (17.96)  & 0.75 \\
SO $(3-2)$ / CF$^+$ $(1-0)$           & 0.74 & 0.37 & (2.33) & 1.25 & (17.93)  & 0.75 \\
CS $(2-1)$ / SO $(3-2)$               & 0.70 & 0.39 & (2.48) & 1.97 & (92.77)  & 0.69 \\
C$_2$H $(1-0)$ / CF$^+$ $(1-0)$       & 0.69 & 0.40 & (2.53) & 2.31 & (204.15) & 0.68 \\
CS $(2-1)$ / C$_2$H $(1-0)$           & 0.69 & 0.41 & (2.55) & 1.28 & (18.86)  & 0.68 \\
SO $(3-2)$ / CN $(1-0)$               & 0.67 & 0.42 & (2.61) & 1.24 & (17.21)  & 0.67 \\
SO $(3-2)$ / HCN $(1-0)$              & 0.66 & 0.42 & (2.64) & 1.39 & (24.56)  & 0.66 \\
HCS$^+$ $(2-1)$ / C$_2$H $(1-0)$      & 0.64 & 0.44 & (2.72) & 1.42 & (26.42)  & 0.62 \\
HCN $(1-0)$ / HNC $(1-0)$             & 0.64 & 0.44 & (2.74) & 1.41 & (25.71)  & 0.63 \\
SO $(3-2)$ / $^{13}$CO $(1-0)$        & 0.63 & 0.44 & (2.77) & 1.41 & (25.64)  & 0.63 \\
H$_2$CS $(3-2)$ / CF$^+$ $(1-0)$      & 0.59 & 0.47 & (2.93) & 1.43 & (26.73)  & 0.59 \\
CS $(2-1)$ / HCS$^+$ $(2-1)$          & 0.58 & 0.47 & (2.94) & 1.50 & (31.41)  & 0.58 \\
C$^{18}$O $(1-0)$ / CN $(1-0)$        & 0.56 & 0.48 & (3.02) & 1.67 & (47.20)  & 0.56 \\
SO $(3-2)$ / HNC $(1-0)$              & 0.53 & 0.50 & (3.14) & 1.33 & (21.52)  & 0.53 \\
HCO$^+$ $(1-0)$ / CN $(1-0)$          & 0.52 & 0.50 & (3.16) & 2.40 & (250.68) & 0.50 \\
SO $(3-2)$ / C$^{18}$O $(1-0)$        & 0.51 & 0.51 & (3.21) & 1.45 & (28.30)  & 0.52 \\
CS $(2-1)$ / CF$^+$ $(1-0)$           & 0.50 & 0.51 & (3.25) & 1.42 & (26.06)  & 0.51 \\
        \hline 
      \end{tabular}
    \end{center}
  \end{table*}


\subsubsection{Cold dense medium}
Tables~\ref{tab:RatiosRankingColumnDensitiesDense} and \ref{tab:RatiosRankingIntensitiesDense} present the ranking we obtain for cold dense medium conditions, respectively for column density ratios and integrated line intensity ratios.

  \begin{table*}
    \caption{Ranking of column density ratios according to their usefulness to predict the ionization fraction in dense cold medium conditions (measured through the $R^2$ of a fitted Random Forest model). 
Additional error measures of the Random Forest model (root mean square error and maximum absolute errors) are also given. 
As these errors concern the logarithm of the ionization fraction, we also provide the equivalent error factors on the ionization fraction.
For comparison, the $R^2$ obtained with the analytical fit described in Sect.~\ref{sect:AnalyticalFits} is also listed in the last column.}
    \label{tab:RatiosRankingColumnDensitiesDense}
    \begin{center}
      \begin{tabular} {c|ccccc|c}
        \hline 
        Column density ratio  & \multicolumn{5}{c|}{Random Forest Model} & Analytical fit \\
                      & $R^2$ & \multicolumn{2}{c}{Root mean square error}  & \multicolumn{2}{c|}{Maximum absolute error} & $R^2$  \\ 
                      &               &  dex  & (equ. factor)  & dex  & (equ. factor) & \\
\hline
CN / N$_2$H$^+$       & 0.92 & 0.23 & (1.70) & 1.49 & (30.76)  & 0.92 \\
HNC / CN              & 0.89 & 0.27 & (1.88) & 1.19 & (15.47)  & 0.89 \\
SO / HCS$^+$          & 0.88 & 0.28 & (1.90) & 1.43 & (26.66)  & 0.83 \\
CN / DCO$^+$          & 0.88 & 0.28 & (1.93) & 1.14 & (13.79)  & 0.88 \\
HCN / CN              & 0.88 & 0.29 & (1.93) & 1.25 & (17.91)  & 0.88 \\
HCO$^+$ / CN          & 0.82 & 0.34 & (2.20) & 1.60 & (39.74)  & 0.82 \\
C$_2$H / N$_2$H$^+$   & 0.82 & 0.34 & (2.20) & 1.44 & (27.71)  & 0.77 \\
C$_2$H / DCO$^+$      & 0.82 & 0.35 & (2.22) & 1.55 & (35.52)  & 0.81 \\
C$^{18}$O / CN        & 0.81 & 0.36 & (2.28) & 1.21 & (16.32)  & 0.81 \\
CF$^+$ / DCO$^+$      & 0.80 & 0.36 & (2.30) & 1.76 & (57.33)  & 0.79 \\
$^{13}$CO / CN        & 0.79 & 0.38 & (2.37) & 1.74 & (55.31)  & 0.79 \\
C$_2$H / CF$^+$       & 0.79 & 0.38 & (2.37) & 1.18 & (15.22)  & 0.78 \\
SO / CN               & 0.78 & 0.38 & (2.40) & 1.26 & (18.41)  & 0.78 \\
C$_2$H / HCO$^+$      & 0.77 & 0.39 & (2.44) & 1.17 & (14.86)  & 0.77 \\
CS / SO               & 0.77 & 0.39 & (2.44) & 1.66 & (45.38)  & 0.76 \\
C$_2$H / C$^{18}$O    & 0.75 & 0.41 & (2.55) & 1.28 & (19.24)  & 0.75 \\
C$_2$H / $^{13}$CO    & 0.75 & 0.41 & (2.55) & 1.31 & (20.30)  & 0.75 \\
SO / C$_2$H           & 0.71 & 0.43 & (2.72) & 1.33 & (21.26)  & 0.71 \\
HCO$^+$ / CF$^+$      & 0.71 & 0.44 & (2.74) & 1.31 & (20.43)  & 0.71 \\
H$_2$CS / C$_2$H      & 0.70 & 0.45 & (2.80) & 1.33 & (21.54)  & 0.69 \\
H$_2$CS / CN          & 0.70 & 0.45 & (2.81) & 1.31 & (20.62)  & 0.70 \\
C$_2$H / HCN          & 0.69 & 0.45 & (2.84) & 1.23 & (17.04)  & 0.68 \\
HCS$^+$ / CN          & 0.69 & 0.45 & (2.84) & 1.62 & (41.78)  & 0.69 \\
C$_2$H / HNC          & 0.68 & 0.46 & (2.89) & 1.28 & (18.88)  & 0.67 \\
HNC / DCO$^+$         & 0.67 & 0.47 & (2.92) & 1.75 & (56.81)  & 0.67 \\
C$^{18}$O / CF$^+$    & 0.67 & 0.47 & (2.94) & 1.42 & (26.28)  & 0.66 \\
$^{13}$CO / CF$^+$    & 0.66 & 0.47 & (2.97) & 1.50 & (31.82)  & 0.66 \\
HCN / DCO$^+$         & 0.65 & 0.48 & (3.00) & 1.80 & (62.41)  & 0.65 \\
CS / CN               & 0.65 & 0.48 & (3.02) & 1.57 & (37.48)  & 0.65 \\
HCS$^+$ / C$_2$H      & 0.65 & 0.48 & (3.03) & 1.55 & (35.61)  & 0.64 \\
HNC / N$_2$H$^+$      & 0.65 & 0.48 & (3.04) & 2.27 & (186.97) & 0.63 \\
CN / CF$^+$           & 0.64 & 0.49 & (3.06) & 2.39 & (243.03) & 0.64 \\
SO / HNC              & 0.62 & 0.50 & (3.15) & 1.53 & (33.96)  & 0.62 \\
SO / HCN              & 0.61 & 0.50 & (3.20) & 1.58 & (37.94)  & 0.61 \\
CS / C$_2$H           & 0.61 & 0.50 & (3.20) & 1.54 & (34.65)  & 0.60 \\
SO / CF$^+$           & 0.61 & 0.51 & (3.20) & 1.70 & (49.64)  & 0.60 \\
HCN / N$_2$H$^+$      & 0.61 & 0.51 & (3.24) & 2.26 & (182.07) & 0.59 \\
HCO$^+$ / DCO$^+$     & 0.57 & 0.53 & (3.40) & 2.41 & (255.07) & 0.57 \\
CF$^+$ / N$_2$H$^+$   & 0.54 & 0.55 & (3.54) & 2.42 & (263.85) & 0.54 \\
HCO$^+$ / HNC         & 0.54 & 0.55 & (3.58) & 2.39 & (245.88) & 0.53 \\
SO / H$_2$CS          & 0.52 & 0.56 & (3.66) & 1.60 & (39.64)  & 0.51 \\
HCO$^+$ / HCN         & 0.51 & 0.57 & (3.70) & 2.43 & (267.69) & 0.50 \\
DCO$^+$ / N$_2$H$^+$  & 0.50 & 0.57 & (3.75) & 2.18 & (151.70) & 0.50 \\
C$^{18}$O / HNC       & 0.50 & 0.58 & (3.78) & 2.02 & (105.54) & 0.50 \\

        \hline 
      \end{tabular}
    \end{center}
  \end{table*}

  \begin{table*}
    \caption{Ranking of line intensity ratios according to their usefulness to predict the ionization fraction in dense cold medium conditions (measured through the $R^2$ of a fitted Random Forest model). 
Additional error measures of the Random Forest model (root mean square error and maximum absolute errors) are also given.
As these errors concern the logarithm of the ionization fraction, we also provide the equivalent error factors on the ionization fraction.
For comparison, the $R^2$ obtained with the analytical fit described in Sect.~\ref{sect:AnalyticalFits} is also listed in the last column.}
    \label{tab:RatiosRankingIntensitiesDense}
    \begin{center}
      \begin{tabular} {c|ccccc|c}
        \hline 
        Line intensity ratio                   & \multicolumn{5}{c|}{Random Forest Model} & Analytical fit \\
                                       &  $R^2$  & \multicolumn{2}{c}{Root mean square error}  & \multicolumn{2}{c|}{Maximum absolute error} & $R^2$ \\ 
         &               &  dex  & (equ. factor)  & dex  & (equ. factor) & \\
\hline
CF$^+$ $(1-0)$ / DCO$^+$ $(1-0)$       & 0.86 & 0.30 & (2.00) & 1.57 & (36.93)  & 0.85  \\
$^{13}$CO $(1-0)$ / HCO$^+$ $(1-0)$    & 0.86 & 0.30 & (2.01) & 1.45 & (27.93)  & 0.86  \\
CN $(1-0)$ / N$_2$H$^+$ $(1-0)$        & 0.86 & 0.30 & (2.01) & 1.83 & (66.88)  & 0.86  \\
C$_2$H $(1-0)$ / N$_2$H$^+$ $(1-0)$    & 0.86 & 0.31 & (2.04) & 1.09 & (12.29)  & 0.81  \\
HCO$^+$ $(1-0)$ / CF$^+$ $(1-0)$       & 0.84 & 0.33 & (2.12) & 1.14 & (13.75)  & 0.83  \\
C$_2$H $(1-0)$ / HCN $(1-0)$           & 0.83 & 0.34 & (2.19) & 1.32 & (20.69)  & 0.81  \\
$^{13}$CO $(1-0)$ / DCO$^+$ $(1-0)$    & 0.81 & 0.35 & (2.24) & 1.92 & (82.34)  & 0.82  \\
C$_2$H $(1-0)$ / HNC $(1-0)$           & 0.81 & 0.35 & (2.24) & 1.35 & (22.14)  & 0.80  \\
C$^{18}$O $(1-0)$ / DCO$^+$ $(1-0)$    & 0.81 & 0.35 & (2.26) & 1.77 & (58.68)  & 0.81  \\
C$_2$H $(1-0)$ / DCO$^+$ $(1-0)$       & 0.80 & 0.36 & (2.30) & 1.58 & (38.36)  & 0.80  \\
CF$^+$ $(1-0)$ / N$_2$H$^+$ $(1-0)$    & 0.80 & 0.37 & (2.32) & 2.09 & (122.49) & 0.77  \\
C$_2$H $(1-0)$ / HCO$^+$ $(1-0)$       & 0.79 & 0.37 & (2.37) & 1.19 & (15.64)  & 0.78  \\
CN $(1-0)$ / DCO$^+$ $(1-0)$           & 0.78 & 0.38 & (2.41) & 1.58 & (37.69)  & 0.78  \\
C$^{18}$O $(1-0)$ / HCO$^+$ $(1-0)$    & 0.76 & 0.40 & (2.52) & 1.30 & (20.13)  & 0.75  \\
HCO$^+$ $(1-0)$ / CN $(1-0)$           & 0.75 & 0.41 & (2.55) & 1.83 & (67.50)  & 0.75  \\
HCN $(1-0)$ / CN $(1-0)$               & 0.75 & 0.41 & (2.56) & 2.18 & (152.46) & 0.75  \\
HNC $(1-0)$ / CN $(1-0)$               & 0.75 & 0.41 & (2.57) & 1.64 & (43.67)  & 0.75  \\
SO $(3-2)$ / C$_2$H $(1-0)$            & 0.73 & 0.43 & (2.66) & 1.36 & (22.71)  & 0.72  \\
SO $(3-2)$ / CN $(1-0)$                & 0.70 & 0.44 & (2.78) & 1.42 & (26.59)  & 0.70  \\
CS $(2-1)$ / C$_2$H $(1-0)$            & 0.70 & 0.44 & (2.79) & 1.41 & (25.43)  & 0.70  \\
SO $(3-2)$ / $^{13}$CO $(1-0)$         & 0.69 & 0.45 & (2.83) & 1.71 & (51.83)  & 0.67  \\
H$_2$CS $(3-2)$ / C$_2$H $(1-0)$       & 0.68 & 0.46 & (2.86) & 1.45 & (28.31)  & 0.68  \\
$^{13}$CO $(1-0)$ / N$_2$H$^+$ $(1-0)$ & 0.65 & 0.48 & (3.00) & 2.30 & (201.04) & 0.65  \\
SO $(3-2)$ / CF$^+$ $(1-0)$            & 0.64 & 0.49 & (3.06) & 1.46 & (29.17)  & 0.63  \\
CS $(2-1)$ / CN $(1-0)$                & 0.64 & 0.49 & (3.06) & 1.62 & (42.00)  & 0.64  \\
C$_2$H $(1-0)$ / CF$^+$ $(1-0)$        & 0.64 & 0.49 & (3.09) & 1.37 & (23.67)  & 0.63  \\
HNC $(1-0)$ / DCO$^+$ $(1-0)$          & 0.60 & 0.52 & (3.28) & 2.00 & (99.15)  & 0.60  \\
SO $(3-2)$ / HNC $(1-0)$               & 0.58 & 0.53 & (3.36) & 1.64 & (44.11)  & 0.52  \\
SO $(3-2)$ / HCN $(1-0)$               & 0.57 & 0.53 & (3.41) & 1.66 & (45.48)  & 0.50  \\
H$_2$CS $(3-2)$ / CN $(1-0)$           & 0.56 & 0.54 & (3.44) & 1.86 & (72.28)  & 0.56  \\
HCN $(1-0)$ / DCO$^+$ $(1-0)$          & 0.56 & 0.54 & (3.45) & 1.99 & (98.14)  & 0.56  \\
C$_2$H $(1-0)$ / $^{13}$CO $(1-0)$     & 0.56 & 0.54 & (3.47) & 1.38 & (23.81)  & 0.54  \\
SO $(3-2)$ / C$^{18}$O $(1-0)$         & 0.56 & 0.54 & (3.47) & 2.15 & (140.92) & 0.54  \\
C$^{18}$O $(1-0)$ / N$_2$H$^+$ $(1-0)$ & 0.55 & 0.54 & (3.48) & 2.40 & (248.83) & 0.55  \\
HNC $(1-0)$ / N$_2$H$^+$ $(1-0)$       & 0.55 & 0.55 & (3.52) & 2.14 & (138.57) & 0.55  \\
HCS$^+$ $(2-1)$ / C$_2$H $(1-0)$       & 0.54 & 0.55 & (3.53) & 1.64 & (43.65)  & 0.52  \\
HCO$^+$ $(1-0)$ / HNC $(1-0)$          & 0.54 & 0.55 & (3.57) & 2.09 & (123.36) & 0.54  \\
C$_2$H $(1-0)$ / C$^{18}$O $(1-0)$     & 0.53 & 0.56 & (3.61) & 1.40 & (25.06)  & 0.50  \\
CS $(2-1)$ / CF$^+$ $(1-0)$            & 0.52 & 0.56 & (3.64) & 2.11 & (127.92) & 0.52  \\
HCO$^+$ $(1-0)$ / DCO$^+$ $(1-0)$      & 0.52 & 0.57 & (3.68) & 2.71 & (509.55) & 0.52  \\
HCO$^+$ $(1-0)$ / HCN $(1-0)$          & 0.51 & 0.57 & (3.69) & 2.16 & (143.66) & 0.51  \\
CS $(2-1)$ / HCS$^+$ $(2-1)$           & 0.51 & 0.57 & (3.73) & 1.79 & (61.10)  & 0.51  \\
        \hline 
      \end{tabular}
    \end{center}
  \end{table*}


\subsection{Analytical fit coefficients}

We provide here the fit coefficients for the analytical fits described in Sect.~\ref{sect:AnalyticalFits}.

\subsubsection{Translucent medium}

Tables~\ref{tab:RatiosColumnDensitiesTranslucentFullFitCoeff} and \ref{tab:RatiosIntensitiesTranslucentFullFitCoeff} list the fit coefficients for translucent medium conditions, respectively for column density ratios and integrated line intensity ratios. 
See Sect.~\ref{sect:AnalyticalFits} for a description of the method used. 
For each ratio, ranked according the results of Sect~\ref{sect:Tracers}, we list the fit coefficients (corresponding to the fit formula given in Eq.~\ref{eq:main_fit_translucent}) for predicting  $\log_{10}(x(\mathrm{e}^-))$, the quality of this fit estimated as the (cross-validated) $R^2$, the root mean square error on $\log_{10}(x(\mathrm{e}^-))$ and corresponding error factor on $x(\mathrm{e}^-)$, the maximum absolute error on $\log_{10}(x(\mathrm{e}^-))$ and corresponding error factor on $x(\mathrm{e}^-)$, the fit coefficients (corresponding to the fit formula given in Eq.~\ref{eq:uncertainty_fit}) for estimating the uncertainty on the prediction, and the limits of the validity range of the fit (given as $\log_{10}$ of the ratio values.

  \begin{sidewaystable*}
  \scriptsize
    \caption{Fit coefficients (for our main fit and scatter fit) and fit quality for column density ratios in translucent medium conditions. 
We list the fit coefficients for predicting $\log_{10}(x(\mathrm{e}^-)$ (according to the fit formula given in Eq.~\ref{eq:main_fit_translucent}), the quality of this fit estimated as the (cross-validated) $R^2$, root mean square error (RMSE) on $\log_{10}(x(\mathrm{e}^-)$ and corresponding error factor on $x(\mathrm{e}^-)$, maximum absolute error factor on $\log_{10}(x(\mathrm{e}^-))$ and corresponding error factor on $x(\mathrm{e}^-)$, the fit coefficients for estimating the uncertainty on the prediction (according to the fit formula given in Eq.~\ref{eq:uncertainty_fit}), and the validity range of the fit (given as $\log_{10}$ of the ratio values).}
    \label{tab:RatiosColumnDensitiesTranslucentFullFitCoeff}
    \begin{center}
      \begin{tabular} {c|cccccccccccc|cccccc|cc}
        \hline 
        Column density ratio &   \multicolumn{7}{c}{Main fit coefficients}  & $R^2$ & \multicolumn{2}{l}{RMSE} & \multicolumn{2}{l}{Max. abs. error} &  \multicolumn{6}{c}{Scatter fit coefficients} &  \multicolumn{2}{l}{Validity limits} \\ 
         & $f_\mathrm{max}$ & $a_0$ & $a_1$ & $a_2$ & $a_3$ & $a_4$ & $a_5$  &           &  dex  & (equ. factor)  & dex  & (equ. factor) & $b_0$ & $b_1$ & $b_2$ & $b_3$ & $b_4$ & $b_5$ & min. & max. \\
\hline
C$_2$H / HCN             & -3.635(0)  & -8.370(-1) & 1.213(0)   & 2.613(-1)  & -7.071(-2) & -4.022(-2) & -3.973(-3)  & 0.96 & 0.15 & (1.42) & 0.82 & (6.54)   & 4.437(-2)  & -2.067(-2) & -3.212(-2) & 3.046(-3)  & 6.837(-3)  & 1.128(-3)  & -3.253 & 0.624  \\
C$_2$H / $^{13}$CO       & -4.031(0)  & 2.614(-1)  & -8.546(0)  & -8.138(0)  & -2.676(0)  & -3.822(-1) & -2.014(-2)  & 0.93 & 0.19 & (1.54) & 1.09 & (12.22)  & 2.720(-1)  & 6.069(-1)  & 5.294(-1)  & 1.990(-1)  & 3.313(-2)  & 2.015(-3)  & -5.636 & -1.402 \\
C$^{18}$O / CF$^+$       & -3.824(0)  & 8.235(0)   & -1.565(1)  & 1.360(1)   & -6.064(0)  & 1.263(0)   & -9.831(-2)  & 0.93 & 0.19 & (1.53) & 0.74 & (5.52)   & 5.740(-3)  & -1.372(-3) & -2.414(-3) & 2.222(-2)  & -1.045(-2) & 1.273(-3)  & 0.310  & 4.291  \\
HCN / CF$^+$             & -3.151(0)  & -4.520(-1) & -7.391(-1) & -3.476(-1) & 5.622(-2)  & 3.622(-2)  & -6.675(-3)  & 0.92 & 0.20 & (1.58) & 0.79 & (6.20)   & -1.630(-2) & 3.703(-2)  & 6.947(-2)  & -1.994(-2) & -1.537(-2) & 4.409(-3)  & 1.454  & 2.698  \\
C$_2$H / C$^{18}$O       & -4.104(0)  & 2.499(0)   & 1.619(0)   & -6.992(-1) & -4.739(-1) & -8.592(-2) & -4.882(-3)  & 0.91 & 0.22 & (1.65) & 1.07 & (11.85)  & 4.621(-2)  & -2.457(-2) & 4.130(-2)  & 5.330(-2)  & 1.611(-2)  & 1.478(-3)  & -4.625 & -0.938 \\
HCN / CN                 & -3.948(0)  & -1.851(0)  & 3.336(-1)  & -4.178(-1) & -1.094(1)  & -1.086(1)  & -3.002(0)   & 0.88 & 0.25 & (1.78) & 1.82 & (65.92)  & 1.486(-1)  & 1.234(-1)  & 2.341(-1)  & 5.713(-1)  & 4.508(-1)  & 1.090(-1)  & -2.119 & 0.092  \\
$^{13}$CO / CF$^+$       & -3.861(0)  & 9.452(1)   & -1.327(2)  & 7.579(1)   & -2.152(1)  & 2.990(0)   & -1.622(-1)  & 0.88 & 0.25 & (1.77) & 1.18 & (15.10)  & -2.722(-1) & 1.420(0)   & -1.649(0)  & 7.481(-1)  & -1.448(-1) & 1.003(-2)  & 1.733  & 5.037  \\
SO / HCS$^+$             & -3.937(0)  & 7.348(0)   & -1.481(1)  & 1.225(1)   & -4.821(0)  & 8.667(-1)  & -5.790(-2)  & 0.85 & 0.28 & (1.91) & 1.10 & (12.67)  & -7.201(-3) & -5.729(-2) & 8.149(-2)  & 1.230(-2)  & -1.518(-2) & 1.998(-3)  & -0.588 & 4.940  \\
C$_2$H / HNC             & -3.525(0)  & -1.265(0)  & 9.225(-1)  & -4.240(-2) & -1.323(-1) & 1.760(-2)  & 1.387(-2)   & 0.82 & 0.31 & (2.03) & 1.61 & (40.76)  & 1.435(-1)  & 7.221(-2)  & -5.232(-2) & -3.427(-2) & 4.612(-3)  & 3.616(-3)  & -2.749 & 2.001  \\
HCO$^+$ / CF$^+$         & -4.062(0)  & 1.013(-1)  & -1.493(0)  & -5.710(-1) & 4.317(-1)  & 2.263(-1)  & -1.113(-1)  & 0.82 & 0.30 & (2.01) & 1.15 & (14.13)  & 1.473(-1)  & 3.230(-2)  & -6.171(-2) & -3.840(-2) & 6.062(-3)  & 6.361(-3)  & -2.187 & 2.343  \\
C$_2$H / CN              & -4.085(0)  & 2.350(0)   & 6.570(-1)  & -5.369(0)  & -4.544(0)  & -1.383(0)  & -1.455(-1)  & 0.81 & 0.31 & (2.05) & 1.12 & (13.27)  & -1.678(-1) & -8.364(-1) & -5.548(-1) & 3.671(-2)  & 9.602(-2)  & 1.703(-2)  & -3.045 & -2.004 \\
$^{13}$CO / C$^{18}$O    & -4.413(0)  & -1.376(3)  & 4.285(3)   & -3.801(3)  & -4.142(2)  & 1.950(3)   & -6.427(2)   & 0.81 & 0.31 & (2.05) & 1.06 & (11.53)  & -2.662(1)  & 9.190(1)   & -1.232(2)  & 8.077(1)   & -2.599(1)  & 3.294(0)   & 1.537  & 1.746  \\
C$_2$H / HCO$^+$         & -4.040(0)  & -7.828(-1) & 8.847(-1)  & 6.397(-2)  & -8.373(-2) & 1.195(-3)  & 6.546(-3)   & 0.77 & 0.35 & (2.22) & 1.29 & (19.71)  & 2.277(-1)  & 4.613(-2)  & -7.111(-2) & -2.400(-3) & 5.612(-3)  & -3.318(-4) & -2.675 & 2.561  \\
SO / C$_2$H              & -3.726(0)  & -1.373(0)  & -5.037(-1) & 2.892(-2)  & 2.025(-2)  & -2.726(-4) & -5.281(-4)  & 0.77 & 0.34 & (2.21) & 1.27 & (18.66)  & 1.601(-1)  & -5.940(-2) & -1.611(-2) & 3.737(-3)  & 4.482(-4)  & -4.508(-5) & -4.428 & 3.991  \\
HNC / CF$^+$             & -3.953(0)  & 3.128(-1)  & -3.844(-1) & -5.504(-1) & -7.870(-1) & 7.526(-1)  & -1.539(-1)  & 0.76 & 0.35 & (2.25) & 1.31 & (20.28)  & 1.545(-1)  & 1.244(-1)  & -5.586(-2) & -4.392(-2) & 5.847(-3)  & 3.900(-3)  & -3.233 & 2.420  \\
H$_2$CS / C$_2$H         & -3.932(0)  & -2.060(0)  & -4.396(-2) & -3.645(-2) & -8.939(-2) & -1.980(-2) & -1.291(-3)  & 0.75 & 0.36 & (2.30) & 1.20 & (15.91)  & -6.965(-2) & 3.752(-2)  & 1.385(-1)  & 4.521(-2)  & 5.197(-3)  & 2.020(-4)  & -7.441 & -0.375 \\
SO / CF$^+$              & -3.911(0)  & -6.385(-1) & -6.249(-1) & -4.971(-2) & 3.138(-2)  & 5.788(-3)  & -1.529(-3)  & 0.74 & 0.37 & (2.33) & 1.27 & (18.42)  & 2.277(-1)  & -1.366(-2) & -3.289(-2) & -8.874(-4) & 1.036(-3)  & 7.759(-5)  & -3.604 & 3.682  \\
CN / CF$^+$              & -3.930(0)  & 1.246(2)   & -3.004(2)  & 2.888(2)   & -1.369(2)  & 3.170(1)   & -2.863(0)   & 0.73 & 0.37 & (2.36) & 1.58 & (37.65)  & 2.958(-1)  & -4.309(-1) & -1.273(0)  & 2.041(0)   & -8.987(-1) & 1.239(-1)  & 1.265  & 2.891  \\
CS / SO                  & -4.138(0)  & 1.013(-2)  & 1.554(0)   & -1.174(0)  & -9.575(-1) & 4.480(-1)  & 3.229(-1)   & 0.69 & 0.40 & (2.52) & 1.77 & (59.26)  & 3.102(-1)  & -1.434(-1) & -2.798(-1) & 1.635(-1)  & 5.832(-2)  & -3.287(-2) & -1.316 & 2.510  \\
C$^{18}$O / CN           & -4.120(-1) & -4.024(0)  & -1.043(0)  & -5.355(-1) & -1.223(-1) & 5.611(-1)  & -1.755(-1)  & 0.68 & 0.41 & (2.56) & 1.24 & (17.32)  & 1.645(-1)  & 6.466(-1)  & -4.403(-1) & -1.157(0)  & 1.123(0)   & -2.616(-1) & -inf   & inf    \\
HCO$^+$ / CN             & 8.833(0)   & -1.516(1)  & 7.408(-1)  & 3.082(0)   & 1.917(0)   & 4.716(-1)  & 4.116(-2)   & 0.66 & 0.42 & (2.64) & 2.14 & (136.69) & -2.239(-2) & 8.260(-1)  & 2.605(0)   & 1.973(0)   & 5.569(-1)  & 5.327(-2)  & -4.482 & 0.377  \\
SO / CN                  & -3.978(0)  & -1.821(0)  & -4.141(-1) & 2.051(-1)  & 2.088(-2)  & -1.389(-2) & -2.321(-3)  & 0.65 & 0.43 & (2.68) & 1.44 & (27.28)  & 8.152(-2)  & -1.765(-1) & -1.356(-3) & 2.063(-2)  & 2.697(-3)  & 1.888(-5)  & -4.865 & 0.783  \\
HNC / CN                 & -4.201(0)  & -1.808(0)  & -8.136(-1) & -1.069(1)  & -2.839(1)  & -2.231(1)  & -5.527(0)   & 0.61 & 0.45 & (2.83) & 1.27 & (18.44)  & -1.931(-2) & 1.551(0)   & 5.046(0)   & 4.541(0)   & 1.585(0)   & 1.917(-1)  & -3.301 & -0.490 \\
CS / C$_2$H              & -3.770(0)  & -1.192(0)  & -7.400(-1) & 7.328(-3)  & 5.782(-2)  & 1.405(-3)  & -2.178(-3)  & 0.63 & 0.44 & (2.75) & 1.33 & (21.20)  & 2.706(-1)  & -9.081(-2) & -4.556(-2) & 1.102(-2)  & 1.754(-3)  & -3.864(-4) & -1.312 & 2.180  \\
CS / HCS$^+$             & -3.648(0)  & 2.920(2)   & -5.013(2)  & 3.417(2)   & -1.149(2)  & 1.896(1)   & -1.228(0)   & 0.64 & 0.44 & (2.74) & 1.48 & (30.19)  & -3.115(0)  & 1.318(1)   & -1.535(1)  & 7.549(0)   & -1.661(0)  & 1.350(-1)  & 2.258  & 3.604  \\
H$_2$CS / HCS$^+$        & -3.991(0)  & -9.394(-1) & -1.776(0)  & 1.009(0)   & 1.073(0)   & -3.221(-1) & -2.743(-1)  & 0.62 & 0.45 & (2.80) & 1.57 & (37.31)  & 3.064(-1)  & 7.900(-2)  & -1.855(-1) & -7.805(-2) & -1.412(-3) & 1.601(-3)  & -1.386 & 0.693  \\
H$_2$CS / CF$^+$         & -3.881(0)  & -1.970(0)  & -2.273(-1) & -4.858(-2) & -7.103(-2) & -1.508(-2) & -9.724(-4)  & 0.62 & 0.45 & (2.82) & 1.31 & (20.40)  & 2.987(-2)  & -9.972(-2) & 9.940(-2)  & 5.239(-2)  & 7.456(-3)  & 3.323(-4)  & -6.112 & -0.673 \\
HCS$^+$ / C$_2$H         & -4.110(0)  & -1.786(0)  & -7.555(-1) & -1.665(0)  & -1.042(0)  & -2.232(-1) & -1.584(-2)  & 0.57 & 0.48 & (2.99) & 1.89 & (77.03)  & 1.135(-1)  & 1.207(-1)  & 1.116(-1)  & 9.235(-2)  & 2.315(-2)  & 1.738(-3)  & -5.545 & -1.885 \\
SO / $^{13}$CO           & -4.194(0)  & -1.879(0)  & -1.026(0)  & -1.531(0)  & -8.797(-1) & -1.806(-1) & -1.243(-2)  & 0.56 & 0.48 & (3.03) & 1.42 & (26.19)  & 1.702(-1)  & 8.542(-1)  & 8.353(-1)  & 2.667(-1)  & 3.510(-2)  & 1.657(-3)  & -5.763 & -1.594 \\
C$_2$H / CF$^+$          & -4.550(0)  & -1.147(0)  & 4.789(-1)  & 2.745(0)   & 4.953(0)   & -2.355(-1) & -2.466(0)   & 0.51 & 0.51 & (3.22) & 2.30 & (201.21) & 2.612(-1)  & 7.017(-1)  & -4.663(-1) & -6.193(-1) & 3.600(-1)  & 1.172(-2)  & 0.299  & 1.473  \\
SO / HNC                 & -4.145(0)  & -8.731(-1) & -1.090(0)  & 9.796(-2)  & 1.584(-1)  & -2.686(-3) & -1.105(-2)  & 0.51 & 0.51 & (3.22) & 1.41 & (25.80)  & 3.250(-1)  & -2.133(-1) & -1.030(-1) & 3.483(-2)  & 9.133(-3)  & -8.541(-4) & -2.371 & 0.647  \\
SO / HCN                 & -4.004(0)  & -1.455(0)  & -8.495(-1) & 2.256(-1)  & 1.154(-1)  & -1.973(-2) & -8.133(-3)  & 0.50 & 0.51 & (3.24) & 1.59 & (38.96)  & 2.539(-1)  & -2.286(-1) & -7.418(-2) & 2.577(-2)  & 9.280(-3)  & 6.229(-4)  & -3.316 & 0.305  \\
CS / CF$^+$              & -3.887(0)  & -4.910(-1) & -6.626(-1) & -1.164(-1) & 3.678(-2)  & 1.710(-2)  & -3.709(-3)  & 0.50 & 0.51 & (3.25) & 1.43 & (26.71)  & 3.976(-1)  & 6.585(-2)  & -8.564(-2) & -1.212(-2) & 3.812(-3)  & 4.385(-4)  & -1.948 & 2.035  \\
        \hline 
      \end{tabular}
    \end{center}
  \end{sidewaystable*}

  \begin{sidewaystable*}
  \scriptsize
    \caption{Fit coefficients (for our main fit and scatter fit) and fit quality for line intensity ratios in translucent medium conditions.
We list the fit coefficients for predicting $\log_{10}(x(\mathrm{e}^-)$ (according to the fit formula given in Eq.~\ref{eq:main_fit_translucent}), the quality of this fit estimated as the (cross-validated) $R^2$, root mean square error (RMSE) on $\log_{10}(x(\mathrm{e}^-)$ and corresponding error factor on $x(\mathrm{e}^-)$, maximum absolute error factor on $\log_{10}(x(\mathrm{e}^-))$ and corresponding error factor on $x(\mathrm{e}^-)$, the fit coefficients for estimating the uncertainty on the prediction (according to the fit formula given in Eq.~\ref{eq:uncertainty_fit}), and the validity range of the fit (given as $\log_{10}$ of the ratio values).}
    \label{tab:RatiosIntensitiesTranslucentFullFitCoeff}
    \begin{center}
      \begin{tabular} {c|cccccccccccc|cccccc|cc}
        \hline 
        Line intensity ratio &   \multicolumn{7}{c}{Main fit coefficients}  & $R^2$ & \multicolumn{2}{l}{RMSE} & \multicolumn{2}{l}{Max. abs. error} &  \multicolumn{6}{c}{Scatter fit coefficients} &  \multicolumn{2}{l}{Validity limits} \\ 
         & $f_\mathrm{max}$ & $a_0$ & $a_1$ & $a_2$ & $a_3$ & $a_4$ & $a_5$  &           &  dex  & (equ. factor)  & dex  & (equ. factor) & $b_0$ & $b_1$ & $b_2$ & $b_3$ & $b_4$ & $b_5$ & min. & max. \\
\hline
C$_2$H $(1-0)$ / HCN $(1-0)$          & -3.474(0) & 6.598(-1)  & 6.384(-1)  & -2.314(0)  & -1.681(0)  & -4.320(-1) & -3.817(-2) & 0.93 & 0.19 & (1.55) & 1.09 & (12.25)  & -1.541(-2) & 3.633(-2)  & 3.417(-1)  & 3.043(-1)  & 9.345(-2)  & 9.524(-3)  & -4.081 & 0.273  \\
C$_2$H $(1-0)$ / $^{13}$CO $(1-0)$    & -3.973(0) & 2.657(-1)  & -7.172(0)  & -6.245(0)  & -1.818(0)  & -2.233(-1) & -9.668(-3) & 0.92 & 0.21 & (1.62) & 1.20 & (15.82)  & 6.962(-1)  & 1.679(0)   & 1.397(0)   & 5.037(-1)  & 8.171(-2)  & 4.908(-3)  & -5.529 & -0.643 \\
C$_2$H $(1-0)$ / C$^{18}$O $(1-0)$    & 1.911(0)  & -5.969(0)  & 4.842(-1)  & 4.559(-1)  & 4.676(-1)  & 1.404(-1)  & 1.303(-2)  & 0.91 & 0.21 & (1.63) & 1.09 & (12.30)  & 1.715(-2)  & -1.452(-2) & 7.638(-2)  & 5.710(-2)  & 1.266(-2)  & 8.749(-4)  & -4.902 & 0.369  \\
C$^{18}$O $(1-0)$ / CF$^+$ $(1-0)$    & -3.840(0) & 1.962(0)   & -4.187(0)  & 5.215(0)   & -4.202(0)  & 1.448(0)   & -1.744(-1) & 0.89 & 0.24 & (1.75) & 1.00 & (10.02)  & 2.952(-3)  & 1.324(-2)  & 8.212(-2)  & -2.444(-3) & -2.985(-2) & 7.030(-3)  & -1.695 & 3.172  \\
C$_2$H $(1-0)$ / HNC $(1-0)$          & -3.623(0) & -3.627(-1) & 8.444(-1)  & -2.051(-1) & -3.929(-2) & 5.314(-2)  & 1.185(-2)  & 0.87 & 0.27 & (1.85) & 1.17 & (14.88)  & 1.068(-1)  & -5.354(-2) & -4.733(-2) & 1.029(-2)  & 6.327(-3)  & 3.901(-4)  & -3.596 & 2.228  \\
SO $(3-2)$ / HCS$^+$ $(2-1)$          & -3.954(0) & 6.092(-1)  & -1.066(0)  & 1.852(0)   & -1.834(0)  & 6.183(-1)  & -6.828(-2) & 0.85 & 0.28 & (1.90) & 1.14 & (13.65)  & 2.658(-2)  & 1.019(-1)  & 4.823(-2)  & -3.205(-2) & -1.002(-2) & 3.634(-3)  & -1.758 & 3.671  \\
$^{13}$CO $(1-0)$ / C$^{18}$O $(1-0)$ & -4.114(0) & -1.185(1)  & 7.634(1)   & -2.225(2)  & 3.014(2)   & -1.835(2)  & 4.095(1)   & 0.84 & 0.29 & (1.94) & 1.08 & (12.08)  & -3.072(0)  & 1.340(1)   & -2.181(1)  & 1.718(1)   & -6.584(0)  & 9.807(-1)  & 0.504  & 1.603  \\
SO $(3-2)$ / C$_2$H $(1-0)$           & -3.909(0) & -7.841(-1) & -5.734(-1) & -7.304(-3) & 2.501(-2)  & 2.095(-3)  & -9.363(-4) & 0.80 & 0.32 & (2.11) & 1.23 & (17.18)  & 1.771(-1)  & -2.883(-2) & -2.118(-2) & 1.454(-3)  & 6.016(-4)  & -3.513(-6) & -5.968 & 4.967  \\
C$_2$H $(1-0)$ / CN $(1-0)$           & 6.415(0)  & -1.077(1)  & 3.854(-1)  & 2.471(-1)  & 8.541(-1)  & 4.505(-1)  & 6.662(-2)  & 0.78 & 0.34 & (2.18) & 1.20 & (15.91)  & 7.188(-2)  & -2.570(-1) & 1.165(-1)  & 4.372(-1)  & 2.106(-1)  & 2.931(-2)  & -3.177 & 0.318  \\
C$_2$H $(1-0)$ / HCO$^+$ $(1-0)$      & -3.984(0) & 5.832(-1)  & 6.705(-1)  & -7.908(-2) & 1.345(-1)  & 7.862(-2)  & 9.885(-3)  & 0.77 & 0.35 & (2.23) & 1.42 & (26.35)  & 1.084(-1)  & -1.346(-1) & 1.495(-2)  & 4.138(-2)  & 4.806(-3)  & -4.845(-4) & -4.399 & 1.597  \\
HCN $(1-0)$ / CF$^+$ $(1-0)$          & -3.864(0) & 1.659(0)   & -5.861(0)  & 1.230(1)   & -1.490(1)  & 7.509(0)   & -1.318(0)  & 0.77 & 0.35 & (2.23) & 1.21 & (16.05)  & 1.260(-3)  & 1.742(-1)  & 2.160(-1)  & -1.679(-1) & -8.601(-2) & 4.790(-2)  & -1.715 & 2.062  \\
HCO$^+$ $(1-0)$ / CF$^+$ $(1-0)$      & -4.154(0) & 9.872(-1)  & -2.395(-1) & -5.629(-1) & -2.091(0)  & 1.806(0)   & -3.754(-1) & 0.76 & 0.35 & (2.25) & 1.34 & (21.88)  & 1.221(-1)  & 8.695(-2)  & 1.088(-1)  & -4.521(-2) & -8.987(-2) & 3.306(-2)  & -1.380 & 2.566  \\
HNC $(1-0)$ / CF$^+$ $(1-0)$          & -2.083(0) & -2.448(0)  & -5.945(-1) & -2.719(-1) & -5.546(-2) & 4.506(-2)  & 1.463(-2)  & 0.76 & 0.36 & (2.27) & 1.35 & (22.48)  & 1.903(-1)  & 9.399(-2)  & -1.069(-1) & -4.576(-2) & 1.550(-2)  & 5.941(-3)  & -1.974 & 2.051  \\
$^{13}$CO $(1-0)$ / CF$^+$ $(1-0)$    & -3.893(0) & 3.384(1)   & -6.341(1)  & 5.033(1)   & -2.025(1)  & 3.971(0)   & -3.016(-1) & 0.76 & 0.36 & (2.26) & 1.56 & (36.03)  & 2.188(-1)  & -8.211(-2) & -8.658(-1) & 9.356(-1)  & -3.158(-1) & 3.412(-2)  & -0.203 & 3.755  \\
CN $(1-0)$ / CF$^+$ $(1-0)$           & -3.853(0) & 4.168(-1)  & -3.398(0)  & 1.588(1)   & -3.864(1)  & 3.277(1)   & -9.017(0)  & 0.75 & 0.37 & (2.32) & 1.40 & (24.87)  & 8.622(-2)  & 3.146(-1)  & 2.333(-2)  & -3.975(-1) & -2.909(-2) & 1.198(-1)  & -1.536 & 1.542  \\
H$_2$CS $(3-2)$ / C$_2$H $(1-0)$      & -3.983(0) & -1.893(0)  & -1.921(-1) & 1.043(-1)  & -1.803(-2) & -8.856(-3) & -7.804(-4) & 0.75 & 0.36 & (2.31) & 1.24 & (17.45)  & 8.301(-3)  & -1.115(-1) & 2.225(-2)  & 1.833(-2)  & 2.440(-3)  & 8.376(-5)  & -7.161 & 1.612  \\
SO $(3-2)$ / CF$^+$ $(1-0)$           & -3.961(0) & -9.262(-1) & -7.186(-1) & 2.040(-2)  & 5.801(-2)  & 4.632(-4)  & -2.885(-3) & 0.75 & 0.37 & (2.32) & 1.26 & (18.32)  & 2.149(-1)  & -6.064(-2) & -4.249(-2) & 2.179(-3)  & 2.409(-3)  & 2.162(-4)  & -6.954 & 3.493  \\
CS $(2-1)$ / SO $(3-2)$               & -2.351(0) & -2.507(0)  & 1.054(0)   & -9.533(-1) & -1.716(-1) & 4.825(-1)  & -1.206(-1) & 0.69 & 0.40 & (2.52) & 1.85 & (70.93)  & 3.143(-1)  & -9.099(-2) & -2.967(-1) & 1.701(-1)  & 2.099(-2)  & -1.772(-2) & -1.102 & 2.641  \\
C$_2$H $(1-0)$ / CF$^+$ $(1-0)$       & -1.031(0) & -3.408(0)  & 5.487(-1)  & -2.625(-1) & 2.109(0)   & 2.149(0)   & 5.333(-1)  & 0.68 & 0.41 & (2.56) & 2.30 & (199.69) & 1.201(-1)  & -4.769(-1) & 2.133(-1)  & 1.137(0)   & 6.275(-1)  & 8.939(-2)  & -1.939 & 0.456  \\
CS $(2-1)$ / C$_2$H $(1-0)$           & -3.945(0) & -3.822(-1) & -7.675(-1) & -9.786(-2) & 5.125(-2)  & 1.053(-2)  & -2.883(-3) & 0.68 & 0.41 & (2.56) & 1.29 & (19.42)  & 2.713(-1)  & -3.205(-3) & -5.462(-2) & -4.568(-4) & 2.535(-3)  & -3.266(-5) & -3.767 & 4.069  \\
SO $(3-2)$ / CN $(1-0)$               & -4.057(0) & -1.310(0)  & -7.640(-1) & 1.827(-1)  & 7.806(-2)  & -1.538(-2) & -5.992(-3) & 0.67 & 0.42 & (2.60) & 1.27 & (18.80)  & 2.024(-1)  & -1.670(-1) & -6.207(-2) & 1.774(-2)  & 7.736(-3)  & 6.638(-4)  & -4.935 & 2.531  \\
SO $(3-2)$ / HCN $(1-0)$              & -4.035(0) & -1.669(0)  & -5.130(-1) & 3.117(-1)  & 1.948(-2)  & -4.429(-2) & -8.056(-3) & 0.66 & 0.42 & (2.64) & 1.39 & (24.50)  & 9.314(-2)  & -2.264(-1) & -9.914(-3) & 5.106(-2)  & 1.414(-2)  & 1.088(-3)  & -5.496 & 1.583  \\
HCS$^+$ $(2-1)$ / C$_2$H $(1-0)$      & -4.002(0) & -1.881(0)  & -4.844(-1) & 4.778(-1)  & 4.116(-2)  & -6.703(-2) & -1.274(-2) & 0.62 & 0.45 & (2.79) & 2.00 & (100.98) & 3.917(-2)  & -1.232(-1) & 6.014(-2)  & 1.073(-2)  & -1.155(-2) & -2.075(-3) & -3.699 & 0.034  \\
HCN $(1-0)$ / HNC $(1-0)$             & -3.986(0) & -1.925(0)  & 7.150(-1)  & 1.800(1)   & -3.703(1)  & 2.622(1)   & -6.001(0)  & 0.63 & 0.44 & (2.76) & 1.39 & (24.67)  & 2.204(-2)  & 6.159(-1)  & 8.799(-1)  & -2.688(0)  & 1.746(0)   & -3.502(-1) & -0.065 & 2.008  \\
SO $(3-2)$ / $^{13}$CO $(1-0)$        & -4.173(0) & -1.791(0)  & -4.388(-1) & -7.894(-1) & -6.634(-1) & -1.676(-1) & -1.352(-2) & 0.63 & 0.44 & (2.78) & 1.39 & (24.52)  & 1.574(-2)  & 3.669(-1)  & 5.426(-1)  & 2.042(-1)  & 2.994(-2)  & 1.537(-3)  & -4.955 & -0.228 \\
H$_2$CS $(3-2)$ / CF$^+$ $(1-0)$      & -4.104(0) & -1.775(0)  & -3.032(-1) & -6.913(-2) & -4.420(-2) & -3.797(-3) & -5.305(-7) & 0.59 & 0.47 & (2.93) & 1.40 & (25.37)  & 2.041(-2)  & -5.507(-2) & 1.255(-1)  & 5.418(-2)  & 6.992(-3)  & 2.863(-4)  & -7.051 & -0.406 \\
CS $(2-1)$ / HCS$^+$ $(2-1)$          & -3.662(0) & 3.512(1)   & -9.714(1)  & 1.078(2)   & -5.878(1)  & 1.542(1)   & -1.553(0)  & 0.58 & 0.47 & (2.93) & 1.47 & (29.45)  & -3.111(0)  & 1.456(1)   & -2.398(1)  & 1.770(1)   & -5.900(0)  & 7.213(-1)  & 0.852  & 2.460  \\
C$^{18}$O $(1-0)$ / CN $(1-0)$        & -3.852(0) & 1.244(0)   & -2.141(0)  & -2.632(-1) & -6.854(-2) & 3.998(-1)  & -1.186(-1) & 0.56 & 0.48 & (3.02) & 1.66 & (46.05)  & 6.330(-2)  & 4.472(-1)  & 1.394(-1)  & -6.256(-1) & 3.127(-1)  & -4.568(-2) & -0.348 & 2.021  \\
SO $(3-2)$ / HNC $(1-0)$              & -4.216(0) & -1.118(0)  & -1.061(0)  & 3.940(-1)  & 1.831(-1)  & -6.152(-2) & -2.416(-2) & 0.53 & 0.50 & (3.13) & 1.37 & (23.28)  & 2.624(-1)  & -3.132(-1) & -1.328(-1) & 6.489(-2)  & 3.354(-2)  & 3.645(-3)  & -3.649 & 1.672  \\
HCO$^+$ $(1-0)$ / CN $(1-0)$          & -1.118(0) & -3.481(0)  & -3.633(-1) & -6.051(-1) & -6.039(-1) & 1.302(-1)  & 1.776(-1)  & 0.50 & 0.51 & (3.25) & 2.61 & (409.10) & 1.382(-1)  & 8.082(-1)  & 7.290(-2)  & -7.536(-1) & -8.144(-2) & 1.482(-1)  & -1.183 & 0.729  \\
SO $(3-2)$ / C$^{18}$O $(1-0)$        & -4.051(0) & -1.883(0)  & -2.375(-1) & 6.632(-1)  & 2.176(-1)  & 1.139(-2)  & -1.918(-3) & 0.52 & 0.50 & (3.20) & 1.41 & (25.55)  & 1.080(-3)  & -2.984(-1) & 1.167(-1)  & 1.399(-1)  & 3.361(-2)  & 2.502(-3)  & -4.359 & 0.330  \\
CS $(2-1)$ / CF$^+$ $(1-0)$           & -3.977(0) & -6.552(-1) & -8.417(-1) & -6.468(-2) & 8.677(-2)  & 1.219(-2)  & -6.424(-3) & 0.51 & 0.51 & (3.23) & 1.43 & (26.99)  & 4.100(-1)  & -1.382(-2) & -1.186(-1) & -1.136(-2) & 7.840(-3)  & 1.231(-3)  & -3.656 & 2.409  \\
        \hline 
      \end{tabular}
    \end{center}
  \end{sidewaystable*}

\subsubsection{Cold dense medium}

Tables~\ref{tab:RatiosColumnDensitiesDenseFullFitCoeff} and \ref{tab:RatiosIntensitiesDenseFullFitCoeff} list the fit coefficients for cold dense medium conditions, respectively for column density ratios and integrated line intensity ratios. 
For each ratio, ranked according the results of Sect~\ref{sect:Tracers}, we list the fit coefficients (corresponding to the fit formula given in Eq.~\ref{eq:main_fit_dense}) for predicting  $\log_{10}(x(\mathrm{e}^-))$, the quality of this fit estimated as the (cross-validated) $R^2$, the root mean square error on $\log_{10}(x(\mathrm{e}^-))$ and corresponding error factor on $x(\mathrm{e}^-)$, the maximum absolute error on $\log_{10}(x(\mathrm{e}^-))$ and corresponding error factor on $x(\mathrm{e}^-)$, the fit coefficients (corresponding to the fit formula given in Eq.~\ref{eq:uncertainty_fit}) for estimating the uncertainty on the prediction, and the limits of the validity range of the fit (given as $\log_{10}$ of the ratio values.

  \begin{sidewaystable*}
  \scriptsize
    \caption{Fit coefficients (for our main fit and scatter fit) and fit quality for column density ratios in cold dense medium conditions.
We list the fit coefficients for predicting $\log_{10}(x(\mathrm{e}^-)$ (according to the fit formula given in Eq.~\ref{eq:main_fit_dense}), the quality of this fit estimated as the (cross-validated) $R^2$, root mean square error (RMSE) on $\log_{10}(x(\mathrm{e}^-)$ and corresponding error factor on $x(\mathrm{e}^-)$, maximum absolute error factor on $\log_{10}(x(\mathrm{e}^-))$ and corresponding error factor on $x(\mathrm{e}^-)$, the fit coefficients for estimating the uncertainty on the prediction (according to the fit formula given in Eq.~\ref{eq:uncertainty_fit}), and the validity range of the fit (given as $\log_{10}$ of the ratio values).}
    \label{tab:RatiosColumnDensitiesDenseFullFitCoeff}
    \begin{center}
      \begin{tabular} {c|ccccccccccc|cccccc|cc}
        \hline 
        Column density ratio &   \multicolumn{6}{c}{Main fit coefficients}  & $R^2$ & \multicolumn{2}{l}{RMSE} & \multicolumn{2}{l}{Max. abs. error} &  \multicolumn{6}{c}{Scatter fit coefficients} &  \multicolumn{2}{l}{Validity limits} \\ 
         & $a_0$ & $a_1$ & $a_2$ & $a_3$ & $a_4$ & $a_5$  &           &  dex  & (equ. factor)  & dex  & (equ. factor) & $b_0$ & $b_1$ & $b_2$ & $b_3$ & $b_4$ & $b_5$ & min. & max. \\
\hline
CN / N$_2$H$^+$      & -7.561(0)  & 7.191(-1)  & 4.245(-2)  & -5.030(-2) & 1.433(-3)  & 1.432(-3)  & 0.92 & 0.23 & (1.69) & 1.44 & (27.85)  & 8.041(-2)  & -7.902(-3) & -3.071(-2) & 1.565(-2)  & -2.670(-3) & 1.433(-4)  & -1.163 & 4.105  \\
HNC / CN             & -6.688(0)  & -1.441(0)  & 2.400(-1)  & 3.289(-1)  & -2.172(-1) & 3.518(-2)  & 0.89 & 0.27 & (1.88) & 1.14 & (13.87)  & 8.695(-2)  & -5.348(-2) & -2.167(-2) & 9.789(-2)  & -5.606(-2) & 7.240(-3)  & -0.867 & 2.167  \\
SO / HCS$^+$         & -1.164(1)  & 4.093(1)   & -3.472(1)  & 1.154(1)   & -1.706(0)  & 9.336(-2)  & 0.83 & 0.33 & (2.16) & 1.77 & (58.35)  & 1.986(1)   & -2.280(1)  & 1.065(1)   & -2.512(0)  & 2.960(-1)  & -1.378(-2) & 3.190  & 5.651  \\
CN / DCO$^+$         & -7.932(0)  & 3.588(-1)  & 4.656(-2)  & 2.160(-2)  & -1.182(-2) & 1.250(-3)  & 0.88 & 0.28 & (1.92) & 1.17 & (14.69)  & 7.834(-2)  & 6.037(-2)  & -9.351(-3) & -1.152(-2) & 3.247(-3)  & -2.249(-4) & -1.046 & 5.488  \\
HCN / CN             & -6.429(0)  & -1.311(0)  & -2.158(-2) & 2.582(-1)  & -6.006(-2) & -3.863(-3) & 0.88 & 0.29 & (1.93) & 1.28 & (19.02)  & 1.110(-1)  & -1.312(-2) & -1.223(-1) & 9.218(-2)  & -1.380(-3) & -7.957(-3) & -0.752 & 2.434  \\
HCO$^+$ / CN         & -7.272(0)  & -5.440(-1) & 5.703(-2)  & 2.188(-2)  & -4.248(-3) & -1.607(-3) & 0.82 & 0.34 & (2.20) & 1.61 & (41.13)  & 1.712(-1)  & 3.642(-3)  & -3.996(-2) & -6.869(-3) & 3.220(-3)  & 8.277(-4)  & -3.149 & 2.640  \\
C$_2$H / N$_2$H$^+$  & -6.105(0)  & 3.568(-1)  & -1.771(-1) & 9.534(-3)  & 1.147(-2)  & -7.543(-5) & 0.77 & 0.39 & (2.43) & 1.25 & (17.67)  & 2.364(-2)  & 1.214(-1)  & -5.300(-2) & -2.543(-2) & 4.519(-3)  & 1.639(-3)  & -3.400 & 2.847  \\
C$_2$H / DCO$^+$     & -6.781(0)  & 5.614(-1)  & -6.852(-2) & -2.464(-2) & 4.068(-3)  & 4.588(-4)  & 0.81 & 0.35 & (2.25) & 1.47 & (29.48)  & 1.484(-1)  & -5.616(-2) & -2.227(-2) & 7.746(-3)  & 1.215(-3)  & -3.250(-4) & -3.210 & 4.432  \\
C$^{18}$O / CN       & -6.033(0)  & -5.206(-1) & -7.988(-2) & 3.017(-3)  & 1.172(-2)  & -1.748(-3) & 0.81 & 0.36 & (2.27) & 1.24 & (17.32)  & 7.942(-2)  & 8.908(-2)  & 1.171(-2)  & -2.295(-2) & 4.440(-3)  & -2.265(-4) & -0.940 & 5.042  \\
CF$^+$ / DCO$^+$     & -6.565(0)  & 8.365(-1)  & -3.588(-1) & -2.491(-2) & 7.078(-2)  & -1.264(-2) & 0.79 & 0.37 & (2.35) & 1.76 & (57.35)  & 1.608(-1)  & -1.133(-1) & -8.819(-2) & 3.772(-2)  & 1.686(-2)  & -5.617(-3) & -1.640 & 3.318  \\
$^{13}$CO / CN       & -5.790(0)  & 1.174(-1)  & -4.659(-1) & 1.035(-1)  & -4.918(-3) & -3.338(-4) & 0.79 & 0.37 & (2.36) & 1.75 & (56.62)  & -1.861(-1) & 4.377(-1)  & -1.777(-1) & 3.415(-2)  & -3.900(-3) & 2.118(-4)  & 0.385  & 5.961  \\
C$_2$H / CF$^+$      & -7.099(0)  & 1.254(0)   & 1.560(-2)  & -3.191(-1) & 1.035(-2)  & 4.254(-2)  & 0.78 & 0.38 & (2.40) & 1.18 & (15.19)  & 1.911(-1)  & -1.025(-1) & -1.367(-1) & 9.891(-2)  & 2.349(-2)  & -1.613(-2) & -1.699 & 1.797  \\
SO / CN              & -6.333(0)  & -3.618(-1) & -6.588(-2) & 1.401(-2)  & 2.745(-3)  & -5.273(-4) & 0.78 & 0.38 & (2.40) & 1.22 & (16.73)  & 1.287(-1)  & 1.902(-2)  & 4.003(-3)  & 8.329(-4)  & -7.636(-4) & 4.753(-5)  & -3.358 & 5.549  \\
C$_2$H / HCO$^+$     & -6.097(0)  & 1.859(-1)  & -1.136(-1) & 5.531(-2)  & 2.500(-2)  & 2.227(-3)  & 0.77 & 0.39 & (2.46) & 1.19 & (15.35)  & 5.314(-2)  & 4.352(-2)  & -8.102(-2) & 1.599(-3)  & 1.375(-2)  & 1.961(-3)  & -4.577 & 1.813  \\
CS / SO              & -5.713(0)  & 3.742(-1)  & -1.150(0)  & -1.653(0)  & -1.125(0)  & -2.430(-1) & 0.76 & 0.40 & (2.49) & 1.79 & (62.25)  & 6.368(-2)  & 2.650(-1)  & 1.343(0)   & 1.135(0)   & 2.714(-1)  & 5.662(-3)  & -2.357 & -0.002 \\
C$_2$H / C$^{18}$O   & -5.166(0)  & 1.292(0)   & 9.232(-1)  & 3.319(-1)  & 4.751(-2)  & 2.334(-3)  & 0.75 & 0.41 & (2.55) & 1.31 & (20.65)  & -1.130(-1) & -3.125(-1) & -1.545(-1) & -1.194(-2) & 2.417(-3)  & 2.642(-4)  & -6.608 & -0.143 \\
C$_2$H / $^{13}$CO   & -1.759(0)  & 5.082(0)   & 2.513(0)   & 6.058(-1)  & 6.581(-2)  & 2.618(-3)  & 0.75 & 0.41 & (2.55) & 1.30 & (20.14)  & 4.727(-1)  & 3.387(-1)  & 1.868(-2)  & -3.289(-2) & -6.941(-3) & -3.864(-4) & -7.591 & -1.542 \\
SO / C$_2$H          & -5.851(0)  & -6.810(-2) & -6.714(-2) & -1.468(-2) & 5.136(-3)  & -3.241(-4) & 0.71 & 0.44 & (2.72) & 1.37 & (23.66)  & 5.503(-2)  & -2.381(-2) & 1.215(-2)  & 9.070(-3)  & -2.495(-3) & 1.532(-4)  & -2.114 & 8.543  \\
HCO$^+$ / CF$^+$     & -5.631(0)  & -6.854(-1) & 3.847(-1)  & -1.052(-1) & -1.078(-1) & 3.214(-2)  & 0.71 & 0.44 & (2.74) & 1.34 & (21.79)  & 6.275(-2)  & -7.457(-2) & -5.731(-1) & 8.260(-1)  & -3.242(-1) & 3.815(-2)  & -0.351 & 2.934  \\
H$_2$CS / C$_2$H     & -6.798(0)  & -5.504(-1) & 3.614(-3)  & 2.847(-2)  & -2.113(-4) & -5.451(-4) & 0.69 & 0.45 & (2.82) & 1.43 & (26.95)  & 2.221(-1)  & 7.335(-2)  & -1.322(-2) & -5.414(-3) & 4.551(-4)  & 7.322(-5)  & -3.475 & 4.518  \\
H$_2$CS / CN         & -7.685(0)  & -2.746(-1) & 1.395(-1)  & -2.263(-2) & -2.562(-2) & -3.222(-3) & 0.70 & 0.45 & (2.80) & 1.37 & (23.30)  & 1.732(-1)  & -2.424(-2) & 3.817(-2)  & -1.308(-2) & -1.419(-2) & -1.907(-3) & -4.966 & 1.603  \\
C$_2$H / HCN         & -5.412(0)  & 4.796(-1)  & 5.142(-1)  & 4.294(-1)  & 1.026(-1)  & 7.489(-3)  & 0.68 & 0.46 & (2.88) & 1.24 & (17.25)  & -2.534(-1) & -1.044(0)  & -1.024(0)  & -3.568(-1) & -5.272(-2) & -2.858(-3) & -4.447 & -0.266 \\
HCS$^+$ / CN         & -8.384(0)  & -9.575(-1) & -7.642(-1) & -3.411(-1) & -5.939(-2) & -3.510(-3) & 0.69 & 0.46 & (2.85) & 1.57 & (37.23)  & -9.782(-2) & -5.297(-1) & -4.655(-1) & -1.963(-1) & -3.530(-2) & -2.185(-3) & -6.784 & -0.246 \\
C$_2$H / HNC         & -5.444(0)  & 5.237(-1)  & 5.572(-1)  & 4.900(-1)  & 1.239(-1)  & 9.559(-3)  & 0.67 & 0.47 & (2.93) & 1.25 & (17.80)  & -1.168(-1) & -7.700(-1) & -9.054(-1) & -3.404(-1) & -5.148(-2) & -2.703(-3) & -4.146 & -0.105 \\
HNC / DCO$^+$        & -7.494(0)  & -1.407(0)  & 1.530(0)   & -5.464(-1) & 9.408(-2)  & -6.335(-3) & 0.67 & 0.47 & (2.92) & 1.76 & (58.10)  & -1.085(-2) & -1.019(-1) & 5.482(-1)  & -3.220(-1) & 6.689(-2)  & -4.698(-3) & 0.764  & 5.131  \\
C$^{18}$O / CF$^+$   & 1.388(1)   & -3.218(1)  & 2.018(1)   & -5.967(0)  & 8.227(-1)  & -4.278(-2) & 0.66 & 0.47 & (2.96) & 1.44 & (27.36)  & -2.665(0)  & 6.118(0)   & -4.845(0)  & 1.697(0)   & -2.677(-1) & 1.557(-2)  & 1.779  & 5.227  \\
$^{13}$CO / CF$^+$   & 1.006(2)   & -1.260(2)  & 5.833(1)   & -1.309(1)  & 1.419(0)   & -5.957(-2) & 0.66 & 0.48 & (2.99) & 1.43 & (26.82)  & 4.435(0)   & 5.410(-1)  & -2.778(0)  & 1.140(0)   & -1.736(-1) & 9.169(-3)  & 3.069  & 6.033  \\
HCN / DCO$^+$        & -6.217(0)  & -3.979(0)  & 3.334(0)   & -1.153(0)  & 1.913(-1)  & -1.224(-2) & 0.65 & 0.48 & (3.01) & 1.81 & (64.85)  & -5.703(-1) & 8.276(-1)  & -1.108(-1) & -7.561(-2) & 2.159(-2)  & -1.533(-3) & 1.009  & 5.207  \\
CS / CN              & -6.863(0)  & -5.631(-1) & -1.615(-2) & 4.874(-2)  & 1.232(-3)  & -2.008(-3) & 0.65 & 0.48 & (3.01) & 1.46 & (29.04)  & 3.074(-1)  & 1.578(-2)  & -3.348(-2) & 4.549(-3)  & 1.180(-3)  & -3.663(-4) & -3.742 & 3.316  \\
HCS$^+$ / C$_2$H     & -7.555(0)  & -2.507(-1) & 1.150(-1)  & -3.825(-3) & -8.251(-3) & -7.183(-4) & 0.64 & 0.49 & (3.08) & 1.57 & (36.90)  & 3.142(-1)  & -7.090(-2) & -4.661(-2) & 5.577(-3)  & 3.154(-3)  & 2.290(-4)  & -5.144 & 2.446  \\
HNC / N$_2$H$^+$     & -6.822(0)  & -3.863(0)  & 5.578(0)   & -2.720(0)  & 5.725(-1)  & -4.307(-2) & 0.63 & 0.49 & (3.10) & 2.29 & (195.89) & -3.782(-1) & 2.183(0)   & -2.288(0)  & 1.006(0)   & -2.056(-1) & 1.618(-2)  & 0.298  & 3.465  \\
CN / CF$^+$          & -8.292(0)  & 4.875(-1)  & 7.130(-2)  & -4.240(-1) & 2.855(-1)  & -4.862(-2) & 0.64 & 0.48 & (3.05) & 2.38 & (240.38) & 1.685(-3)  & 8.331(-1)  & -1.956(0)  & 1.580(0)   & -4.845(-1) & 4.920(-2)  & -0.065 & 3.401  \\
SO / HNC             & -6.355(0)  & -5.639(-1) & -1.313(-1) & 6.174(-2)  & 1.050(-2)  & -3.664(-3) & 0.62 & 0.50 & (3.18) & 1.54 & (34.41)  & 2.206(-1)  & 1.144(-1)  & -4.656(-3) & -1.157(-2) & -3.395(-4) & 3.928(-4)  & -2.846 & 3.755  \\
SO / HCN             & -6.466(0)  & -6.205(-1) & -9.568(-2) & 7.446(-2)  & 7.301(-3)  & -4.182(-3) & 0.61 & 0.51 & (3.21) & 1.53 & (33.98)  & 2.492(-1)  & 1.161(-1)  & -1.390(-2) & -1.235(-2) & 2.439(-4)  & 3.870(-4)  & -2.974 & 3.415  \\
CS / C$_2$H          & -6.028(0)  & -3.536(-1) & -9.409(-2) & 1.801(-2)  & 2.176(-3)  & -3.588(-4) & 0.60 & 0.52 & (3.28) & 1.59 & (39.23)  & 5.485(-2)  & 1.789(-1)  & 9.862(-3)  & -1.936(-2) & 3.054(-3)  & -1.403(-4) & -2.127 & 5.830  \\
SO / CF$^+$          & -5.813(0)  & 3.743(-2)  & 4.544(-2)  & -1.101(-1) & 2.504(-2)  & -1.587(-3) & 0.60 & 0.52 & (3.28) & 1.75 & (56.20)  & 4.402(-2)  & -2.032(-2) & 5.045(-3)  & 2.516(-2)  & -6.924(-3) & 4.744(-4)  & -0.735 & 7.064  \\
HCN / N$_2$H$^+$     & -6.045(0)  & -5.187(0)  & 5.600(0)   & -2.231(0)  & 3.893(-1)  & -2.431(-2) & 0.59 & 0.52 & (3.30) & 2.37 & (231.91) & -8.023(-1) & 2.766(0)   & -2.301(0)  & 8.246(-1)  & -1.379(-1) & 8.904(-3)  & 0.506  & 3.553  \\
HCO$^+$ / DCO$^+$    & 8.131(0)   & -4.927(1)  & 5.478(1)   & -2.793(1)  & 6.803(0)   & -6.392(-1) & 0.57 & 0.53 & (3.41) & 2.42 & (264.65) & -9.194(0)  & 2.540(1)   & -2.786(1)  & 1.525(1)   & -4.070(0)  & 4.175(-1)  & 1.074  & 2.103  \\
CF$^+$ / N$_2$H$^+$  & -5.896(0)  & 4.777(-1)  & -2.300(-1) & 3.024(-2)  & 2.751(-2)  & -1.222(-3) & 0.54 & 0.55 & (3.55) & 2.13 & (134.90) & 4.794(-2)  & -6.315(-1) & 2.671(-1)  & 3.412(-1)  & -6.066(-2) & -4.304(-2) & -2.535 & 1.680  \\
HCO$^+$ / HNC        & -7.496(0)  & -4.396(-1) & 7.013(-2)  & -1.139(-1) & -3.859(-2) & 9.718(-4)  & 0.53 & 0.56 & (3.61) & 2.48 & (302.40) & 3.875(-1)  & -1.868(-1) & -2.885(-1) & -8.864(-2) & -6.456(-3) & 1.098(-3)  & -2.711 & 0.660  \\
SO / H$_2$CS         & 1.413(1)   & -4.647(1)  & 4.143(1)   & -1.749(1)  & 3.456(0)   & -2.577(-1) & 0.51 & 0.57 & (3.68) & 1.81 & (64.14)  & -1.221(1)  & 2.515(1)   & -1.995(1)  & 7.625(0)   & -1.397(0)  & 9.857(-2)  & 1.502  & 3.882  \\
HCO$^+$ / HCN        & -7.578(0)  & -4.133(-1) & 5.389(-2)  & -2.386(-2) & 3.404(-2)  & 1.376(-2)  & 0.50 & 0.57 & (3.73) & 2.55 & (357.99) & 3.530(-1)  & -3.288(-1) & -4.007(-1) & -1.360(-1) & -1.345(-2) & 1.319(-3)  & -2.769 & 0.427  \\
DCO$^+$ / N$_2$H$^+$ & -8.336(0)  & -6.084(-2) & 1.634(0)   & -3.723(-2) & -6.371(-1) & -1.646(-1) & 0.50 & 0.57 & (3.75) & 2.16 & (146.16) & -1.468(-2) & 1.147(0)   & 4.358(0)   & 4.314(0)   & 1.752(0)   & 2.586(-1)  & -1.747 & 0.004  \\
C$^{18}$O / HNC      & -5.902(0)  & -8.434(-1) & -5.550(-1) & 6.929(-1)  & -2.584(-1) & 3.151(-2)  & 0.50 & 0.58 & (3.76) & 1.96 & (90.47)  & -2.582(-3) & 4.299(-1)  & 5.236(-1)  & -7.059(-1) & 2.471(-1)  & -2.807(-2) & -0.418 & 2.894  \\
        \hline 
      \end{tabular}
    \end{center}
  \end{sidewaystable*}

  \begin{sidewaystable*}
  \scriptsize
    \caption{Fit coefficients (for our main fit and scatter fit) and fit quality for line intensity ratios in cold dense medium conditions.
We list the fit coefficients for predicting $\log_{10}(x(\mathrm{e}^-)$ (according to the fit formula given in Eq.~\ref{eq:main_fit_dense}), the quality of this fit estimated as the (cross-validated) $R^2$, root mean square error (RMSE) on $\log_{10}(x(\mathrm{e}^-)$ and corresponding error factor on $x(\mathrm{e}^-)$, maximum absolute error factor on $\log_{10}(x(\mathrm{e}^-))$ and corresponding error factor on $x(\mathrm{e}^-)$, the fit coefficients for estimating the uncertainty on the prediction (according to the fit formula given in Eq.~\ref{eq:uncertainty_fit}), and the validity range of the fit (given as $\log_{10}$ of the ratio values).}
    \label{tab:RatiosIntensitiesDenseFullFitCoeff}
    \begin{center}
      \begin{tabular} {c|ccccccccccc|cccccc|cc}
        \hline 
        Line intensity ratio &   \multicolumn{6}{c}{Main fit coefficients}  & $R^2$ & \multicolumn{2}{l}{RMSE} & \multicolumn{2}{l}{Max. abs. error} &  \multicolumn{6}{c}{Scatter fit coefficients} &  \multicolumn{2}{l}{Validity limits} \\ 
         & $a_0$ & $a_1$ & $a_2$ & $a_3$ & $a_4$ & $a_5$  &           &  dex  & (equ. factor)  & dex  & (equ. factor) & $b_0$ & $b_1$ & $b_2$ & $b_3$ & $b_4$ & $b_5$ & min. & max. \\
\hline
CF$^+$ $(1-0)$ / DCO$^+$ $(1-0)$       & -6.135(0)  & 3.377(-1)  & -2.109(-1) & 1.007(-1)  & 1.967(-2)  & -9.665(-3) & 0.85 & 0.31 & (2.05) & 1.56 & (36.47)  & -2.594(-2) & 8.830(-2)  & -1.892(-2) & -2.429(-2) & 1.676(-3)  & 2.274(-3)  & -2.595 & 2.655  \\
$^{13}$CO $(1-0)$ / HCO$^+$ $(1-0)$    & -7.891(0)  & 1.249(0)   & 2.200(0)   & -3.222(0)  & 1.556(0)   & -2.556(-1) & 0.86 & 0.31 & (2.03) & 1.48 & (30.31)  & 1.278(-1)  & 4.071(-2)  & -2.886(-1) & 2.660(-1)  & -9.483(-2) & 1.212(-2)  & -0.256 & 2.245  \\
CN $(1-0)$ / N$_2$H$^+$ $(1-0)$        & -6.192(0)  & 5.913(-1)  & -2.320(-1) & 2.217(-2)  & 4.920(-2)  & 5.057(-3)  & 0.86 & 0.30 & (2.01) & 1.76 & (57.06)  & -7.328(-3) & 9.392(-2)  & -6.054(-2) & -3.017(-2) & 1.027(-2)  & 3.238(-3)  & -2.724 & 1.569  \\
C$_2$H $(1-0)$ / N$_2$H$^+$ $(1-0)$    & -5.845(0)  & 6.220(-2)  & 1.010(-1)  & 1.101(-1)  & 1.754(-2)  & 5.489(-4)  & 0.81 & 0.36 & (2.28) & 1.53 & (34.06)  & -3.781(-2) & 8.024(-2)  & 6.650(-2)  & 2.219(-3)  & -2.629(-3) & -2.851(-4) & -4.812 & 0.850  \\
HCO$^+$ $(1-0)$ / CF$^+$ $(1-0)$       & -4.345(0)  & -4.877(0)  & 5.810(0)   & -3.156(0)  & 7.136(-1)  & -5.632(-2) & 0.83 & 0.34 & (2.17) & 1.20 & (15.83)  & 1.481(-1)  & -2.119(-1) & -2.246(-2) & 1.454(-1)  & -5.678(-2) & 6.267(-3)  & 0.394  & 3.633  \\
C$_2$H $(1-0)$ / HCN $(1-0)$           & -1.549(0)  & 7.331(0)   & 5.080(0)   & 1.688(0)   & 2.482(-1)  & 1.316(-2)  & 0.81 & 0.35 & (2.25) & 1.62 & (41.39)  & -1.069(0)  & -1.616(0)  & -9.046(-1) & -2.325(-1) & -2.917(-2) & -1.464(-3) & -5.536 & -1.255 \\
$^{13}$CO $(1-0)$ / DCO$^+$ $(1-0)$    & -7.370(0)  & -2.501(0)  & 3.053(0)   & -1.165(0)  & 1.927(-1)  & -1.172(-2) & 0.82 & 0.35 & (2.23) & 1.86 & (73.12)  & -7.237(-1) & 1.605(0)   & -9.307(-1) & 2.022(-1)  & -1.423(-2) & -1.181(-4) & 0.756  & 5.025  \\
C$_2$H $(1-0)$ / HNC $(1-0)$           & -1.028(-2) & 9.830(0)   & 6.576(0)   & 2.094(0)   & 2.993(-1)  & 1.558(-2)  & 0.80 & 0.36 & (2.32) & 1.47 & (29.19)  & -1.383(0)  & -2.176(0)  & -1.200(0)  & -2.985(-1) & -3.579(-2) & -1.708(-3) & -5.182 & -1.288 \\
C$^{18}$O $(1-0)$ / DCO$^+$ $(1-0)$    & -7.844(0)  & -4.830(-1) & 1.819(0)   & -9.231(-1) & 1.932(-1)  & -1.474(-2) & 0.81 & 0.35 & (2.26) & 1.74 & (54.66)  & 2.188(-2)  & 5.469(-1)  & -5.232(-1) & 1.525(-1)  & -9.905(-3) & -9.963(-4) & 0.092  & 3.865  \\
C$_2$H $(1-0)$ / DCO$^+$ $(1-0)$       & -6.040(0)  & 1.684(-1)  & -7.619(-2) & 2.643(-2)  & 8.393(-3)  & 4.276(-4)  & 0.80 & 0.37 & (2.33) & 1.51 & (32.15)  & 1.618(-2)  & 5.016(-2)  & -3.621(-2) & -3.914(-3) & 3.470(-3)  & 5.088(-4)  & -5.195 & 2.316  \\
CF$^+$ $(1-0)$ / N$_2$H$^+$ $(1-0)$    & -5.718(0)  & 6.358(-2)  & -8.785(-2) & 1.745(-1)  & 2.604(-2)  & -5.295(-3) & 0.77 & 0.39 & (2.45) & 1.85 & (70.03)  & 6.231(-2)  & -9.758(-2) & -1.812(-1) & 4.513(-2)  & 6.121(-2)  & 1.077(-2)  & -3.235 & 0.988  \\
C$_2$H $(1-0)$ / HCO$^+$ $(1-0)$       & -5.226(0)  & 1.683(0)   & 1.393(0)   & 5.332(-1)  & 8.166(-2)  & 4.315(-3)  & 0.78 & 0.38 & (2.39) & 1.13 & (13.61)  & -1.098(-1) & -2.630(-1) & -1.013(-1) & 1.043(-2)  & 6.635(-3)  & 5.457(-4)  & -6.029 & -0.182 \\
CN $(1-0)$ / DCO$^+$ $(1-0)$           & -6.975(0)  & 5.955(-1)  & 1.245(-2)  & -2.396(-2) & -5.134(-4) & 1.198(-3)  & 0.78 & 0.38 & (2.40) & 1.59 & (38.68)  & 2.178(-1)  & -6.718(-2) & -6.819(-2) & 2.032(-2)  & 6.423(-3)  & -1.715(-3) & -2.563 & 3.009  \\
C$^{18}$O $(1-0)$ / HCO$^+$ $(1-0)$    & -7.081(0)  & 1.597(0)   & 3.871(-1)  & -8.281(-1) & -1.822(-1) & 2.346(-1)  & 0.75 & 0.40 & (2.53) & 1.33 & (21.38)  & 2.098(-1)  & -1.281(-1) & -1.887(-1) & 1.840(-1)  & 6.022(-2)  & -6.323(-2) & -1.069 & 1.247  \\
HCO$^+$ $(1-0)$ / CN $(1-0)$           & -6.004(0)  & -7.336(-1) & -1.856(-1) & 1.466(-1)  & -3.290(-2) & 2.461(-3)  & 0.75 & 0.41 & (2.54) & 1.85 & (70.06)  & 7.670(-2)  & 1.316(-1)  & 6.624(-2)  & -5.588(-2) & 2.769(-4)  & 1.978(-3)  & -0.758 & 3.627  \\
HCN $(1-0)$ / CN $(1-0)$               & -1.018(1)  & 1.609(1)   & -2.134(1)  & 1.173(1)   & -2.950(0)  & 2.796(-1)  & 0.75 & 0.41 & (2.54) & 2.12 & (130.92) & -3.482(0)  & 9.487(0)   & -8.900(0)  & 3.847(0)   & -7.821(-1) & 6.019(-2)  & 0.709  & 3.254  \\
HNC $(1-0)$ / CN $(1-0)$               & -1.461(1)  & 2.959(1)   & -3.630(1)  & 1.954(1)   & -4.900(0)  & 4.671(-1)  & 0.75 & 0.41 & (2.55) & 1.65 & (44.57)  & -1.763(0)  & 4.808(0)   & -4.132(0)  & 1.530(0)   & -2.379(-1) & 1.035(-2)  & 0.790  & 3.186  \\
SO $(3-2)$ / C$_2$H $(1-0)$            & -5.883(0)  & 1.034(-1)  & 3.381(-2)  & -9.485(-2) & 1.943(-2)  & -1.108(-3) & 0.72 & 0.43 & (2.69) & 1.33 & (21.50)  & 9.812(-2)  & -5.068(-2) & -4.320(-2) & 4.111(-2)  & -8.104(-3) & 4.591(-4)  & -1.179 & 6.475  \\
SO $(3-2)$ / CN $(1-0)$                & -5.963(0)  & -3.237(-1) & -3.231(-1) & 2.195(-2)  & 4.249(-2)  & -8.218(-3) & 0.70 & 0.45 & (2.80) & 1.52 & (33.48)  & 1.028(-1)  & 8.595(-2)  & 4.488(-2)  & -1.190(-2) & -8.902(-3) & 1.850(-3)  & -2.035 & 3.640  \\
CS $(2-1)$ / C$_2$H $(1-0)$            & -5.861(0)  & 3.386(-2)  & -6.629(-2) & -6.863(-2) & 2.034(-2)  & -1.455(-3) & 0.70 & 0.45 & (2.79) & 1.42 & (26.59)  & 6.795(-2)  & -1.756(-2) & 2.382(-3)  & 2.284(-2)  & -6.713(-3) & 4.937(-4)  & -1.434 & 6.536  \\
SO $(3-2)$ / $^{13}$CO $(1-0)$         & -7.396(0)  & -1.359(0)  & 5.844(-1)  & 1.010(0)   & 3.542(-1)  & 3.815(-2)  & 0.67 & 0.47 & (2.94) & 1.51 & (32.26)  & 2.849(-1)  & 7.080(-2)  & -2.698(-1) & -2.670(-1) & -8.877(-2) & -9.795(-3) & -3.128 & 0.524  \\
H$_2$CS $(3-2)$ / C$_2$H $(1-0)$       & -6.136(0)  & -4.376(-1) & -1.052(-1) & 2.297(-2)  & 4.201(-3)  & -7.259(-4) & 0.68 & 0.46 & (2.90) & 1.48 & (30.37)  & 8.699(-2)  & 1.545(-1)  & -3.579(-3) & -1.589(-2) & 2.899(-3)  & -1.406(-4) & -2.233 & 5.002  \\
$^{13}$CO $(1-0)$ / N$_2$H$^+$ $(1-0)$ & -7.687(0)  & -3.644(-1) & 1.591(0)   & -6.888(-1) & 1.063(-1)  & -4.774(-3) & 0.65 & 0.48 & (3.02) & 2.24 & (172.53) & 9.971(-2)  & 2.723(-1)  & 3.714(-1)  & -5.808(-1) & 2.048(-1)  & -2.165(-2) & -0.090 & 3.179  \\
SO $(3-2)$ / CF$^+$ $(1-0)$            & -5.838(0)  & 2.436(-1)  & -1.341(-1) & -1.769(-1) & 5.994(-2)  & -4.735(-3) & 0.63 & 0.49 & (3.12) & 1.51 & (32.21)  & 5.810(-2)  & -6.434(-3) & 1.630(-2)  & 7.035(-3)  & 2.821(-3)  & -1.160(-3) & -0.838 & 3.612  \\
CS $(2-1)$ / CN $(1-0)$                & -6.269(0)  & -4.353(-1) & -2.154(-1) & 3.254(-2)  & 2.412(-2)  & -5.177(-3) & 0.64 & 0.49 & (3.06) & 1.61 & (41.05)  & 2.456(-1)  & 1.704(-1)  & -6.298(-2) & -2.131(-2) & 7.203(-3)  & -3.129(-4) & -2.125 & 3.381  \\
C$_2$H $(1-0)$ / CF$^+$ $(1-0)$        & -5.877(0)  & 7.235(-1)  & -1.044(0)  & -1.034(0)  & -3.814(-1) & -4.969(-2) & 0.63 & 0.49 & (3.11) & 1.67 & (46.37)  & 4.812(-2)  & -3.066(-2) & 4.094(-1)  & 1.477(-1)  & -6.908(-2) & -2.336(-2) & -2.231 & 0.380  \\
HNC $(1-0)$ / DCO$^+$ $(1-0)$          & -7.445(0)  & -9.709(-1) & 1.543(0)   & -5.990(-1) & 1.047(-1)  & -7.187(-3) & 0.60 & 0.52 & (3.27) & 1.96 & (91.56)  & 7.342(-2)  & 2.833(-1)  & 2.507(-1)  & -3.250(-1) & 9.346(-2)  & -8.171(-3) & 0.152  & 3.964  \\
SO $(3-2)$ / HNC $(1-0)$               & -7.068(0)  & -1.115(0)  & 4.761(-1)  & 4.592(-1)  & -5.606(-3) & -2.580(-2) & 0.52 & 0.56 & (3.66) & 1.93 & (85.05)  & 3.942(-1)  & 7.041(-2)  & -2.838(-1) & -2.959(-2) & 9.907(-2)  & 2.645(-2)  & -1.408 & 1.023  \\
SO $(3-2)$ / HCN $(1-0)$               & -7.078(0)  & -1.083(0)  & 5.829(-1)  & 4.300(-1)  & -7.650(-2) & -4.289(-2) & 0.50 & 0.57 & (3.74) & 1.78 & (60.45)  & 4.082(-1)  & 1.217(-1)  & -2.744(-1) & -9.133(-2) & 6.756(-2)  & 2.311(-2)  & -1.864 & 1.181  \\
H$_2$CS $(3-2)$ / CN $(1-0)$           & -7.083(0)  & -6.115(-1) & 2.391(-2)  & 7.134(-2)  & -3.526(-3) & -4.954(-3) & 0.56 & 0.54 & (3.43) & 1.77 & (58.42)  & 4.373(-1)  & -1.234(-1) & -1.023(-1) & 4.550(-2)  & 5.825(-3)  & -3.198(-3) & -3.134 & 2.190  \\
HCN $(1-0)$ / DCO$^+$ $(1-0)$          & -7.472(0)  & -3.954(-1) & 5.704(-1)  & -1.092(-2) & -4.384(-2) & 6.173(-3)  & 0.56 & 0.54 & (3.45) & 1.96 & (92.06)  & 1.076(-1)  & 3.150(-1)  & 2.249(-1)  & -3.267(-1) & 9.620(-2)  & -8.474(-3) & 0.057  & 3.979  \\
C$_2$H $(1-0)$ / $^{13}$CO $(1-0)$     & -2.520(1)  & -2.705(1)  & -1.383(1)  & -3.247(0)  & -3.614(-1) & -1.540(-2) & 0.54 & 0.55 & (3.54) & 1.37 & (23.61)  & -9.851(-1) & -9.986(-1) & -5.750(-1) & -2.002(-1) & -3.112(-2) & -1.668(-3) & -5.006 & -2.101 \\
SO $(3-2)$ / C$^{18}$O $(1-0)$         & -6.758(0)  & -1.072(0)  & -1.279(-1) & 2.362(-1)  & 6.525(-2)  & -2.967(-4) & 0.54 & 0.55 & (3.55) & 2.14 & (138.67) & 3.792(-1)  & 2.370(-1)  & -3.064(-1) & -1.103(-1) & 6.841(-2)  & 2.218(-2)  & -2.133 & 1.288  \\
C$^{18}$O $(1-0)$ / N$_2$H$^+$ $(1-0)$ & -7.435(0)  & 5.433(-1)  & 4.916(-1)  & 4.396(-2)  & -1.683(-1) & 3.367(-2)  & 0.55 & 0.54 & (3.49) & 2.35 & (223.45) & 3.622(-1)  & 3.660(-1)  & -3.032(-1) & -2.749(-1) & 2.035(-1)  & -3.178(-2) & -0.869 & 2.399  \\
HNC $(1-0)$ / N$_2$H$^+$ $(1-0)$       & -7.479(0)  & -1.201(-1) & 2.498(0)   & -1.729(0)  & 4.207(-1)  & -2.895(-2) & 0.55 & 0.54 & (3.50) & 2.03 & (107.25) & 9.170(-2)  & 1.183(0)   & 1.686(-1)  & -2.126(0)  & 1.303(0)   & -2.215(-1) & -0.178 & 2.560  \\
HCS$^+$ $(2-1)$ / C$_2$H $(1-0)$       & -6.910(0)  & -4.969(-1) & 9.186(-2)  & 1.739(-2)  & -8.200(-3) & 4.164(-4)  & 0.52 & 0.56 & (3.66) & 1.72 & (52.43)  & 4.306(-1)  & 4.465(-2)  & -7.944(-2) & 3.195(-3)  & 4.950(-3)  & -6.888(-4) & -2.785 & 3.211  \\
HCO$^+$ $(1-0)$ / HNC $(1-0)$          & -7.297(0)  & -1.051(0)  & 1.048(0)   & 6.128(-1)  & -3.643(-1) & -1.887(-1) & 0.54 & 0.55 & (3.56) & 2.03 & (107.01) & 3.815(-1)  & 1.841(-1)  & -2.198(-1) & -2.694(-1) & -1.498(-1) & -3.694(-2) & -1.683 & 0.721  \\
C$_2$H $(1-0)$ / C$^{18}$O $(1-0)$     & -7.648(0)  & -5.022(0)  & -3.999(0)  & -1.225(0)  & -1.670(-1) & -8.382(-3) & 0.50 & 0.57 & (3.73) & 1.41 & (25.63)  & 4.605(-1)  & 5.719(-1)  & 2.782(-1)  & 1.040(-1)  & 1.839(-2)  & 1.092(-3)  & -4.824 & -1.091 \\
CS $(2-1)$ / CF$^+$ $(1-0)$            & -5.922(0)  & -1.527(-1) & -2.376(-1) & -5.469(-2) & 4.886(-2)  & -6.188(-3) & 0.52 & 0.56 & (3.64) & 2.05 & (111.81) & 1.150(-2)  & 2.021(-1)  & 8.627(-2)  & -4.137(-2) & -2.773(-3) & 1.295(-3)  & -1.449 & 3.578  \\
HCO$^+$ $(1-0)$ / DCO$^+$ $(1-0)$      & -1.911(0)  & -2.238(1)  & 3.036(1)   & -1.799(1)  & 5.000(0)   & -5.278(-1) & 0.52 & 0.57 & (3.67) & 2.53 & (341.17) & -1.131(0)  & 3.237(0)   & -3.531(0)  & 2.419(0)   & -8.588(-1) & 1.124(-1)  & 0.735  & 1.721  \\
HCO$^+$ $(1-0)$ / HCN $(1-0)$          & -7.307(0)  & -9.731(-1) & 1.345(0)   & 6.279(-1)  & -5.966(-1) & -2.808(-1) & 0.51 & 0.57 & (3.69) & 2.06 & (115.42) & 4.043(-1)  & 1.953(-1)  & -2.770(-1) & -2.153(-1) & -2.202(-2) & 5.842(-3)  & -1.646 & 0.759  \\
CS $(2-1)$ / HCS$^+$ $(2-1)$           & -2.476(1)  & 4.266(1)   & -3.509(1)  & 1.376(1)   & -2.687(0)  & 2.108(-1)  & 0.51 & 0.57 & (3.72) & 1.86 & (72.47)  & 1.533(1)   & -3.793(1)  & 3.431(1)   & -1.445(1)  & 2.903(0)   & -2.257(-1) & 1.659  & 3.594  \\
        \hline 
      \end{tabular}
    \end{center}
  \end{sidewaystable*}


\section{Analytical model visualization}
\label{app:fit_figures}

We present in this section figures of the RF models and analytical fits for the best six tracers in each case.

\subsection{Translucent medium}

\FigAnalyticalmodelBestSixTranslucentColDenRatios

\FigAnalyticalmodelBestSixTranslucentIntensityRatios

Figure~\ref{fig:AnalyticalmodelBestSixTranslucentColDenRatios} presents scatter plots of the ionization fraction versus each of the best six column density ratios in translucent medium conditions. 
Superimposed are the RF model (red line), the analytical fit (solid black line), the uncertainty fit (dashed lines), and the validity range (blue vertical lines).
Figure~\ref{fig:AnalyticalmodelBestSixTranslucentIntensityRatios} similarly shows scatter plots of  the ionization fraction versus each of the best six integrated line intensity ratios in translucent medium conditions. 

\subsection{Cold Dense medium}

Figure~\ref{fig:AnalyticalmodelBestSixDenseColDenRatios} and \ref{fig:AnalyticalmodelBestSixDenseIntensityRatios} similarly present the scatter plots (for column density ratios and integrated line intensity ratios  respectively) for the best six tracers in cold dense medium conditions.

\FigAnalyticalmodelBestSixDenseColDenRatios

\FigAnalyticalmodelBestSixDenseIntensityRatios

\end{document}